\documentclass{aa}
\usepackage[colorlinks=true,citecolor=blue,urlcolor=red,linkcolor=blue]{hyperref}
\usepackage{graphicx}
\usepackage{mathtools}
\usepackage{amsmath}
\usepackage[final]{pdfpages}
\usepackage{caption}
\usepackage{longtable}
\usepackage{threeparttable}
\usepackage{booktabs}
\usepackage{pdflscape}

\def\Hano{\hbox{H\,{$\alpha$}}}

\def\HeIInono{\hbox{He\,{\sc ii}}}
\def\Lya{\hbox{Lyman\,{$\alpha$}\,$\lambda$1216\AA}}
\def\Lyano{\hbox{Ly\,{$\alpha$}}}
\def\Lybno{\hbox{Ly\,{$\beta$}}}

\def\OIVnono{\hbox{O\,{\sc iv}}}

\def\NVnono{\hbox{N\,{\sc v}}}

\def\CII{\hbox{[C\,{\sc ii}]\,158\,$\mu$m}}
\def\CIIfull{\hbox{[C\,{\sc ii}]$\lambda$157.7\,$\mu$m\,($\rm ^2P_{3/2}\,\rightarrow\,^2P_{1/2}$)}}
\def\CIIno{\hbox{[C\,{\sc ii}]}}
\def\CIIsub{\hbox{[C\,{\sc ii}]$_{158}$}}

\def\LCIIno{\hbox{$L_{\rm [C\,{\sc II}]}$}}

\def\CIVnono{\hbox{C\,{\sc iv}}}

\def\OIIIcd{\hbox{[O\,{\sc iii}]$\lambda\lambda$4959,5007\AA}}

\def\H2{\hbox{H$_{2}$}}

\def\FEE{\hbox{$f_{\rm ext}$}}

\def\deg{$^{\circ}$}
\def\mic{$\mu$m}

\def\MHH{\hbox{$M_{\rm H_2}$}}

\def\Lsun{\hbox{$L_\odot$}}
\def\Msun{\hbox{$M_{\odot}$}}
\def\Zsun{\hbox{$Z_{\odot}$}}
\def\Mdyn{\hbox{$M_{\rm dyn}$}}

\def\MBHdot{\hbox{$M_{\bullet}$}}
\def\Msunny{\hbox{$M_{\odot}\,\rm yr^{-1}$}}
\def\Mstar{\hbox{$M_{\star}$}}
\def\Lbol{\hbox{$L_{\rm bol}$}}
\def\LEdd{\hbox{$L_{\rm Edd}$}}
\def\tSalp{\hbox{$t_{\rm Salp}$}}
\def\lamEdd{\hbox{$\lambda_{\rm Edd}$}}
\def\LIR{\hbox{$L_{\rm IR}$}}

\def\LX{\hbox{$L_{\rm X}$}}
\def\SigmaIR{\hbox{$\Sigma_{\rm IR}$}}

\def\SigmaCII{\hbox{$\Sigma_{\rm [CII]}$}}

\def\SFRIR{\hbox{SFR$_{\rm IR}$}}

\def\LFIR{\hbox{$L_{\rm FIR}$}}

\def\Lbol{\hbox{$L_{\rm bol}$}}

\def\Zsun{\hbox{$Z_{\odot}$}}

\def\nMpc{\hbox{Mpc$^{-1}$}}

\def\kmns{\hbox{km$\,$s$^{-1}$}}

\def\lsd{\hbox{$L{_\odot}\,$kpc$^{-2}$}}

\def\Rro23{\hbox{R$_{\tauro=2/3}$}}
\def\R23{\hbox{R$_{2/3}$}}

\def\T23{\hbox{T$_{2/3}$}}

\begin{document}

\title{Kinematics and Star Formation of High-Redshift\\
Hot Dust-Obscured Quasars as Seen by ALMA}
\titlerunning{Kinematics and Star Formation of Extremely Luminous Hot DOGs}
\authorrunning{D\'{\i}az-Santos et al.}


\author{Tanio~D\'{\i}az-Santos\inst{1,2,3},
Roberto~J.~Assef\inst{1}, Peter~R.~M.~Eisenhardt\inst{4}, Hyunsung~D.~Jun\inst{5}, Gareth.~C.~Jones\inst{6,7}, Andrew~W.~Blain\inst{8}, Daniel~Stern\inst{4}, Manuel~Aravena\inst{1}, Chao-Wei~Tsai\inst{9}, Sean~E.~Lake\inst{9}, Jingwen~Wu\inst{10,9}, and Jorge~Gonz\'alez-L\'opez\inst{1}
}


\institute{
$^1$ N\'ucleo de Astronom\'ia de la Facultad de Ingenier\'ia y Ciencias, Universidad Diego Portales, Av. Ej\'ercito Libertador 441, Santiago, 8320000, Chile.\\
\email{tanio@ia.forth.gr}\\
$^2$ Institute of Astrophysics, Foundation for Research and Technology--Hellas (FORTH), Heraklion, GR-70013, Greece.\\
$^3$ Chinese Academy of Sciences South America Center for Astronomy (CASSACA), National Astronomical Observatories, CAS, Beijing 100101, China.\\
$^4$ Jet Propulsion Laboratory, California Institute of Technology, 4800 Oak Grove Dr., Pasadena, CA 91109, USA.\\
$^5$ School of Physics, Korea Institute for Advanced Study, 85 Hoegiro, Dongdaemun-gu, Seoul 02455, Korea.\\
$^6$ Cavendish Laboratory, University of Cambridge, 19 J. J. Thomson Ave., Cambridge CB3 0HE, UK.\\
$^7$ Kavli Institute for Cosmology, University of Cambridge, Madingley Road, Cambridge CB3 0HA, UK.\\
$^8$ University of Leicester, Physics and Astronomy, University Road, Leicester LE1 7RH, UK.\\
$^9$ National Astronomical Observatories, Chinese Academy of Sciences, 20A Datun Road, Chaoyang District, Beijing 100012, China.\\
$^{10}$ University of Chinese Academy of Sciences, Beijing 100049, China.\\
}

\abstract{
Hot, dust-obscured galaxies (Hot DOGs) are a population of hyper-luminous obscured quasars identified by \textit{WISE}. We present ALMA observations of the \CII\, fine-structure line and underlying dust continuum emission in a sample of seven of the most extremely luminous (EL; \Lbol\,$\geq$\,10$^{14}$\,\Lsun) Hot DOGs, at redshifts \textit{z}\,$\simeq$\,3.0--4.6. The \CIIno\, line is robustly detected in four objects, tentatively in one, and likely red-shifted out of the spectral window in the remaining two based on additional data. On average, \CIIno\, is red-shifted by $\simeq$\,780\,\kmns\, from rest-frame ultraviolet emission lines. EL Hot DOGs exhibit consistently very high ionized gas surface densities, with \SigmaCII\,$\simeq$\,1--2\,$\times$\,10$^9$\,\lsd; as high as the most extreme cases seen in other high-redshift quasars. As a population, EL Hot DOG hosts seem to be roughly centered on the main-sequence of star forming galaxies, but the uncertainties are substantial, and individual sources can fall above and below. The average, intrinsic \CIIno\, and dust continuum sizes (FWHMs) are $\simeq$\,2.1\,kpc and $\simeq$\,1.6\,kpc, respectively, with a very narrow range of line-to-continuum size ratios, 1.61\,$\pm$\,0.10, suggesting they could be linearly proportional. The \CIIno\, velocity fields of EL Hot DOGs are diverse: from barely rotating structures, to resolved hosts with ordered, circular motions, to complex, disturbed systems that are likely the result of ongoing mergers. In contrast, all sources display large line-velocity dispersions, FWHM$_{\rm [CII]}$\,$\gtrsim$\,500\,\kmns, which on average are larger than optically and IR-selected quasars at similar or higher redshifts. We argue that one possible hypothesis for the lack of a common velocity structure, the systematically large dispersion of the ionized gas, and the presence of nearby companion galaxies may be that, rather than a single event, the EL Hot DOG phase could be recurrent. The dynamical friction from the frequent in-fall of neighbor galaxies and gas clumps, along with the subsequent quasar feedback, would contribute to the high turbulence of the gas within the host in a process that could potentially trigger not only one continuous EL, obscured event, but instead a number of recurrent, shorter-lived episodes as long as external accretion continues.
}

\keywords{galaxies: formation --- galaxies: evolution --- galaxies: high-redshift --- quasars: super-massive black holes --- galaxies: ISM --- infrared: galaxies}

\maketitle

\section{Introduction}\label{s:intro}


Accumulated evidence over the past two decades suggests that there is a broad relationship between the growth of massive galaxies and their central super-massive black holes (SMBHs); that is, a connection between the histories of star formation rate (SFR) in the former and the accretion rate of the latter. This is evidenced by the approximately synchronous co-evolution of these two galaxy properties over cosmic time \citep[][and references therein]{Madau2014}, which suggests that some sort of self-regulating process must be in action. Indeed, the high-end of the galaxy mass function displays a steep slope \citep{Muzzin2013, Davidzon2017} that requires a mechanism able to quench massive galaxies on short time scales. Feedback from luminous active galactic nuclei (AGN), also regularly labeled as ``quasar feedback'', is currently the preferred physical process with which theoretical models of galaxy evolution and cosmological simulations attempt to reproduce observations (\citealt{Croton2006, AA2017b}; cf., \citealt{Muzzin2012}). Quasar-mode feedback, which manifests itself in the form of galactic gas outflows, has been identified in a large number of \textit{z}\,$\gtrsim$\,2 optically and infrared-selected quasars (OpQs and IRQs, respectively) via fast-moving ($v_{\rm out}$\,$\geq$\,1000\,\kmns), ionized-gas (\OIIIcd) winds \citep{Zakamska2016, LaMassa2017, Bischetti2017, Wu2018, Perrotta2019, Temple2019, Jun2020a, Finnerty2020}.

Access to the kinematics of the interstellar medium (ISM) is important not only to identify the potential evacuation of material due to powerful quasar feedback, but also to characterize the dynamical state of the host. Investigating how gas is transferred and (re)cycled in high-redshift galaxies, as it is accreted from the cosmic web and expelled by the central AGN, is key to establishing whether outflows are actually able to quench ---or at least delay--- further star formation. Moreover, it allows us to assess the role of turbulence as an additional heating mechanism of the ISM gas in dense environments, where galaxy collisions may be frequent.

The Atacama Large Millimeter/sub-millimeter Array (ALMA) enables these studies by providing access to a range of the electro-magnetic spectrum of galaxies that is not affected by dust extinction. Thus, the physical properties and kinematics of the ionized, neutral, and molecular phases of the ISM can be studied without the degeneracies introduced by self-absorption \citep[with the exception of very extreme cases; e.g.,][]{Scoville2015}. This has allowed the systematic detection and analysis of the ISM of high-redshift quasars and star-forming galaxies, mostly via the fine-structure \CIIfull\, (\CIIsub) emission line and its underlying dust continuum, which are red-shifted into the ALMA highest frequencies bands at \textit{z}\,$\gtrsim$\,1 \citep[e.g.,][among many others]{WangR2013, Decarli2018, Venemans2018, Bischetti2018, DS2018, Hodge2019, Leung2019, JonesG2020}.

Within the general framework of galaxy evolution, nearby quasars are thought to be the product of gas-rich, massive galaxy mergers \citep{Sanders1988a, Hopkins2008a} ---a short-lived, ultra-luminous phase lasting a few tens of Myrs \citep{Hopkins2005, Trainor2013}. This phase is subsequently divided into two stages, separated by the expulsion of the gas and dust that has been accreted onto the central engine: obscured quasars harbor SMBHs still embedded in a dusty cocoon that absorbs most of their UV and optical light and re-emits it in the IR, while optically visible quasars have already cleared out the surrounding medium through the action of SMBH feedback and are visible in the optical.

An important population of high-redshift IRQs are hot, dust-obscured galaxies (Hot DOGs). Hot DOGs were discovered by the \textit{Wide-field Infrared Survey Explorer} \citep[\textit{WISE};][]{Wright2010}, selected to be strongly detected at 12 and 22\,$\mu$m, but weakly or not detected at 3.4 and 4.6\,$\mu$m \citep{Eisenhardt2012, Wu2012}. Most of their extreme bolometric luminosity, \Lbol\,$\gtrsim$\,10$^{13}$\,\Lsun, is produced by accretion onto their central SMBH. The host galaxies are on average less massive than what would be expected from such hyper-luminous AGN, suggesting they are radiating at ---or even above--- the Eddington limit \citep{Assef2015, Wu2018, Tsai2018}.

Statistically, Hot DOGs seem to be located in galaxy over-densities \citep{Jones2014, Assef2015} and host powerful ionized outflows \citep{Wu2018, Jun2020a, Jun2020b, Finnerty2020}, in agreement with the merger-driven model of quasars and their expected feedback, and also suggesting that they may be at the centers of proto-clusters. Further evidence comes from ALMA \CIIsub\, observations of the most luminous obscured quasar and Hot DOG known (\Lbol\,=\,3.5\,$\times$\,10$^{14}$\,\Lsun; \citealt{Tsai2015}), WISE~J224607.55--052634.9 (W2246--0526 hereafter; at \textit{z}\,=\,4.601), which revealed a highly turbulent ISM, with a FWHM$_{\rm [CII]}$\,$\geq$\,500\,\kmns\, spread across the entire host galaxy, over $\sim$\,2.5\,kpc \citep{DS2016}. Furthermore, recent deep ALMA imaging of the dust continuum at $\lambda_{\rm rest}$\,=\,212\,$\mu$m of W2246--0526 shows streamers and bridges of dust extending up to $\sim$\,35\,kpc in projected size, which seem to connect at least three companion galaxies with the central Hot DOG. These results suggest the most luminous obscured quasars may be interacting systems ---the result of ongoing merger-driven peaks of SMBH accretion and massive galaxy assembly in the early Universe \citep{DS2018}.

In this work we study a sample of seven of the most luminous Hot DOGs, at redshifts \textit{z}\,$\sim$\,3.0--4.6, observed with ALMA in the \CIIsub\, line. In Section~\ref{s:obs} we introduce the ALMA observations, ancillary data, and comparison samples of other high-\textit{z} OpQs and IRQs. In Section~\ref{s:analysis} we describe the analysis of the data, and in Section~\ref{s:integrated} we present results from the integrated, line and continuum measurements, including the assessment of the global ISM conditions. In Section~\ref{s:kinematics} we introduce the results from the kinematic analysis and resolved morphology, and in Section~\ref{s:ms} we place the Hot DOG hosts in the context of the star-formation galaxy main sequence. Finally, in Section~\ref{s:discussion} we speculate on the implications of these results and discuss the effect of the SMBH on the physics of the surrounding gas and dust, and how Hot DOGs fit within the broader, merger-driven quasar paradigm. The conclusions are summarized in Section~\ref{s:summary}.

Throughout the paper, the following flat cosmology is used: $\Omega_{\rm M}$\,=\,0.28, $\Omega_{\rm \Lambda}$\,=\,0.72, H$_{\rm 0}$\,=\,70\,\kmns\,\nMpc.

\section{Observations}\label{s:obs}

\subsection{The sample}\label{ss:sample}

Our sources are drawn from a catalog of over 2,000 Hot DOGs, photometrically selected from the \textit{WISE} All-Sky Data Release \citep{Cutri2012} using the selection criteria presented by \citet{Eisenhardt2012} (see also \citealt{Assef2015} for details). A subset of the sample have secure redshifts from rest-UV spectroscopic observations that will be presented by \citet{Eisenhardt2020}. We obtained ALMA observations of seven targets selected from this catalog that have declinations accessible to ALMA and are at rest-UV redshifts that put the \CIIsub\ emission line (\CIIno, hereafter) within the frequency range of ALMA's band 8 and 7, in regions of high atmospheric transparency. This pre-selection implies the targets probe the high-redshift end of the Hot DOG population (i.e., concentrated in the 3\,$\lesssim$\,\textit{z}\,$\lesssim$\,4.6 range; see Fig. 1 of \citealt{Assef2015}) and thus also the high end of the luminosity distribution (in the ``extremely'' luminous, EL, regime), as they all have \Lbol\,$\geq$\,10$^{14}$\,\Lsun\, \citep{Tsai2015}. Basic information about the ALMA Hot DOG sample can be found in Table~\ref{t:sample}. The observations for the most distant and most luminous Hot DOG currently known, W2246--0526, have already been presented in \citet{DS2016}.

\begin{figure}
\includegraphics[width=\hsize]{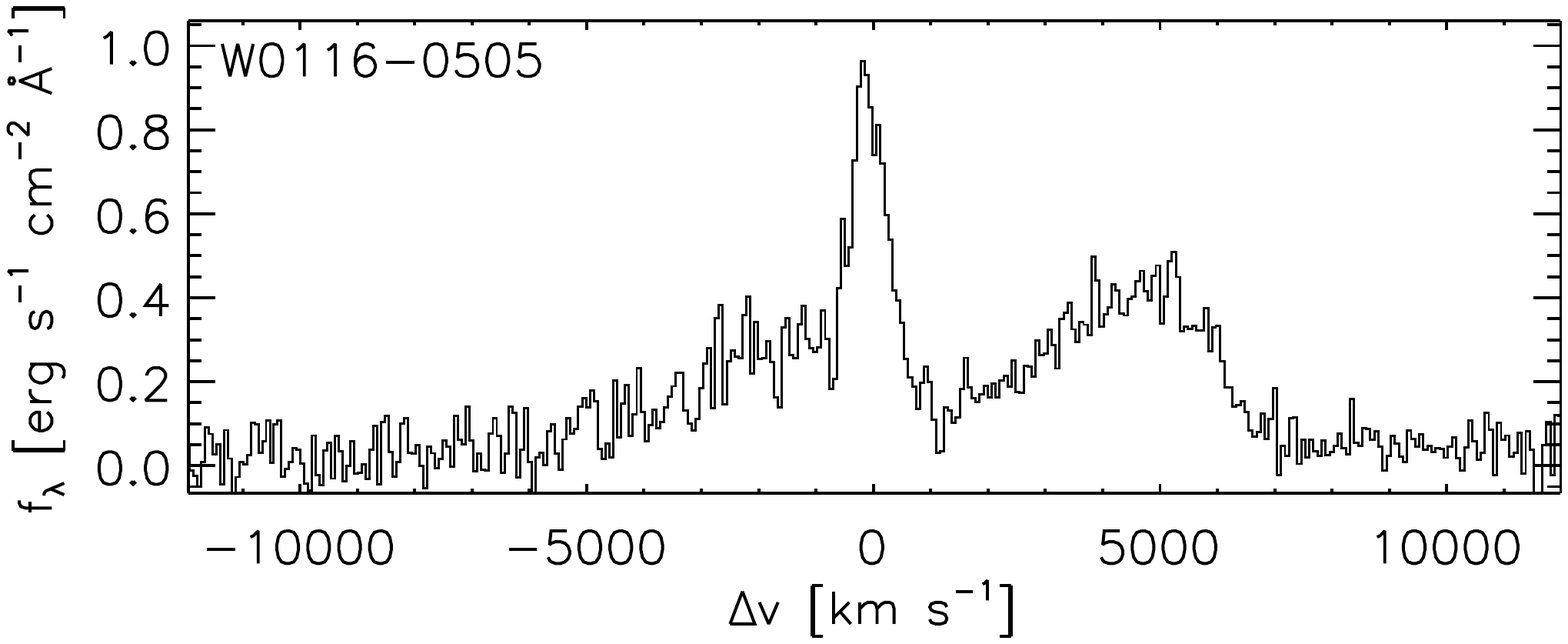}
\vspace{-8.3cm}

\includegraphics[width=\hsize]{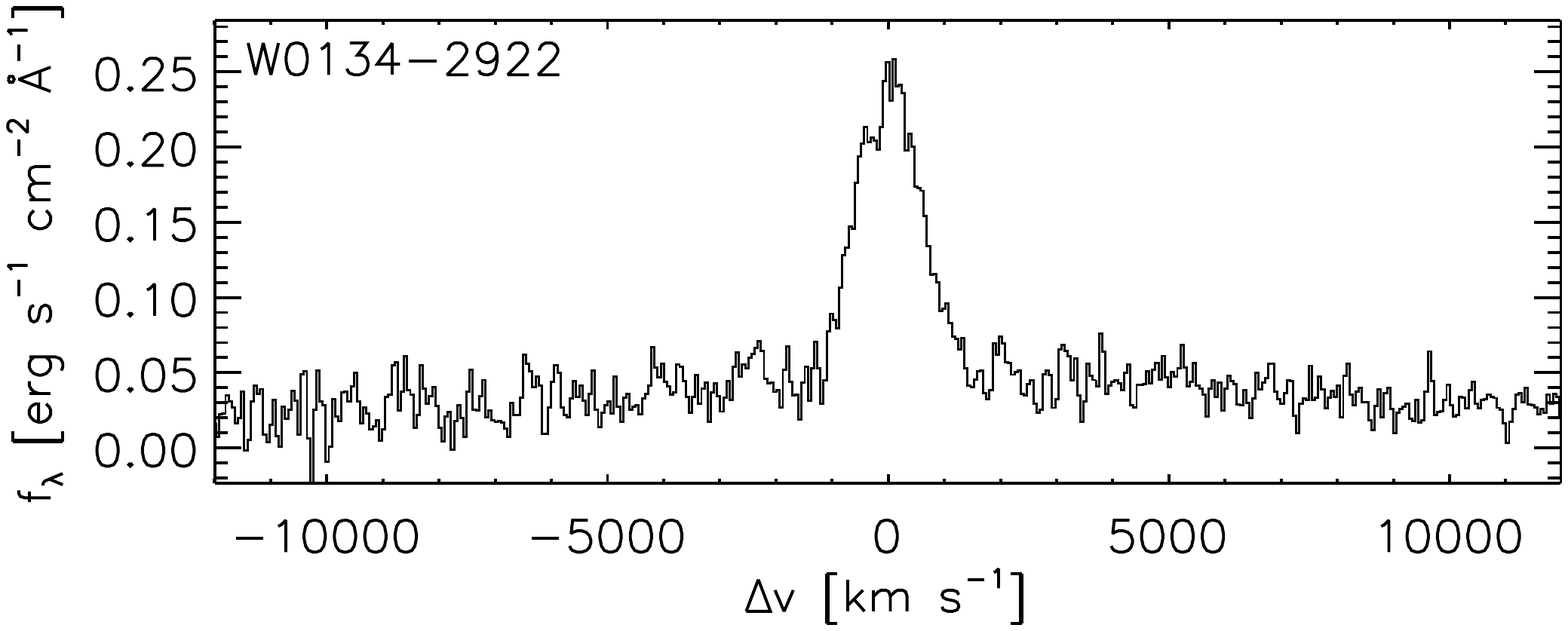}
\vspace{-8.3cm}

\includegraphics[width=\hsize]{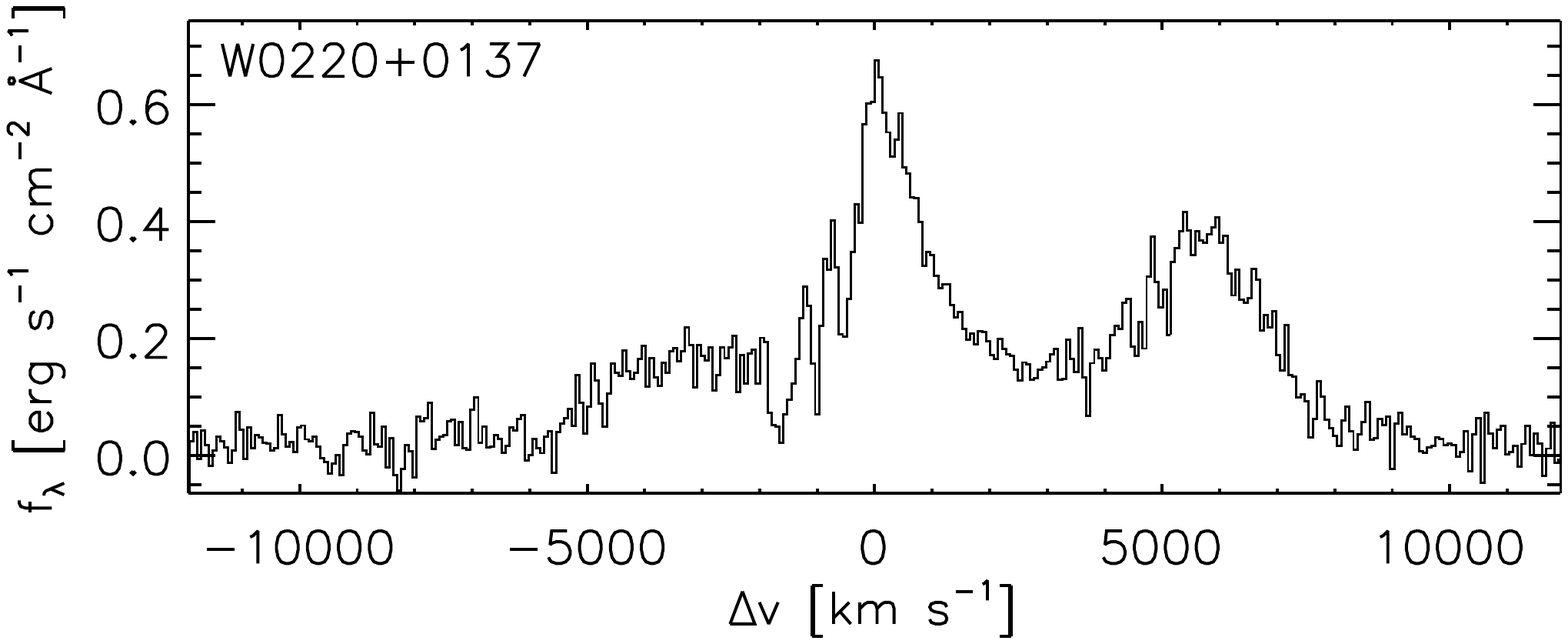}
\vspace{-8.3cm}

\includegraphics[width=\hsize]{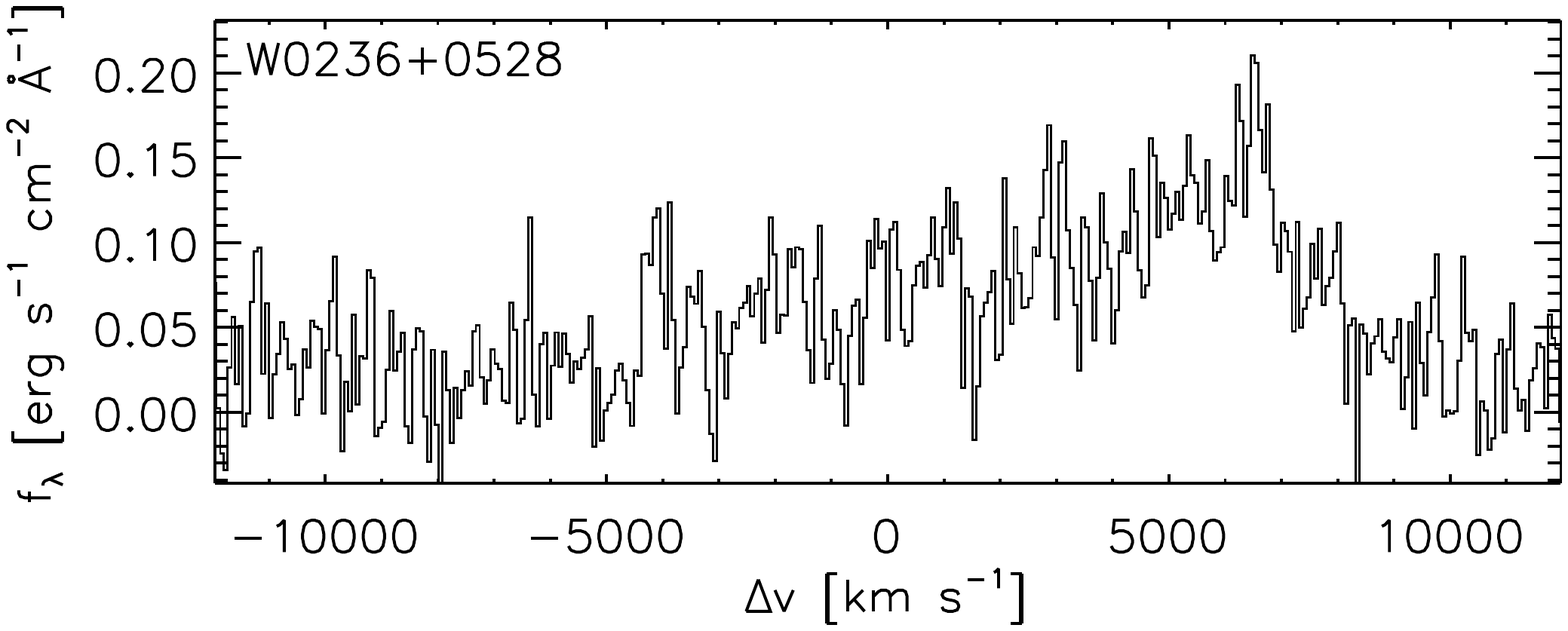}
\vspace{-8.3cm}


\includegraphics[width=\hsize]{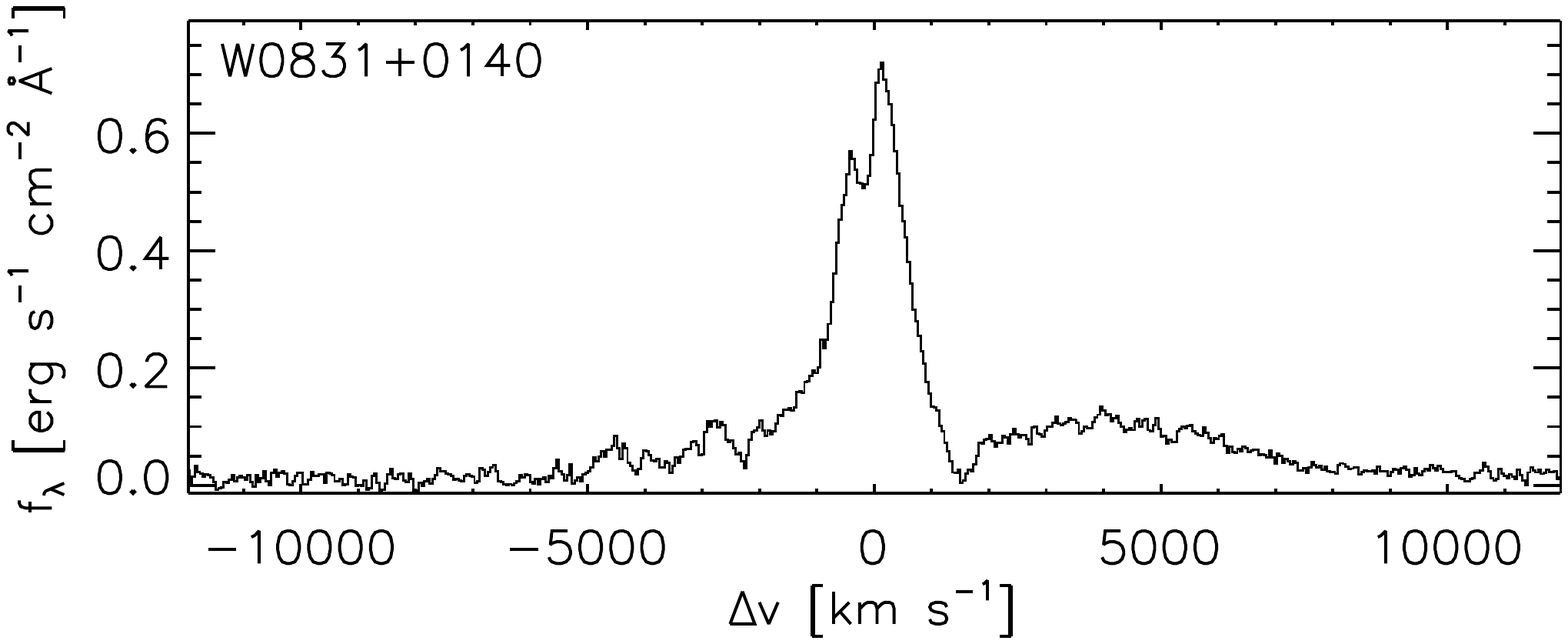}
\vspace{-8.3cm}

\includegraphics[width=\hsize]{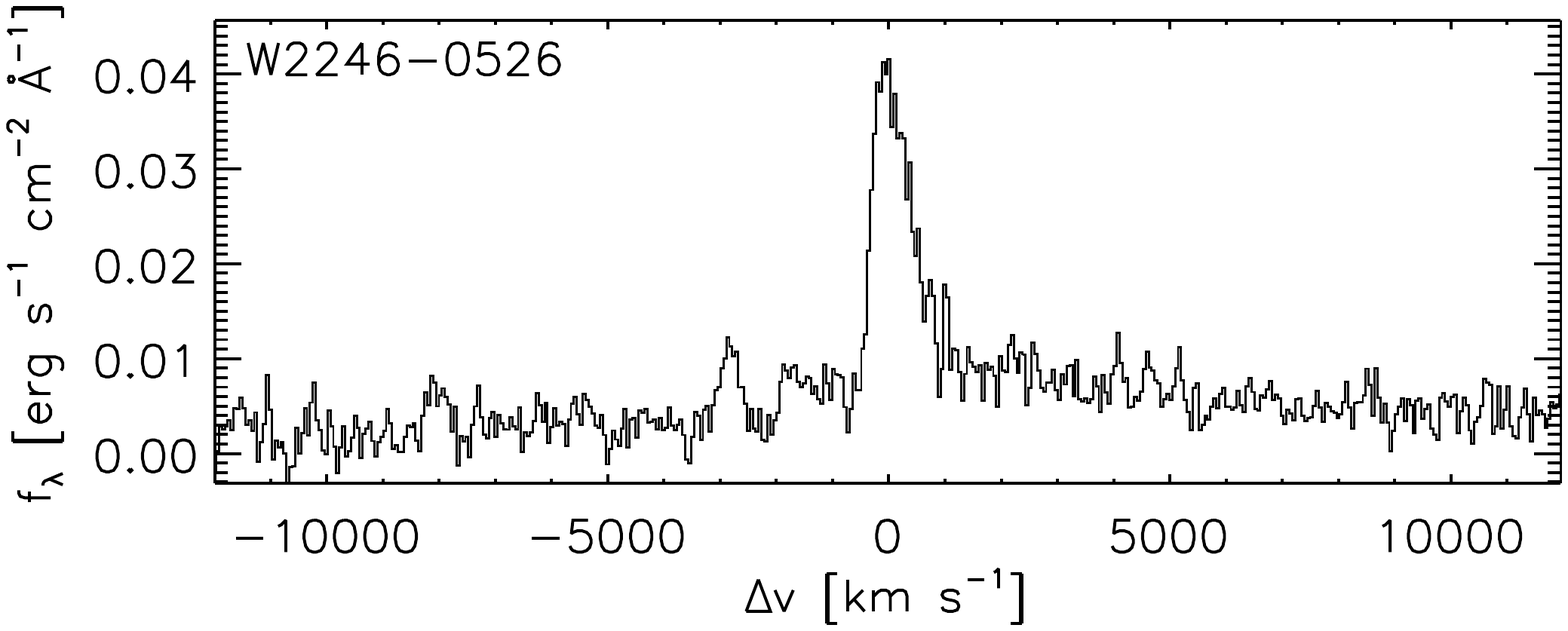}
\vspace{-8.1cm}
\caption{\footnotesize Optical spectra of the Hot DOG sample around the Ly-$\alpha$ emission line, whose rest-frame is set at zero velocity using \textit{z}$_{\rm ALMA}$ (Table~\ref{t:sample}). The feature located at $\sim$\,5000\,\kmns\, is the N\,V$\lambda$1240\AA\, emission line. The spectrum of W0410--0913 is not shown since Ly-$\alpha$ landed in a gap between detector chips.}
\label{f:optspec}
\end{figure}

\begin{table*}
\caption{Extremely luminous Hot DOG sample: Basic properties}
\scriptsize
\centering
\label{t:sample}
\begin{tabular}{cccccccccc}
\hline\hline
Name & R.A & Dec. & $z_{\rm UV}$ & $z_{\rm ALMA}$ & $\Delta v$ & Ang. Scale & \LIR & \Lbol \\ 
 & (J2000) & (J2000) &  &  & [\kmns] & [kpc/\arcsec] & [10$^{13}$\,\Lsun] & [10$^{13}$\,\Lsun] \\
(1) & (2) & (3) & (4) & (5) & (6) & (7) & (8) & (9) \\
\hline
W0116--0505 & 01h16m01.41s & --05d05m04.1s & 3.173$^a$ & 3.1904$^c$       & 1250 & 7.71 & 8.2  & 11.7 \\
W0134--2922 & 01h34m35.69s & --29d22m45.4s & 3.0579$^b$ & 3.0574           & --44 & 7.82 & 6.2  & 11.3 \\
W0220+0137  & 02h20m52.12s & +01d37m11.6s  & 3.122$^a$ & 3.1356           &  989 & 7.76 & 9.6  & 12.9 \\
W0236+0528  & 02h36m31.59s & +05d28m03.1s  & 2.958$^b$ & 2.9559$^\dagger$ &  \dots & 7.88 & \dots  & \dots  \\
W0410--0913 & 04h10m10.60s & --09d13m05.2s & 3.592$^a$ & 3.6301$^d$       & 2487 & 7.38 & 11.3 & 16.8 \\
W0831+0140  & 08h31m53.25s & +01d40m10.8s  & 3.9128$^b$ & 3.9136           &   98 & 7.17 & 12.0 & 18.0 \\
W2246--0526 & 22h46m07.55s & --05d26m35.0s & 4.6021$^b$ & 4.6009           &  423 & 6.68 & 22.1 & 34.9 \\
\hline\hline
\end{tabular}
\tablefoot{\scriptsize (1) Galaxy name; (2) Right ascension; (3) Declination; (4) Redshift derived from rest-frame UV emission lines: $^a$ \cite{Wu2012}, $^b$ \cite{Eisenhardt2020}; (5) Redshift derived from \CIIno, except $^{c}$ CO(4$\rightarrow$3) and $^{d}$ CO(6$\rightarrow$5); (6) Velocity shift between \textit{z$_{\rm ALMA}$} (either \textit{z$_{\rm CO}$} or \textit{z$_{\rm [CII]}$}) and \textit{z$_{\rm UV}$}, with $\Delta v$\,=\,$c(\textit{z$_{\rm IR}$}-\textit{z$_{\rm UV}$})/(1+\textit{z$_{\rm UV}$})$; (7) Angular scale; (8) Infrared luminosity (8-1000\,\mic); (9) Bolometric luminosity.\\
$\dagger$ The line in W0236+0528 is a tentative detection (see Section~\ref{ss:lineoffsets}).}
\end{table*}

\subsection{ALMA observations and data processing}\label{ss:almaobs}

ALMA targeted the \CIIno\, emission line for the sample of seven Hot DOGs under programs 2013.1.00576.S (W0116--0505, W0134--2922, W0220+0137 and W2246--0526) and 2015.1.00612.S (W0236+0528, W0410--0913 and W0831+0140), both PI: R. Assef. The observations were carried out following the standard observatory calibration procedure under good weather conditions (i.e., a typical precipitable water vapor, PWV\,$\simeq$\,0.4--0.6\,mm). More details regarding these observations can be found in Table~\ref{t:observations}. For all sources except W0236+0528, the data were acquired in a single execution block (EB). For W0236+0528, two EBs were obtained, which were concatenated into a single measurement set.

Owing to their redshifts, W2246--0526 was observed in band 7 and the remaining galaxies in band 8. While the requested sensitivity of the observations was 1\,mJy\,beam$^{-1}$ over 100\,\kmns\, channels for the band 8 targets and 1\,mJy\,beam$^{-1}$ over 20\,\kmns\, channels for W2246--0526, the final data were deeper in most cases. Similarly, the requested angular resolution was 0.24\arcsec\, for the band 8 targets and 0.32\arcsec\, for W2246--0526, but higher resolutions ($<$\,0.2\arcsec) were achieved for W0236+0528 and W0831+0140 (see Tables~\ref{t:linefluxes} and \ref{t:contfluxes}). 

The center of the reference spectral window (SPW) was tuned to match the expected observed (redshifted) frequency of \CIIno\, for each source. The redshifts were based on optical spectroscopic (rest-frame UV) observations, whose reduction and analysis will be presented and discussed in \cite{Eisenhardt2020}. The spectral region around the \Lya\, line is shown in Figure~\ref{f:optspec} for all seven galaxies except W0410--0913, for which \Lya\, landed in the gap between detector chips of the instrument (Keck/DEIMOS). The rest-UV redshfits, \textit{z$_{\rm UV}$}, were estimated based on the central wavelength(s) of the available line(s), which were fitted using a Gaussian profile. The lines used depend on the source spectrum, but for the Hot DOGs included here are generally based on \Lyano, the \Lybno/\OIVnono\, blend, and \NVnono. \CIVnono\, was excluded except for W0410--0913, whose UV redshift is also based on the \HeIInono\, line (see discussion in Section~\ref{ss:lineoffsets}). The reference SPW was normally the reddest of the two within the side-band, which in turn was the reddest of the two side-bands. However, adjustments to this configuration were made due to strong absorption lines and the cut-off of the atmospheric band. The SPWs not covering the \CIIno\, line were used to sample the dust continuum.

The scripts provided by ALMA as part of the data delivery were executed using the appropriate version of the Common Astronomy Software Application \citep[CASA;][]{McMullin2007} to process each specific raw dataset and calibrate the visibilities. The resulting measurement sets were inspected to check for possible issues in the amplitude, band-pass and phase calibrations. All four SPWs were usable for all sources and no additional flagging was needed.

We also used CASA (v5.4.0-70) to process and clean the ALMA products. The cleaning algorithm was run using three different Briggs weighting schemes for the \textit{uv} visibility plane: (1) robust parameter set to 0.5 (nearly half-way between uniform and natural weighting); (2) robust parameter set to 2 (similar to natural weighting); and (3) robust parameter set to 2 and a \textit{uv}-taper with a resolution approximately twice that provided by scheme (2). The cleaning area was usually a circle of 1\arcsec\, diameter centered on the object, but elliptical shapes of varying sizes were used when necessary. The angular size (FWHM) of the restored, synthesized beams at the frequency of the \CIIno\, line range from $\sim$\,0.35\arcsec\, for W2246--0526 to $\sim$\,0.15\arcsec\, for W0236+0528 and W0831+0140 (considering the weighting scheme (2)). The spaxel size of the cubes was selected such that the beam is sampled across by 5--10 spaxels. The native channels were averaged to a width of $\sim$\,50\,\kmns\, to maximize the signal-to-noise ratio (SNR) while keeping the final spectral elements sufficiently small to derive kinematic information. The line and continuum cubes were cleaned down to a level of 2\,$\sigma_{\rm rms}$ over the general background\footnote{W0410--0913 and W0831+0140 had to be cleaned down to a levels of 0.5\,$\sigma_{\rm rms}$, respectively, as patterns from the dirty beam were still clearly identifiable in the residual images when cleaning down to 2.0\,$\sigma_{\rm rms}$.}, where $\sigma_{\rm rms}$ is the average r.m.s. per channel of the respective SPW\footnote{The average r.m.s. used to clean the reference SPW containing the line was normally that of the SPW used to probe the dust continuum adjacent to the line.}, calculated after masking out the object with the same aperture used later for the cleaning process. More technical details about the observations can be found in Table~\ref{t:observations}.

\begin{table*}
\caption{ALMA observations}
\centering
\scriptsize
\label{t:observations}
\begin{tabular}{ccccccccc}
\hline\hline
Galaxy & Date & Antennas & Corr. & Time & Pixel & Channel & Spec. depth & Cont. depth \\ 
 &  &  & mode & on-source & scale & width & [mJy] & [$\mu$Jy] \\
 & [yyyy/mm/dd] &  &  & [min] & [\arcsec] & [\kmns] & beam$^{-1}$]  & beam$^{-1}$]  \\
(1)  & (2)  & (3)  & (4)  & (5)  & (6)  & (7)  & (8)  & (9) \\
\hline
W0116--0505 & 2015/06/08 & 38 & FDM & 35.57 & 0.025 & 52 & 1.1 & 127 \\
W0134--2922 & 2015/06/13 & 37 & FDM & 32.97 & 0.025 & 50 & 1.4 & 162 \\
W0220+0137  & 2016/06/14 & 38 & FDM & 20.13 & 0.025 & 51 & 2.5 & 340 \\
W0236+0528  & 2016/06/18 & 37 & TDM & 31.88 & 0.025 & 53 & 1.1 & 149 \\
            & 2016/07/22 & 42 & TDM & 23.67 &       &    &     &     \\
W0410--0913 & 2016/06/30 & 40 & TDM & 11.30 & 0.025 & 57 & 0.9 & 122 \\
W0831+0140  & 2016/08/25 & 39 & TDM & 33.45 & 0.025 & 61 & 0.8 & 106 \\
W2246--0526 & 2014/06/29 & 32 & FDM & 17.50 & 0.050 & 35 & 0.6 &  68 \\
\hline\hline
\end{tabular}
\tablefoot{\scriptsize (1) Galaxy name; (2) UT date of the observations; (3) Number of antennas used during the observations; (4) Correlator mode; (5) Integrated time on source; (6) Pixel scale of the cube; (7) Channel width; (8) Depth (r.m.s.) per channel; (9) Depth (r.m.s.) of the collapsed data cube.}
\end{table*}

\subsection{Comparison samples}\label{ss:compdata}

\subsubsection{High-redshift Quasars}\label{sss:compqsos}

For comparisons with high-redshift optically selected quasars (OpQs) we use the \CIIno\, data compiled by \citet{Decarli2018}, which includes originally published sources, supplemented with OpQs from the literature, all at \textit{z}\,$\gtrsim$\,6, including \citet{Maiolino2005}, \citet{Walter2009}, \citet{Venemans2012}, \citet{WangR2013}, \citet{Willott2013}, \citet{Banados2015}, \citet{Willott2015}, \citet{Venemans2016}, \citet{WangR2016}, \citet{Mazzucchelli2017}, \citet{Venemans2017} and \citet{Willott2017}. In addition, we use the results from the dust continuum data analysis published in \cite{Venemans2020} for the same sample of OpQs .

We also use the sample of high-redshift IR quasars (IRQs) from \cite{Trakhtenbrot2017}, which has been recently expanded and updated by \cite{Nguyen2020}. We note that while these are also optically identified quasars found in the SDSS, before the sample was observed by ALMA, the sources were selected \textit{a priori} to have significant (if not strong) far-IR continuum emission as detected by \textit{Herschel}. Most of these IRQs are located at \textit{z}\,$\simeq$\,4.8.

\subsubsection{High-redshift star-forming galaxies}

To put the Hot DOG hosts in context with purely star-forming galaxies, we use the ALMA Large Program to INvestigate \CIIno\, at Early times survey \citep[ALPINE;][and references therein]{LeFevre2020}, which observed more than a hundred galaxies at 4.4\,$\lesssim$\,\textit{z}\,$\lesssim$\,5.9 in \CIIno\, and dust continuum emission. ALPINE sources are representative of the star-forming galaxy main-sequence (MS) at their respective redshifts.

\subsubsection{GOALS: Nearby IR galaxies and AGN}\label{sss:lowzdata}

We use the Great Observatories All-sky LIRG Survey \citep[GOALS;][]{Armus2009} as the reference sample for nearby galaxies. GOALS is a complete sample of Luminous and Ultra-luminous Infrared Galaxies ((U)LIRGs; \LIR\,=\,10$^{11-12}$ and $\geq$\,10$^{12}$\,\Lsun, respectively) in the nearby Universe (\textit{z}\,$\leq$\,0.1). The sample has extensive multi-frequency photometric and spectroscopic coverage from space- and ground-based facilities, from X-rays \citep{Iwasawa2011, TA2018}, through ultraviolet \citep{Howell2010}, optical \citep{Rich2015}, near-IR \citep{Inami2018}, mid-IR \citep{Stierwalt2013, Stierwalt2014, DS2010b, DS2011}, far-IR \citep{DS2017}, and sub-mm \citep{HI2019}, to radio wavelengths \citep{BM2017, Linden2017}. GOALS bridges the gap between MS galaxies and the most powerful starbursts and AGN in the local Universe, and therefore it is useful to compare with any IR galaxy population identified at high redshift.

\subsubsection{Data homogenization}

Throughout the paper, physical areas are calculated as $\pi$\,$R^2_{\rm 160\mu m, eff}$, where $R_{\rm 160\mu m, eff}$ is the intrinsic (PSF-corrected) 160\,\mic\, dust continuum emission half-light radius (=\,FHWM/2, for a 2D Gaussian). These were obtained from \textit{Herschel} photometry \citep{Lutz2016, Chu2017} for the local (U)LIRG sample, and from ALMA observations for the IRQ sample of \citet{Nguyen2020} and the Hot DOG sample presented here.

\citet{Venemans2018} first reported dust continuum sizes for the OpQ sample of \cite{Decarli2018}. However, most sources were unresolved and/or the data had limited SNR with which to calculate an accurate size of the emitting region. We thus use the updated values recently published in \citet{Venemans2020}, which are based on higher angular resolution data. The continuum sizes reported in the latter work tend to be significantly smaller than those listed in the former (sometimes by a factor of $\sim$\,3), likely reflecting the improved resolution of the more recent observations and the fact that a fraction of the more extended continuum emission may have been resolved out. Due to a lack of data coverage at shorter IR wavelengths than the ALMA observations, only \LFIR\, values are available for some of the OpQs. In these cases, we use the average \LIR/\LFIR\,$\simeq$\,1.7 ratio found for nearby (U)LIRGs to estimate their \LIR.

IR luminosities are available for only the 23 sources in the ALPINE survey that were detected in dust continuum \citep{Bethermin2020}. In order to plot the remainder of the sources on our figures, we estimate a “virtual" \LIR\, from their UV-based SFRs using the standard conversion given in \cite{Murphy2009}. We emphasize that these are not IR luminosities \textit{per se}, but rather represent what the \LIR\, of these galaxies would be if all their star formation was obscured. \cite{Fujimoto2020} do not provide sizes for the dust continuum detections, and therefore we use the \CIIno\, sizes to calculate luminosity surface density quantities. Since these are mostly star-forming galaxies, no correction is performed to their \CIIno\, sizes, such that the line and continuum sizes are assumed to be the same.

\begin{sidewaystable}
\caption{[CII]$_{\rm 158\mu m}$ measurements}
\centering
\scriptsize
\label{t:linefluxes}
\begin{tabular}{ccccccccccccc}
\hline\hline
Galaxy & $L_{\rm [C\,II]}^{\rm data}$ & $L_{\rm [C\,II]}^{\rm model}$ & $\Delta v$ & FWHM$_{\rm spec}$ & $|v_{\rm max}|_{\rm m1}$ & Beam & PA & Intrinsic Size & \multicolumn{2}{c}{Circularized Size} & PSF & Extended \\ 
\cline{10-11}
 & [10$^9$\,\Lsun] & [10$^9$\,\Lsun] & [\kmns] & [\kmns] & [\kmns] & [\arcsec] & [\deg] & [\arcsec] & [\arcsec] & [kpc] & Scale & Fraction \\ 
(1) & (2) & (3) & (4) & (5) & (6) & (7) & (8) & (9) & (10) & (11) & (12) & (13) \\ 
\hline
W0116--0505 &  \dots                        & \dots                        & 1251$^\diamond$ & \dots                      & \dots & \dots                  & \dots & \dots                  & \dots               & \dots              & \dots  & \dots\\
W0134--2922 &  2.78\,$\pm$\,0.40            & 2.86\,$\pm$\,0.33            & --47            & 587\,$\pm$\,100            &    75 & 0.260\,$\times$\,0.209 & --85  & 0.273\,$\times$\,0.203 & 0.235\,$\pm$\,0.087 & 1.84\,$\pm$\,0.68  & 52\%   & 78\% \\
W0220+0137  &  4.79\,$\pm$\,0.38            & 5.21\,$\pm$\,0.41            &  992            & 570\,$\pm$\,68             &   162 & 0.263\,$\times$\,0.248 & --70  & 0.293\,$\times$\,0.190 & 0.236\,$\pm$\,0.043 &  1.83\,$\pm$\,0.33 & 77\%   & 58\% \\
W0236+0528  &  0.48\,$\pm$\,0.20$^\dagger$  & 0.61\,$\pm$\,0.68$^\dagger$  & \dots           & 313\,$\pm$\,255$^\dagger$  & \dots & \dots                  & \dots & \dots                  & \dots               & \dots              & \dots  & \dots\\
W0410--0913 &  1.19\,$\pm$\,0.37$^\ast$     & 23.0\,$\pm$\,7.0$^\ast$      & 2489            & 730\,$\pm$\,213$^\ast$     & \dots & \dots                  & \dots & \dots                  & \dots               & \dots              & \dots  & \dots\\
W0831+0140  &  9.90\,$\pm$\,0.95            & 11.3\,$\pm$\,0.65            &   89            & 979\,$\pm$\,169            &   200 & 0.163\,$\times$\,0.147 & --24  & 0.383\,$\times$\,0.252 & 0.311\,$\pm$\,0.063 & 2.23\,$\pm$\,0.45  & 26\%   & 96\% \\
W2246--0526 &  5.94\,$\pm$\,0.17            & 6.35\,$\pm$\,0.13            & --64            & 601\,$\pm$\,27             &   100 & 0.385\,$\times$\,0.355 &  48   & 0.380\,$\times$\,0.324 & 0.351\,$\pm$\,0.042 & 2.35\,$\pm$\,0.28  & 84\%   & 53\% \\
\hline\hline
\end{tabular}
\tablefoot{\scriptsize (1) Galaxy name; (2) Luminosity of the \CIIno\, line, calculated by directly integrating over the continuum-subtracted spectrum in the $\pm$\,2.5\,$\sigma$ range; (3) Luminosity of the \CIIno\, line derived from the Gaussian fitting; (4) Velocity difference between the redshifts derived from the \CIIsub\, and rest-UV emission lines, where $\Delta v$\,=\,c\,$\Delta z$/(1+$z_{\rm UV}$), and $\Delta z$\,=\,$z_{\rm [CII]}$--$z_{\rm UV}$; (5) FWHM of the line obtained from the extracted spectrum; (6) Absolute maximum projected velocity obtained from visual inspection of the moment 1 maps of the line (see Section~\ref{s:kinematics}), with a typical 1-$\sigma$ uncertainty of $\simeq$\,75\,\kmns; (7) Reference SPW-averaged size (FWHM) of the synthesized Gaussian beam; (8) Position angle of the beam; (9) Intrinsic size (FWHM) of the source line emission; (10-11) Circularized intrinsic size (FWHM); (12) PSF scaling factor for unresolved component subtraction; (13) Fraction of extended emission (\FEE) after PSF-scaled subtraction from the source image.\\ 
$\diamond$ Expected redshift based on the CO(4$\rightarrow$3) line detection (Gonzalez-Lopez et al., in prep.).\\
$\dagger$ The line in W0236+0528 is only considered as a tentative detection (see Section~\ref{ss:lineoffsets}) and it is excluded from the results and discussion sections.\\
$\ast$ The line in W0410--0913 is redshifted out of the frequency range covered by the observations presented here, and only the bluest part of the profile is detected (see Figure~\ref{f:spectra} and details in Section~\ref{ss:lineoffsets}). Therefore, column (2) is the flux accounted by the small portion of the observed line profile, while column (3) is the total flux of the Gaussian fit, considering the assumptions described in Section~\ref{s:analysis}.
}
\end{sidewaystable}

\begin{table*}
\caption{Dust Continuum measurements}
\centering
\scriptsize
\label{t:contfluxes}
\begin{tabular}{cccccccccc}
\hline\hline
Galaxy & $f_{\rm \nu}^{\rm data}$ & $f_{\rm \nu}^{\rm model}$ & Beam & PA & Intrinsic Size & \multicolumn{2}{c}{Circularized Size} & PSF & Extended \\
\cline{7-8}
 & [mJy] & [mJy] & [\arcsec] & [\deg] & [\arcsec] & [\arcsec] & [kpc] & Scale & Fraction \\
(1) & (2) & (3) & (4) & (5) & (6) & (7) & (8) & (9) & (10) \\
\hline
W0116--0505 &  11.7\,$\pm$\,0.4 & 11.9\,$\pm$\,0.3 & 0.269\,$\times$\,0.224 &   43 & \dots                  & $<$\,0.035          & $<$0.26           & 96\% & 20\% \\
W0134--2922 &   5.5\,$\pm$\,0.6 &  5.5\,$\pm$\,0.4 & 0.256\,$\times$\,0.207 & --84 & 0.147\,$\times$\,0.136 & 0.141\,$\pm$\,0.048 & 1.10\,$\pm$\,0.38 & 92\% & 38\% \\
W0220+0137  &  19.9\,$\pm$\,1.1 & 19.1\,$\pm$\,0.6 & 0.258\,$\times$\,0.243 & --65 & 0.193\,$\times$\,0.118 & 0.151\,$\pm$\,0.027 & 1.17\,$\pm$\,0.21 & 85\% & 42\% \\
W0236+0528  &   9.3\,$\pm$\,0.6 &  9.0\,$\pm$\,0.1 & 0.195\,$\times$\,0.172 & --63 & 0.247\,$\times$\,0.207 & 0.226\,$\pm$\,0.026 & 1.78\,$\pm$\,0.20 & 78\% & 73\% \\
W0410--0913 &  43.8\,$\pm$\,0.6 & 43.8\,$\pm$\,0.5 & 0.347\,$\times$\,0.267 &   84 & 0.400\,$\times$\,0.307 & 0.350\,$\pm$\,0.005 & 2.59\,$\pm$\,0.04 & 81\% & 72\% \\
W0831+0140  &  24.4\,$\pm$\,0.5 & 24.4\,$\pm$\,0.2 & 0.167\,$\times$\,0.150 & --15 & 0.249\,$\times$\,0.208 & 0.227\,$\pm$\,0.008 & 1.63\,$\pm$\,0.06 & 68\% & 81\% \\
W2246--0526 &   7.2\,$\pm$\,0.1 &  7.2\,$\pm$\,0.1 & 0.376\,$\times$\,0.348 &   52 & 0.178\,$\times$\,0.202 & 0.189\,$\pm$\,0.023 & 1.27\,$\pm$\,0.15 & 89\% & 30\% \\
\hline\hline
\end{tabular}
\tablefoot{\scriptsize (1) Galaxy name; (2) Averaged (moment 0), observed-frame continuum flux density using the entire line-subtracted cube (line-free SPWs plus green line in Figure~\ref{f:spectra}); (3) Continuum flux density derived from the polynomial fitting at the average wavelength; (4) Cube-averaged size (FWHM) of the synthesized Gaussian beam; (5) Position angle of the beam; (6) Intrinsic size (FWHM) of the source continuum emission; (7-8) Circularized intrinsic size (FWHM); (9) PSF scaling factor for unresolved component subtraction; (10) Fraction of extended emission (\FEE) after PSF-scaled subtraction from the source image.}
\end{table*}


\section{Analysis}\label{s:analysis}

\subsection{Measurement of line and continuum fluxes}

\begin{figure*}
\includegraphics[scale=0.42]{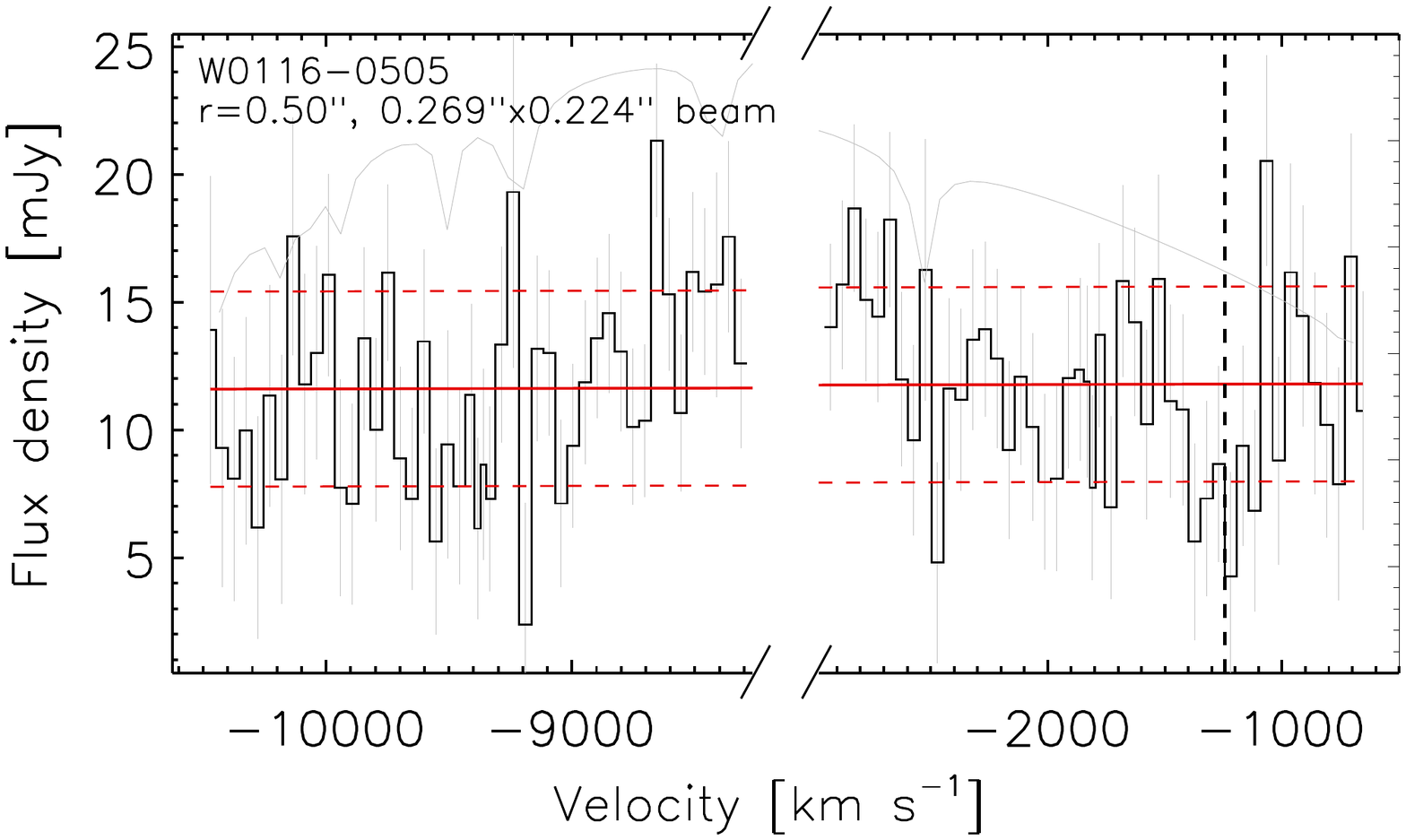}
\includegraphics[scale=0.42]{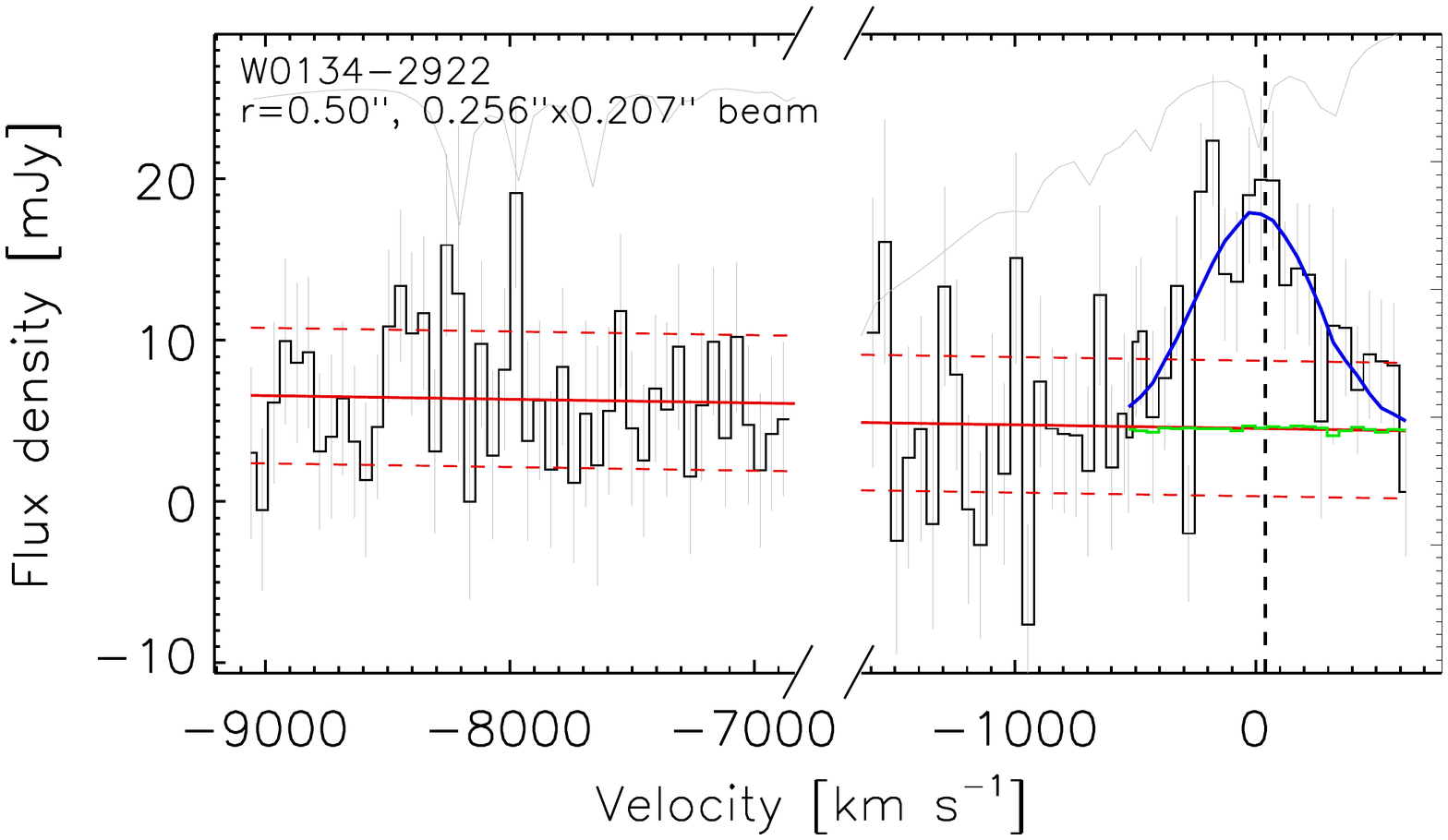}
\vspace{-6.3cm}

\includegraphics[scale=0.42]{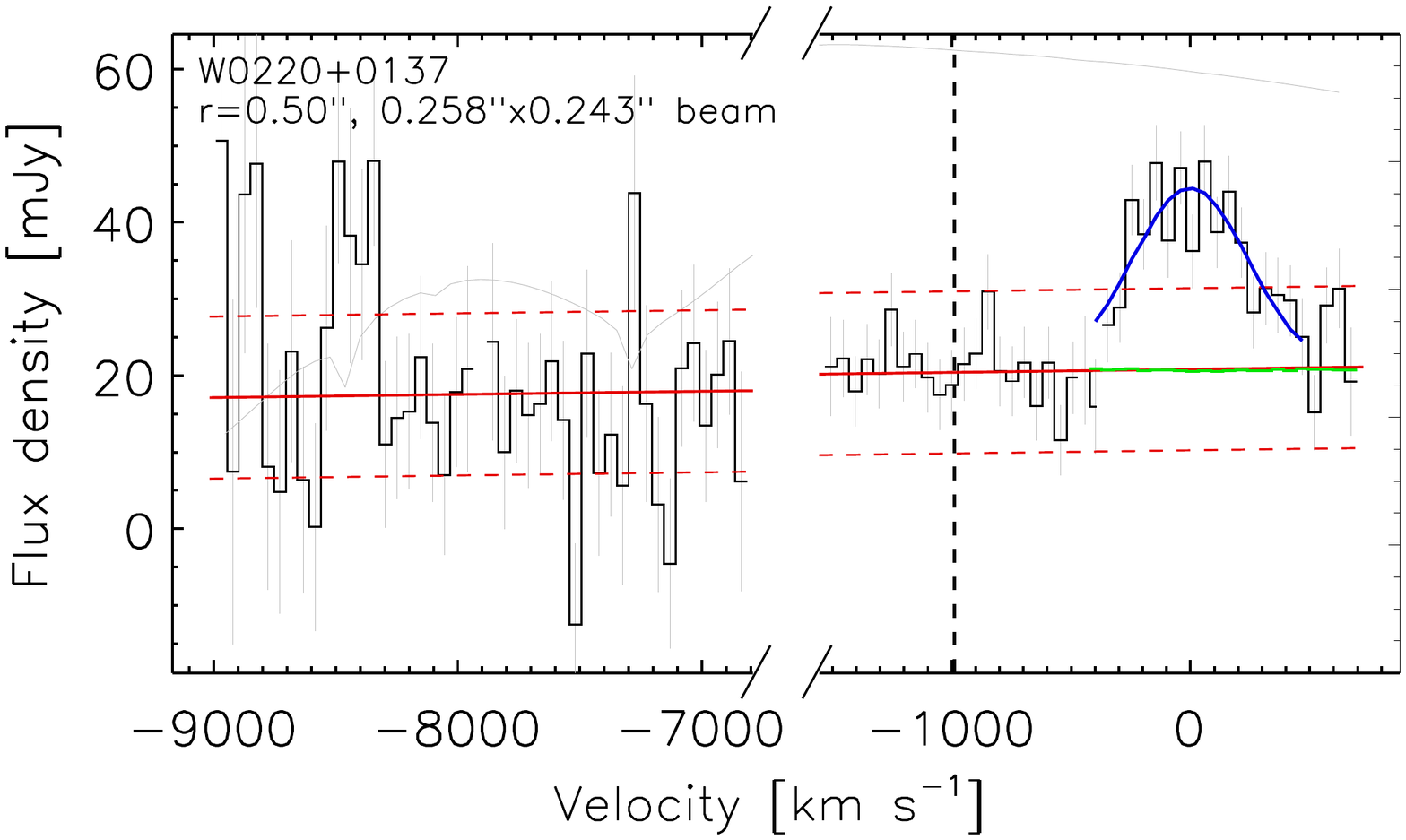}
\includegraphics[scale=0.42]{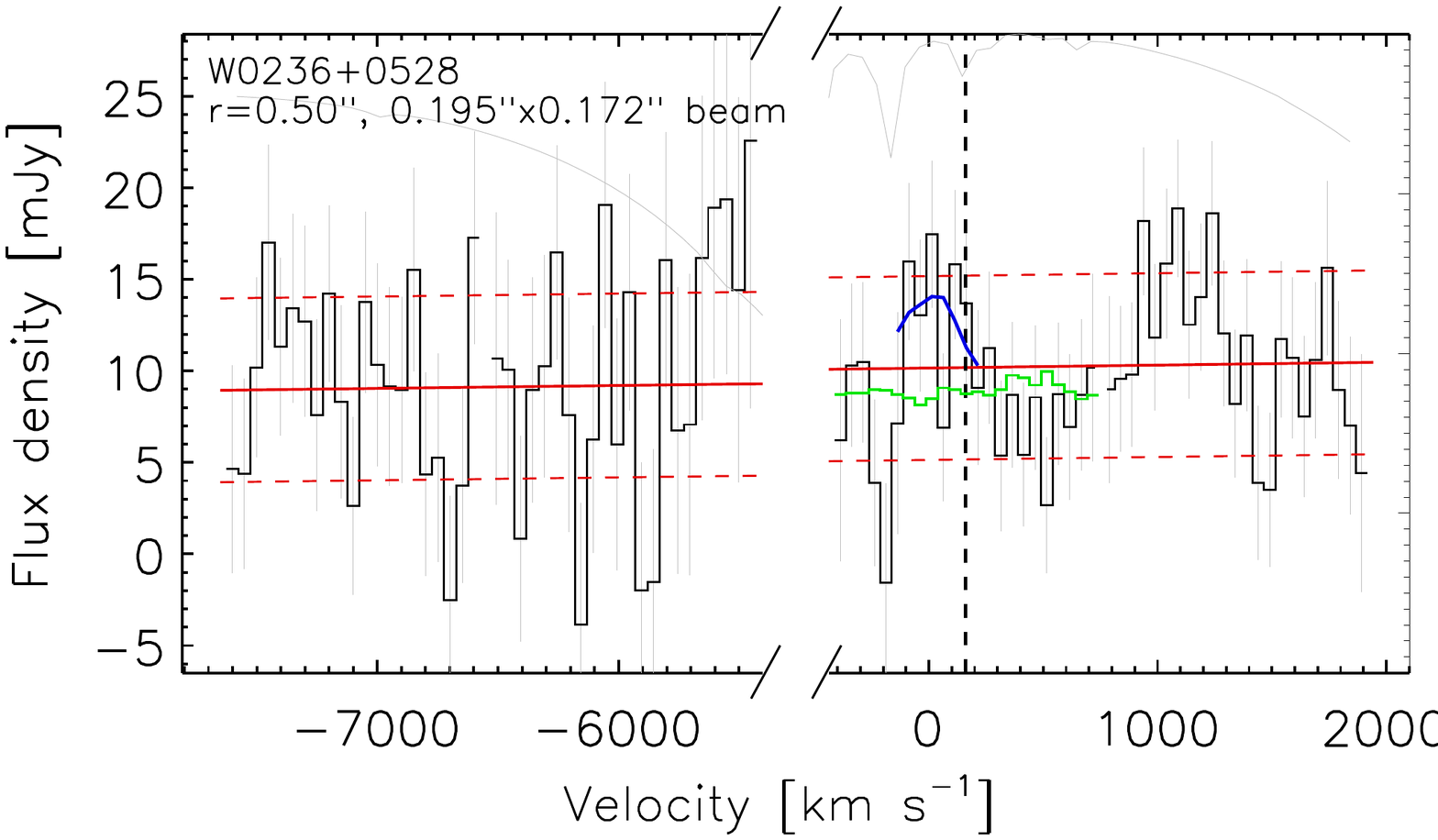}
\vspace{-6.3cm}

\includegraphics[scale=0.42]{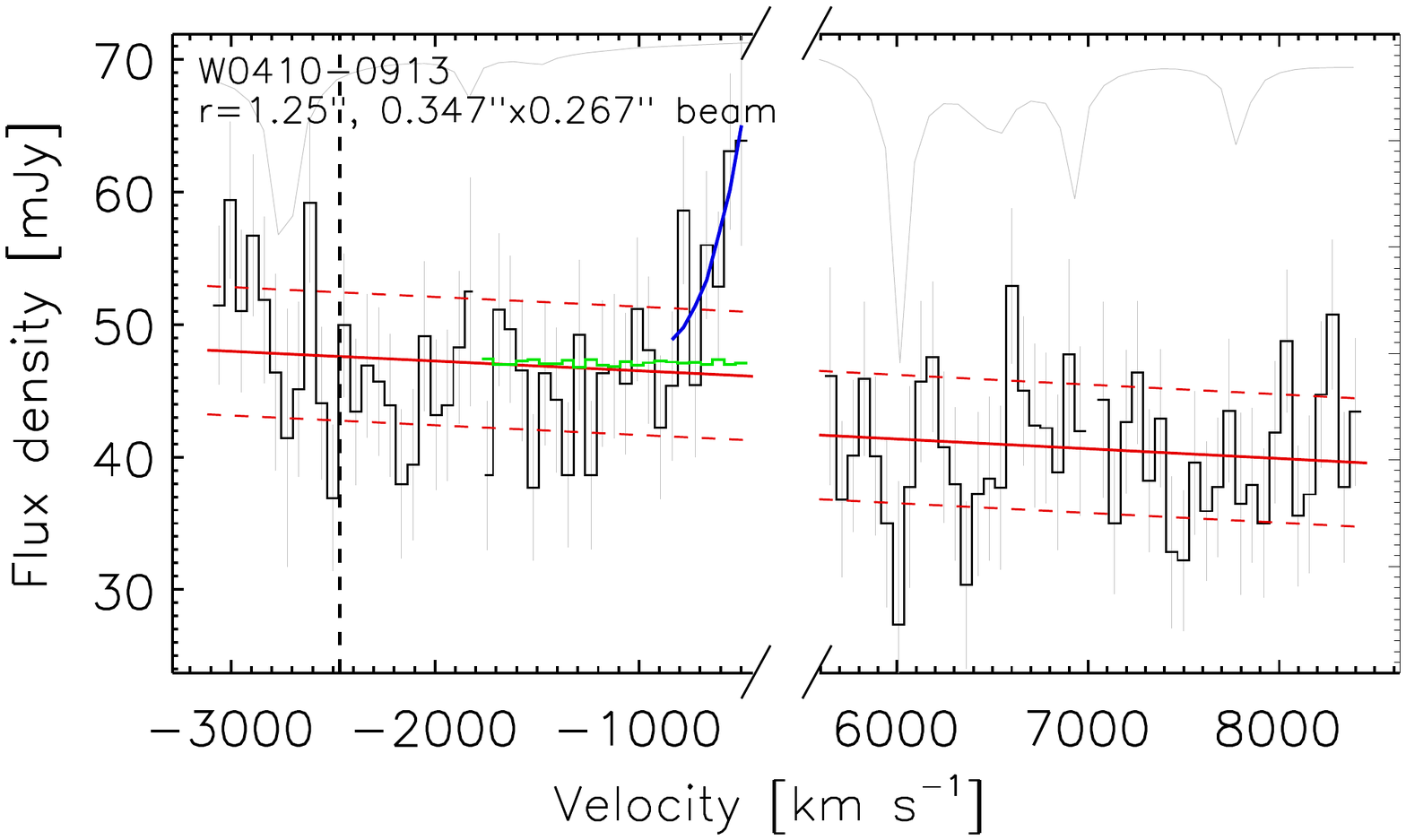}
\includegraphics[scale=0.42]{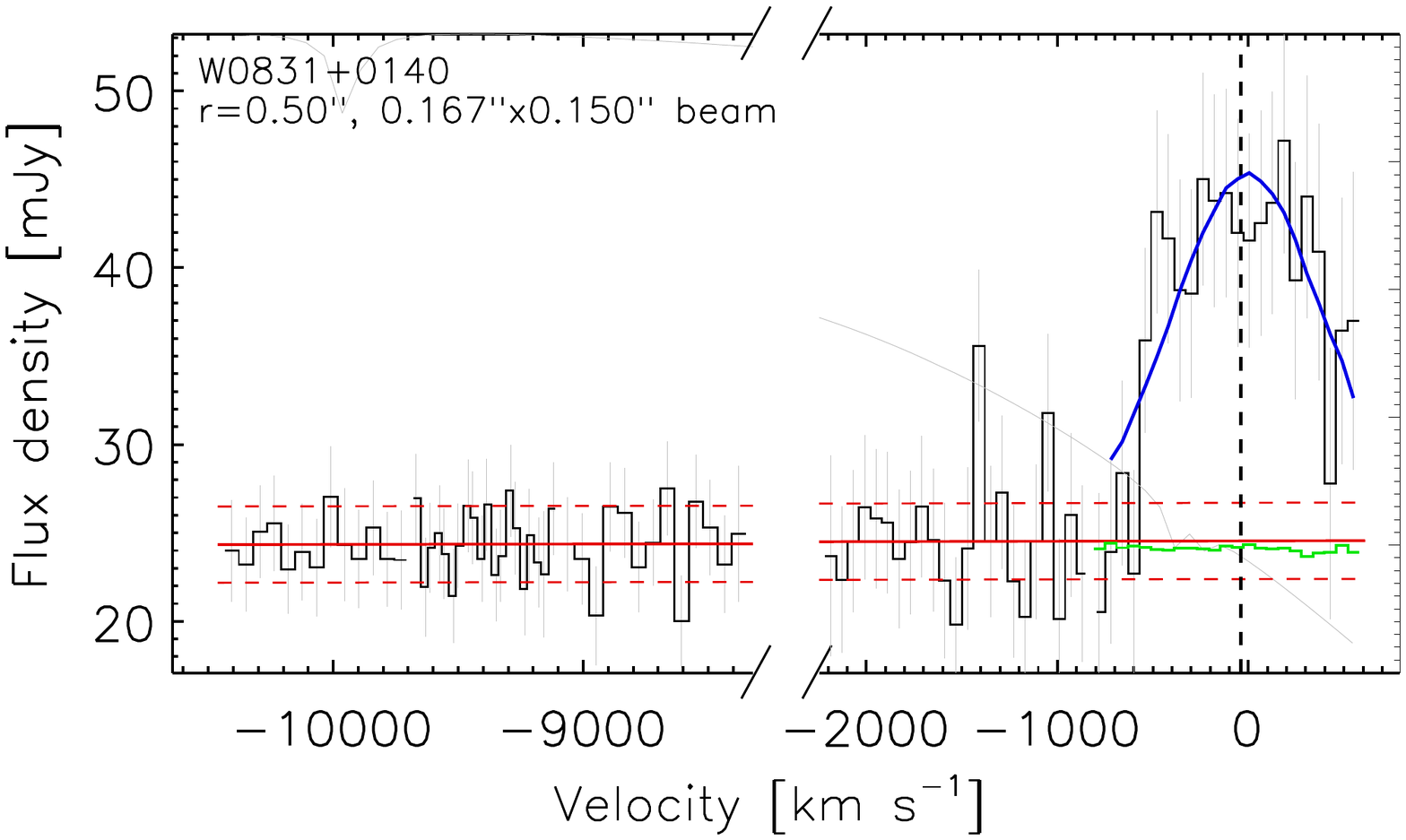}
\vspace{-6.3cm}

\hspace{4.5cm}\includegraphics[scale=0.4]{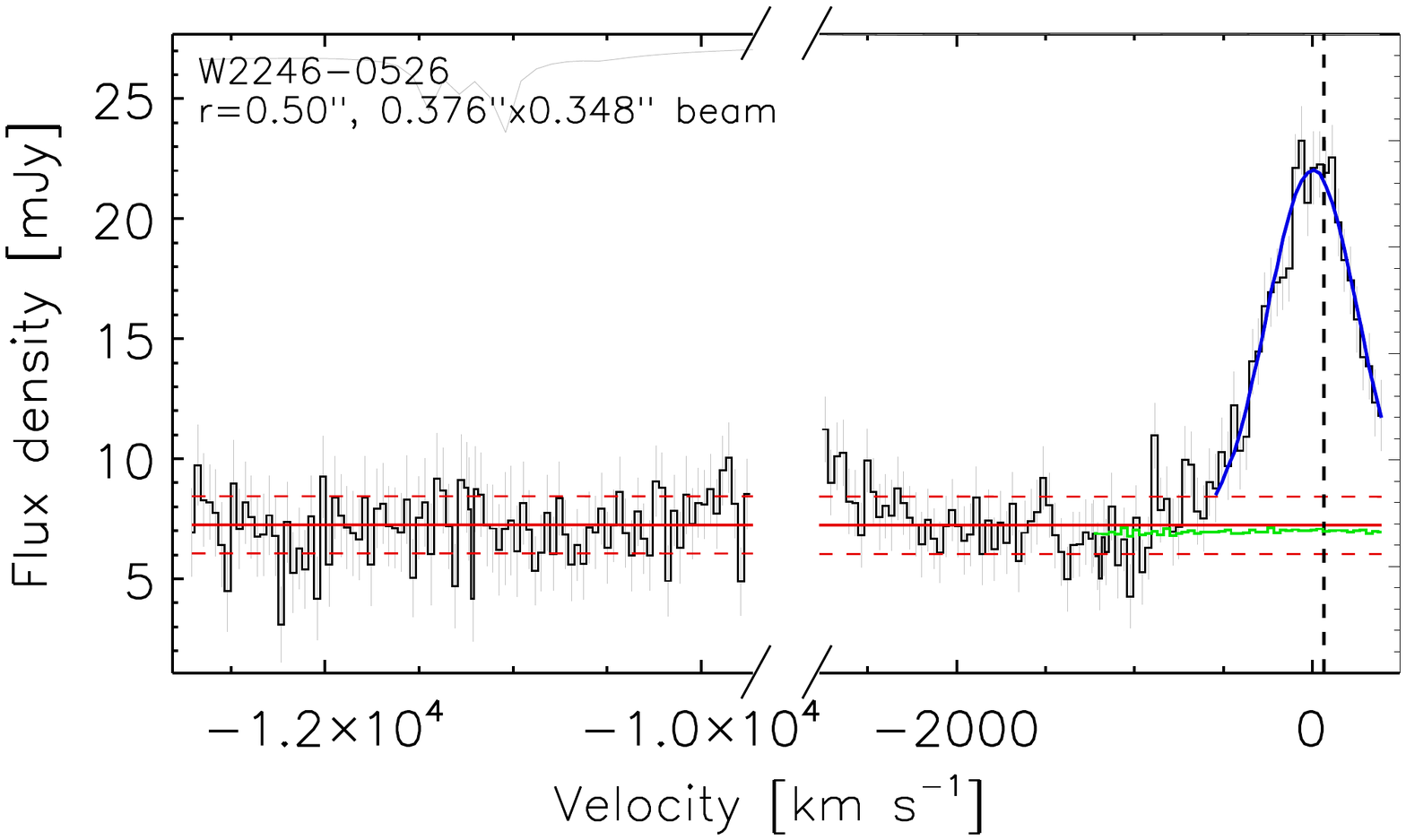}
\vspace{-6cm}
\caption{\footnotesize{ALMA spectra of the luminous Hot DOG sample, including the four SPWs (black histogram). The average size of the synthesized beam of the cube is noted in the legend at the top left, as well as the radius of the aperture used for the extraction of the spectra, which is also shown as a dashed circumference in Figure~\ref{f:contflux}. The continuum spectrum (obtained from the line-subtracted cube) underlying the emission line in each source is shown in green, and the Gaussian fit to the \CIIno\, line is shown in blue. The polynomial fit to the overall line-subtracted continuum as well as its $\pm$1$\sigma$ uncertainty are shown in red, with solid and dashed lines, respectively. The solid gray lines are the atmospheric transmission curve for a precipitable water vapor of 1\,mm\protect\footnotemark, where the maximum transmission within the displayed spectral range has been normalized to the maximum y-axis value of the whole velocity range displayed. The vertical dashed line shows the expected redshift of the line based on the rest-UV spectrum, \textit{z$_{\rm UV}$}.}}
\label{f:spectra}
\end{figure*}

To calculate the line fluxes of our Hot DOGs, the cubes were continuum-subtracted in the \textit{uv} plane using representative frequency ranges from the continuum SPW adjacent to the reference SPW, and in some cases from the SPWs of the other side-band, when no slope due to the Rayleigh-Jeans continuum from dust was identified. Both the spectrum of the dust continuum and that of the continuum-subtracted line were extracted using an aperture of 1\arcsec\, diameter centered at the peak of the galaxy emission (an aperture of 2.5\arcsec\, was used for W0410--0913 due to its large extent). Figure~\ref{f:spectra} shows the spectra for all four SPWs.

\footnotetext{https://almascience.eso.org/about-alma/atmosphere-model}

The emission line profile was fitted by a Gaussian function with the central position, peak intensity and FWHM as free parameters. The luminosity was obtained from the analytic expression of the Gaussian using the best-fit parameters, as well as by directly integrating the observed continuum-subtracted spectrum over a frequency range defined by $\pm\,$2.5$\sigma$ from the line's peak frequency as obtained from the fit, where $\sigma$ is the standard deviation of the Gaussian. Both line luminosities are reported in Table~\ref{t:linefluxes}. In the case of W0410--0913, the line center is redshifted out of the frequency range covered by the observations and only the bluest wing of the line is detected (see Figure~\ref{f:spectra}). In this particular case, the fit to the line was performed using the available portion of the profile and fixing the central position of the Gaussian to the redshift obtained from the CO emission line (discussed in Section~\ref{ss:lineoffsets}).

We also report two observed-frame continuum flux density measurements for each object. The first is calculated by obtaining the moment 0 of the entire cube after subtracting the emission line cube from the reference SPW cube. The second measurement is derived from a first-order polynomial fit to the extracted spectrum of the entire cube and interpolated to the central wavelength (see Figure~\ref{f:spectra}). These measurements can be found in Table~\ref{t:contfluxes} and are all consistent within the uncertainties.

The velocity and dispersion maps of the line were obtained by computing the moment 1 and 2 statistics of the continuum-subtracted cubes. For consistency and completeness, we re-analyze here the data of W2246--0526 that was first presented in \citet{DS2016}. All the newly derived quantities are within the uncertainties of those given in the previous work. The maps of the continuum flux density and line flux, as well as those of the velocity field and dispersion of the line (when detected; see next section) are presented in Figures~\ref{f:contflux}, \ref{f:lineflux}, \ref{f:velfield} and \ref{f:veldisp}, respectively.

\subsection{Sizes and extended structure}\label{ss:size}

The intrinsic sizes of the line and continuum emission were measured by fitting a 2D-Gaussian to the sources, from which the FWHM of the beam is subtracted in quadrature. The values are reported in Tables~\ref{t:linefluxes} and \ref{t:contfluxes}, respectively. We note that because of the limited number of degrees of freedom provided by a Gaussian function to describe the complex spatial variations seen in the flux distribution of some of our sources, these luminosity-weighted Gaussian fits are representative only of the \textit{core emission} of the Hot DOGs. That is, even though the cores of some of the galaxies are compact and/or close to be unresolved, there may still be significant low surface brightness emission that could extend over a region several times larger than the core sizes derived from the Gaussian fitting (see Figures~\ref{f:contflux} and \ref{f:lineflux}).

In order to quantify the contribution of these potential extended structures surrounding and/or underlying the central, unresolved point source of each Hot DOG, we estimated the fraction of extended emission (\FEE) of both the line and the dust continuum. This procedure is performed in the plane of the sky only for those sources with robust line/continuum detections (see below) to take advantage of the high fidelity of the ALMA images. The \FEE\, is calculated by iteratively scaling the well-defined, clean PSF image of the observations to the \CIIno\, and dust continuum images until finding the ratio that minimizes the dispersion in the central region of the residual map, within an aperture defined by an ellipse with semi-major and minor axes equal to those of the PSF. After the optimal scaling is found, the unresolved emission is subtracted from the source image and the \FEE\, is defined as the fraction of the flux remaining in the residual image, measured in a fixed circular aperture equal to that used in the original image. The aperture is set manually for each Hot DOG with the goal of encompassing the total flux (dashed circumferences in Figure~\ref{f:contflux}). The only exception is W2246--0526, for which we defined an aperture that purposefully avoids the companion galaxy located 1\arcsec\, north-east of the central Hot DOG. The \FEE\, values are provided in Tables~\ref{t:linefluxes} and \ref{t:contfluxes} for the line and continuum emission, respectively.

Finally, we also created radial profiles of the line and continuum flux surface densities for each source. The radial profiles are computed by measuring the emission within increasingly expanding ellipse-shaped rings with the same major-to-minor axis ratio and position angle (P.A.) as the best 2D-Gaussian fit to the source. Profiles are calculated for the source, PSF, scaled PSF, and residual images (see above). The profiles of the most compact and extended Hot DOGs in line and continuum emission are presented in Figure~\ref{f:spatprofs}.

\section{Integrated emission properties and morphology}\label{s:integrated}

\subsection{Line detection, velocity shifts and luminosities}\label{ss:lineoffsets}

The \CIIno\, line is robustly detected (SNR\,$\geq$\,20) in four out of the seven Hot DOGs in the sample: W0134--2922, W0220+0137, W0831+0140, and W2246--0526, the latter of which was already reported in \citet{DS2016}. Figure~\ref{f:spectra} shows the spectra of all sources in all four SPWs. We know that for two sources, W0116--0505 and W0410--0913, the line is redshifted completely and partially, respectively, out of the frequency range covered by the reference SPW (and side-band) used to target \CIIno. This is based on additional ALMA observations later obtained for these two sources, which targeted the mid-J CO(4$\rightarrow$3) and CO(6$\rightarrow$5) transitions \citep{GL2021}, respectively, as well as on the shallower observations of CO(4$\rightarrow$3) analyzed in \citet{Fan2018} for W0410--0913. In addition, JVLA observations of the CO(1$\rightarrow$0) line of both sources have also been recently presented in \citet{Penney2020}. The detection of the mid-J CO transitions with ALMA owes to the particular set up of these observations, in which the reference SPW dedicated to the line was not the reddest but instead the bluest of the side-band. Moreover, the lower frequency of these lines compared to \CIIno\, enables a larger bandwidth in velocity space. For both galaxies, the mid-J CO lines were clearly detected in the reddest SPW, which was originally dedicated to probe the dust continuum, allowing for a new redshift determination based on these additional far-IR data, \textit{z$_{\rm CO}$}. The velocity offsets of \textit{z$_{\rm CO}$} with respect to \textit{z$_{\rm UV}$} for W0116--0505 and W0410--0913 are $\Delta v$\,$\simeq$\,1250\,$\pm$\,73 and 2487\,$\pm$\,66\,\kmns, respectively. The redshifts derived from the \CIIno\, line, \textit{z$_{\rm [CII]}$}, for those galaxies in which the line is identified and the profile fully sampled have velocity offsets with respect to \textit{z$_{\rm UV}$} of $\Delta v$\,$\simeq$\,--37\,$\pm$\,75, 989\,$\pm$\,74, 49\,$\pm$\,62 and --64\,$\pm$\,55\,\kmns\, for W0134--2922, W0220+0137, W0831+0140, and W2246--0526, respectively (see Table~\ref{t:linefluxes}). Therefore, most of the newly derived redshifts for our Hot DOG sample indicate positive systemic velocities with respect to those inferred from the optical emission lines, with an average offset of $\Delta v$\,$\simeq$\,779\,$\pm$\,412\,\kmns.

\begin{figure}
\includegraphics[width=\hsize]{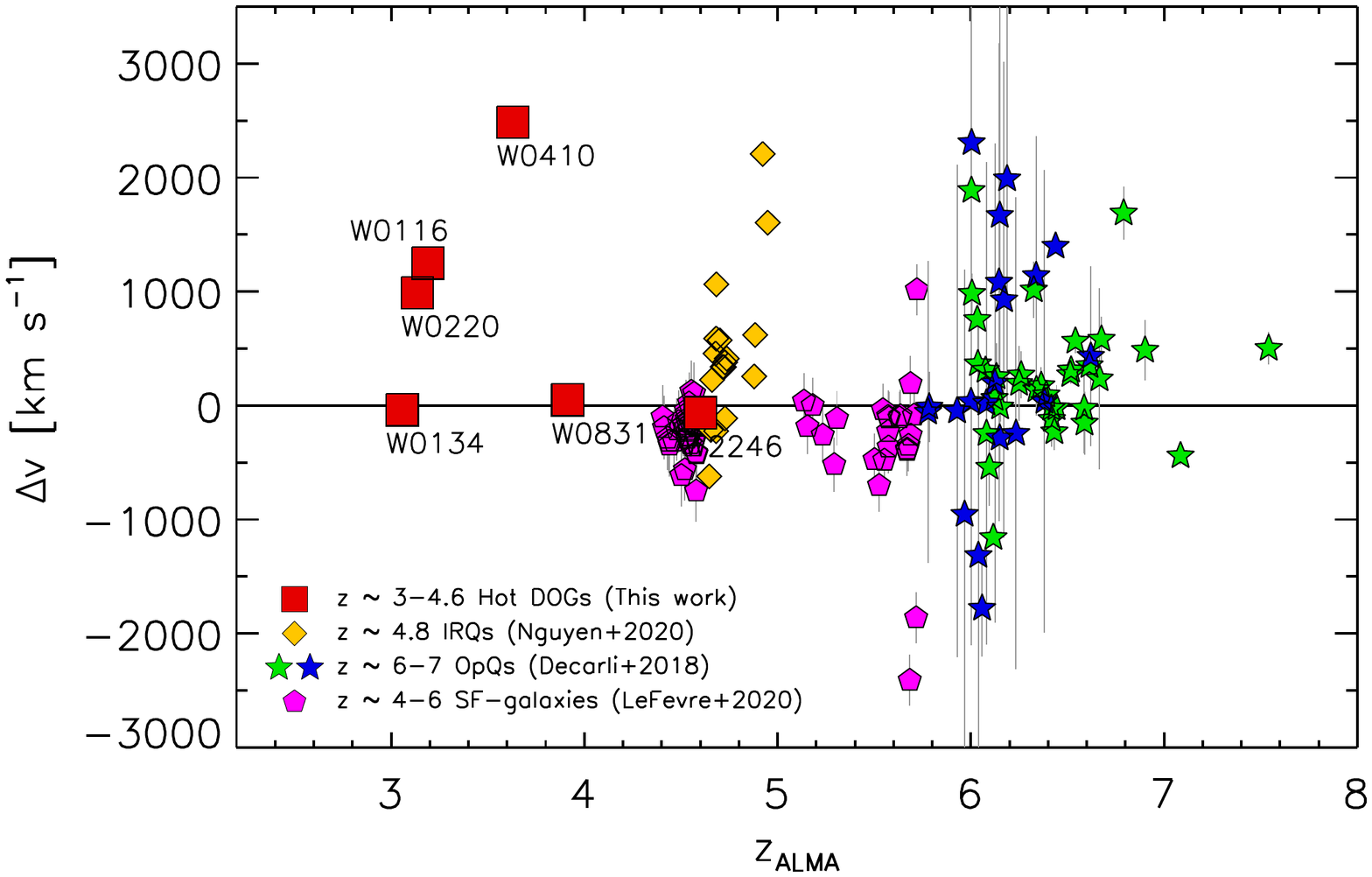}
\vspace{-6cm}
\caption{Velocity shift of \textit{z$_{\rm ALMA}$} with respect to \textit{z$_{\rm UV}$}, $\Delta v$, as a function of \textit{z$_{\rm ALMA}$} (see Table~\ref{t:linefluxes}). Hot DOGs are displayed as red squares. The \textit{z}\,$\simeq$\,4.7 IRQs from \citet{Trakhtenbrot2017} and \citet{Nguyen2020} are represented as yellow pentagons. The compilation of \textit{z}\,$\gtrsim$\,6 OpQs from \citet{Decarli2018} are displayed as green stars when the optical redshift is based on the Mg\,II emission line, and as blue stars when other lines or methods are used. Star-forming galaxies at 4\,$<$\,\textit{z}\,$<$\,6 from the ALPINE survey \citep{Cassata2020} are shown as pink pentagons.}
\label{f:deltav}
\end{figure}

We tentatively detect the \CIIno\, line in the spectrum of W0236+0528 at roughly the frequency (--161\,\kmns) where it was expected to appear, based on the \textit{z$_{\rm UV}$} (see Figure~\ref{f:spectra}). However, there seems to be another spectral feature in emission detected at a SNR\,$\simeq$\,4 in the adjacent, redder SPW, with a velocity offset of $\simeq$\,950\,\kmns. Given the mostly positive velocity offsets found for other Hot DOGs in the sample, we cannot discard that this feature could actually be the \CIIno\, line. Unfortunately there is no additional spectroscopic information for this source, and the noise in the entire side-band is large. We attempted to create an image of the line using two velocity ranges encompassing the frequencies of the two possible feature identifications, but no source could be found in either. However, using a \textit{uv}-tapering of 0.5\arcsec\, to clean the cube and generating the same line images delivers a SNR\,$\simeq$\,3 detection at the SPW where the line was expected, while nothing can be found in the redder SPW. Given the large uncertainties, we thus consider this only as a tentative detection and provide the line luminosity in Table~\ref{t:linefluxes}, but refrain from including this source in the figures or discussion throughout the remainder of the paper.

Redshifted far-IR emission lines with respect to rest-frame UV emission features have also been observed in a significant fraction of high-\textit{z} OpQs \citep[see][for an optical line emission perspective]{Vietri2018}. \citet{Decarli2018} showed that the average of the velocity offsets between the \CIIno\, and optical lines including all samples is 620\,$\pm$\,8\,\kmns\, (i.e., \CIIno\, is also frequently redshifted). We compare our velocity shifts with those of OpQs as well as with the IRQs from \citet{Nguyen2020} in Figure~\ref{f:deltav}. For reference, we also show the velocity offsets of normal star-forming galaxies at \textit{z}\,$\sim$\,4--6. While the \CIIno\, line is, on average, redshifted with respect to \textit{z}$_{\rm UV}$ in all the quasar samples, Hot DOGs and the IRQ sample seem to lack sources that are significantly blueshifted, in contrast to what is found in a number of OpQs or in general in normal star-forming galaxies, which tend to be slightly blueshifted, with $\Delta v$ between $\simeq$\,0 and --500\,\kmns.

The integrated \CIIno\, luminosities of the Hot DOG sample range between $\sim$\,3\,$\times$\,10$^{9}$\,\Lsun\, and $\sim$\,10$^{10}$\,\Lsun\, (see Table~\ref{t:linefluxes})\footnote{We do not include W0236+0528 and W0410--0913 in this statistic. \CIIno\, is only detected tentatively in the former, and the uncertainty of the luminosity is large in the latter, as the flux was obtained from a Gaussian fit to the blue wing of the emission line. However, if true, W0410--0913 would be the most luminous \CIIno\, source detected at high redshift, with a $L_{\rm [CII]}$\,=\,2.3\,$\times$\,10$^{10}$\,\Lsun.}. This is roughly similar to the range exhibited by OpQs at \textit{z}\,$\gtrsim$\,4.7 \citep[e.g.,][]{Trakhtenbrot2017, Willott2017} and typical, UV-selected star-forming galaxies at \textit{z}\,$\sim$\,5.1--5.7 \citep[e.g.,][]{Capak2015, LeFevre2020}, and it is several orders of magnitude larger than normal and sub-$L^*$ star-forming galaxies at \textit{z}\,$\sim$\,6--7 \citep{Ouchi2013, Ota2014, Maiolino2015, Knudsen2016}. The line FWHMs in the Hot DOG sample are always in excess of $\sim$\,500\,\kmns, which is usually larger than the FWHMs measured in other high-redshift sources (see discussion in Section~\ref{s:kinematics}), perhaps with the exception of a few OpQs \citep{Bischetti2019, Izumi2019} and the massive dusty star-forming galaxy HXMM05 at \textit{z}\,=\,2.985 studied in \citet{Leung2019}, which shows a comparable line width to the Hot DOG population with FWHM\,$\sim$\,600\,\kmns. The large, $\sim$\,730\,\kmns\, FWHM measured in W0410--0913, even if estimated using a small portion of the blue wing of the line, seems to be in good agreement with that calculated by \citet{Fan2018}, of $\sim$\,750\,\kmns, based on a clear detection of the CO(4$\rightarrow$3) emission line.

\subsection{\CIIno\, luminosities and Ly-$\alpha$ equivalent widths}

\begin{figure}
\includegraphics[width=\hsize]{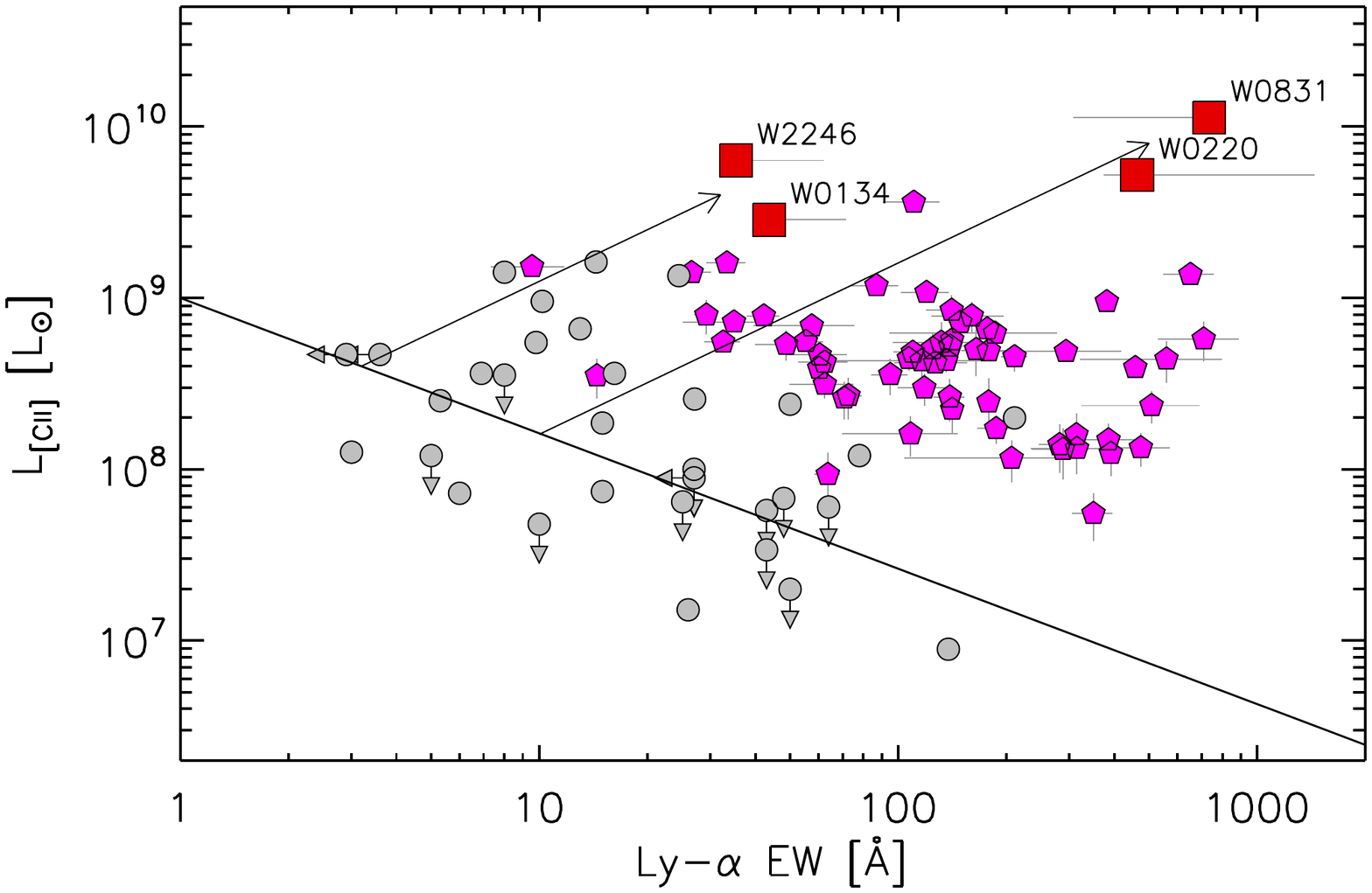}
\vspace{-6cm}
\caption{\CIIno\, luminosity as a function of the rest-frame equivalent width (EW) of Ly-$\alpha$. Hot DOGs are displayed as red squares. Normal star-forming galaxies at \textit{z}\,$\sim$\,4--6 are shown as pink diamonds \citep{LeFevre2020}. The sample of Ly-$\alpha$ emitters (LAEs) at \textit{z}\,$\simeq$\,5--7 from \cite{Harikane2018} are displayed as gray circles, together with the fit to the correlation they found using the Ly-$\alpha$ EW corrected for inter-galactic medium absorption (their equation (25)). While we do not attempt such a correction in our sample of luminous Hot DOGs, the average increase in the EW of LAEs due to this effect is less than a factor of two. Hot DOGs deviate from the correlation by more than one order of magnitude, towards larger \LCIIno\, for a given Ly-$\alpha$ (or vice versa). The arrows indicate a proportionally equal increase in the luminosity of both emission lines by a factor of 10 and 500 from the correlation towards W0134--2922 and W2246--052, and towards W0220+0137 and W0831--0140, respectively.}
\label{f:lyaew}
\end{figure}

\cite{Harikane2018} reported an anti-correlation between the luminosity of \CIIno\, and the equivalent width (EW) of the Ly-$\alpha$ line for a large sample of Lyman alpha emitters (LAEs) at \textit{z}\,$\sim$\,5--7. While the authors do not provide a physical interpretation for this anti-correlation, Figure~\ref{f:lyaew} shows that luminous Hot DOGs lie well above the trend found for the LAEs (gray circles; the best fit is represented by a solid line), in a distinctively different region of the parameter space, displaying more than an order of magnitude higher \LCIIno\, for a given Ly-$\alpha$ EW (or vice versa). No Ly-$\alpha$ EW measurements are published for the OpQ and IRQ samples of \cite{Decarli2018} and \cite{Nguyen2020}. W0220+0137 and W0831--0140 are the two most extreme Hot DOGs, where \LCIIno\, shows an excess of almost three orders of magnitude. Interestingly, Figure~\ref{f:optspec} shows that these two Hot DOGs, in addition to a narrow Ly-$\alpha$ component with FWHM\,$\sim$\,500--1000\,\kmns, also present an extremely broad component with FWHM\,$\sim$\,7500--10000\,\kmns\, underlying the narrow component. W0116--0505 also displays a narrow+broad Ly-$\alpha$ profile, but unfortunately the \CIIno\, line was likely red-shifted out of the SPW for this Hot DOG. W0236+0528, in which \CIIno\, is tentatively detected, only displays a broad Ly-$\alpha$ component, while W2246--0526 and W0134--2922 only show narrow emission, with weak or no broad component.

Figure~\ref{f:lyaew} also shows that the star-forming galaxies of the ALPINE survey lie in the region of the parameter space between the LAEs and the Hot DOGs. This is interesting because of the redshift of each galaxy population. The LAE population is at \textit{z}\,$\simeq$\,5--7, the ALPINE survey is at \textit{z}\,$\simeq$\,4--6, and our luminous Hot DOG sample lies at \textit{z}\,$\simeq$\,3--4.6 (with only W2246--0526 being at \textit{z}\,$>$\,4). These differences could be related to the average escape fraction of Ly-$\alpha$ and/or ionizing photons \citep{Harikane2018}. Unfortunately there is no information regarding this quantity for the ALPINE galaxies or our Hot DOG sample and therefore no comparisons can be made among the samples.

\subsection{Rotation or dispersion supported?}\label{ss:vsigma}

The $|V_{\rm max}|$/$\sigma_V$ ratio, where $|V_{\rm max}|$ is the maximum projected rotational velocity of the galaxy and $\sigma_V$ its overall velocity dispersion, is a first order measurement that indicates whether a system can be gravitationally supported by rotation, when $|V_{\rm max}|$/$\sigma_V$\,$>$\,1, or by dispersion/turbulence, when $|V_{\rm max}|$/$\sigma_V$\,$<$\,1.

To calculate the quantities $|V_{\rm max}|$ and $\sigma_V$, we perform rigorous kinematic modeling of the two sources for which the angular resolution and quality of the data allow such a detailed characterization: W0220+0137 and W2246--0526. A description of the modeling, using the tool $^{\rm 3D}$Barolo, together with the obtained results can be found in Appendix~\ref{a:kinmod}. The $|V_{\rm max}|$ (=\,$V_{\rm rot}$) and $\sigma_V$ (=\,FWHM$_{\rm mod}$/2.355) values obtained from this analysis are reported in Table~\ref{t:linefluxes}. We also attempted to model the kinematics of W0134--2922 and W0831+0140, but the analysis resulted in an unconstrained morphological inclination and position angle for the former, and in position-velocity diagrams of the minor and major axis for the latter that indicates clear non-rotation. Thus, for these sources we use $|V_{\rm max}|$ and $\sigma_V$ values derived from visual inspection of the velocity field maps (Figure~\ref{f:velfield}) and from the FWHM measured directly in the spectra (Figure~\ref{f:spectra}), respectively. Specifically, $|V_{\rm max}|$ is calculated as $(|V_{\rm +}|$\,+\,$|V_{\rm -}|)$/2, where $V_{\rm +}$ and $V_{\rm -}$ are the maximum positive and negative projected mean velocities per resolution element identified in the moment 1 line maps, evaluated at the SNR\,=\,3--4 contours; $\sigma_V$ is calculated as FWHM$_{\rm [CII]}$/2.355. These values are also tabulated in Table~\ref{t:linefluxes}.

Considering this, we report $|V_{\rm max}|$/$\sigma_V$ ratios for each of the rings used in the kinematic modeling of W0220+0137 and W2246--0526. For W0220+0137, the set of \{$d_{\rm r}$, $|V_{\rm max}|$/$\sigma_V$\} values are $d_{\rm r}$\,=\,[0.38, 1.14, 1.90]\,kpc and $|V_{\rm max}|$/$\sigma_V$\,=\,[1.15\,$\pm$\,0.24, 1.90\,$\pm$\,0.35, 2.11$\pm$\,0.71], where $d_{\rm r}$ is the distance of the ring to the kinematic center of the system. For W2246--0526, the set of values are $d_{\rm r}$\,=\,[0.46, 1.37]\,kpc and $|V_{\rm max}|$/$\sigma_V$\,=\,[0.88\,$\pm$\,0.14, 0.71\,$\pm$\,0.13]. The measured $|V_{\rm max}|$/$\sigma_V$ ratios for W0134--2922 and W0831+0140 are 0.30\,$\pm$\,0.42 and 0.48\,$\pm$\,0.27, respectively. We warn that since kinematic modeling was not possible for these two sources given the current data, beam smearing can be significant. As an example, while the FWHMs derived from the modeling of W0220+0137 and W2246--0526 are similar to those obtained directly from the spectra, the projected $|V_{\rm max}|$ measured in the moment 1 maps seem to be systematically underestimated with respect to the modeled rotation velocities by a factor of $\sim$\,2. Therefore, the $|V_{\rm max}|$/$\sigma_V$ ratios for W0134--2922 and W0831+0140 are likely to be at least $\gtrsim$\,0.6.

\citet{Tan2019} observed 8 nearby (\textit{z}\,$<$\,0.19) IR QSOs with ALMA in the CO(1$\rightarrow$0) emission line, a tracer of the cold molecular gas component in galaxies. They modeled the kinematics of the four isolated QSO host galaxies in the sample displaying velocity gradients and show that three of these sources are rotation dominated, with $|V_{\rm max}|$/$\sigma_V$\,=\,4--6, whereas the other one shows evidence for a disturbed disk with turbulent molecular gas. The remaining four QSOs display complex kinematics, three of them likely due to ongoing interactions with companion galaxies (see discussion in Section~\ref{s:discussion}). The $|V_{\rm max}|$/$\sigma_V$ ratios found for the two Hot DOGs for which a kinematic modeling was possible are significantly lower, below $\sim$\,2. This range is also at the lower end of the distribution of values measured in star-forming galaxies at \textit{z}\,$\sim$\,2 \citep{FS2006, FS2009}, Lyman Break Galaxies (LBGs) at \textit{z}\,$\sim$\,3–3.5 \citep{Gnerucci2011}, \Hano-selected galaxies from \textit{z}\,$\sim$\,0.4 to \textit{z}\,$\sim$\,3.3 \citep{Wisnioski2015, Gillman2019}, and \textit{H}-band selected galaxies in the CANDELS field at \textit{z}\,$\sim$\,1.5--3.2 \citep{Price2019}. All these samples have average $|V_{\rm max}|$/$\sigma_V$ ratios in the range of $\simeq$\,1--3, and, in particular, \citet{Gillman2019} and \cite{Price2019} showed that the ratio decreases as a function of redshift, which would agree with the relatively low values found for our Hot DOG sample.

\begin{figure*}
\includegraphics[scale=0.425]{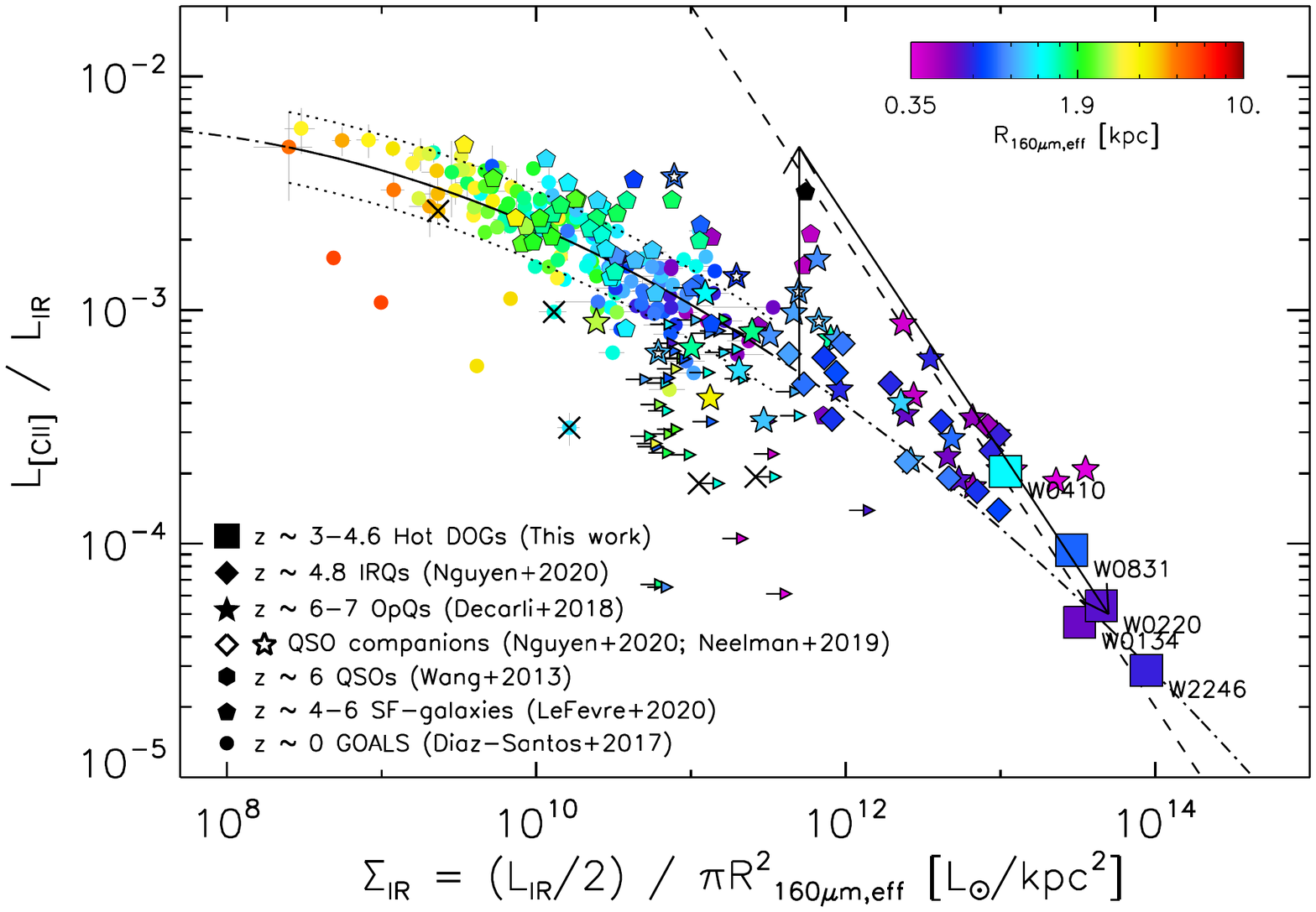}\hspace{-0.25cm}
\includegraphics[scale=0.425]{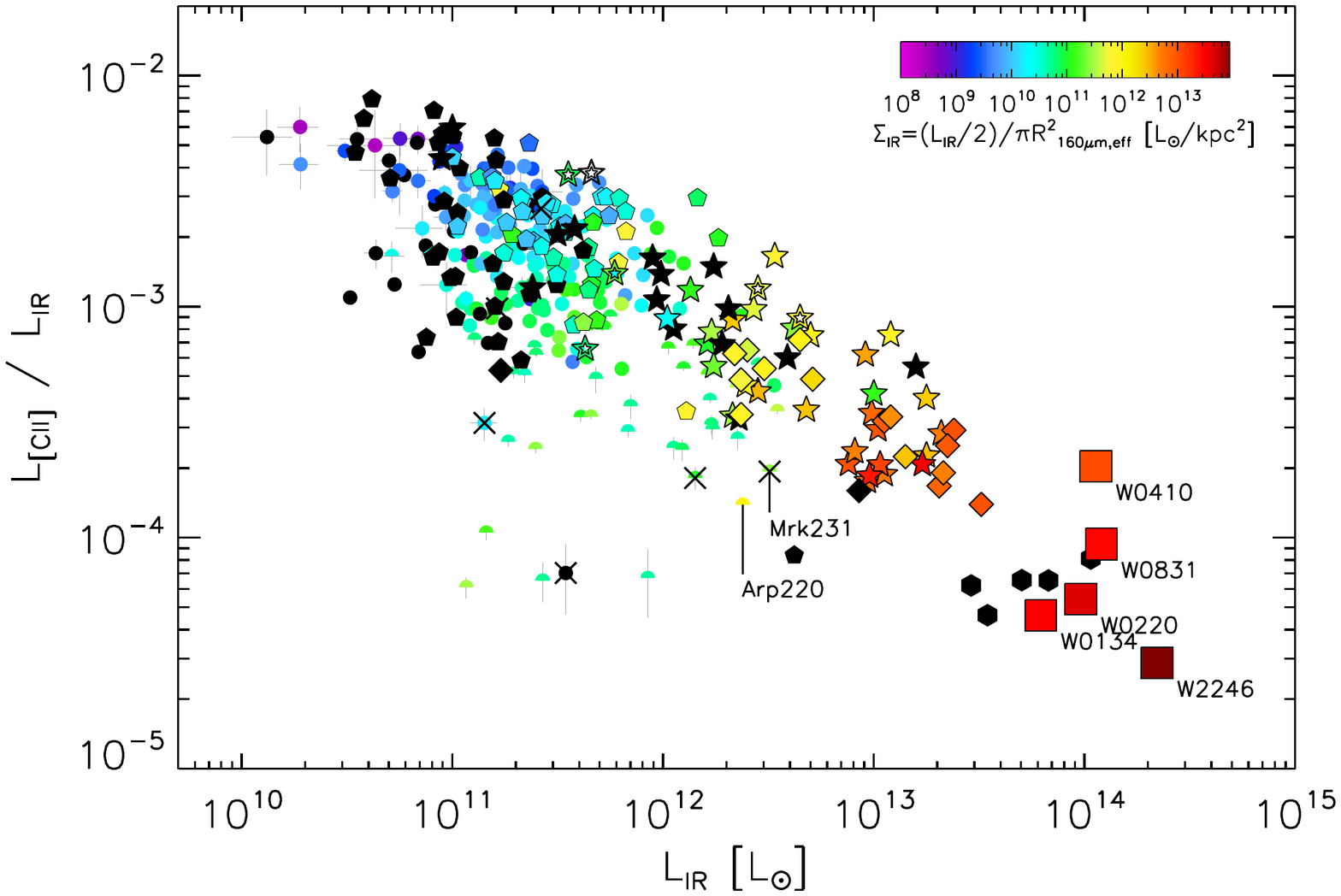}
\vspace{-5.5cm}

\includegraphics[scale=0.425]{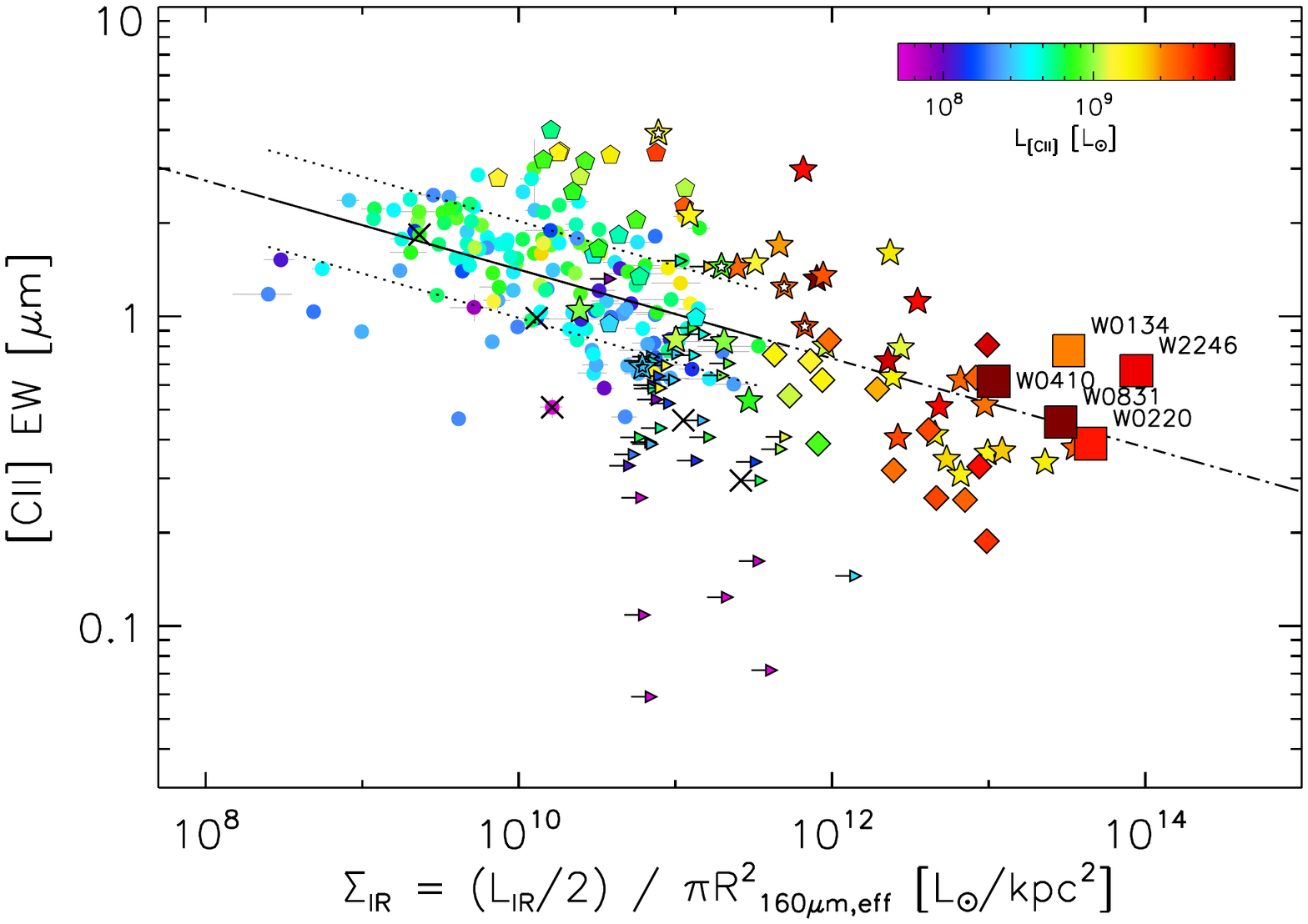}\hspace{-0.25cm}
\includegraphics[scale=0.425]{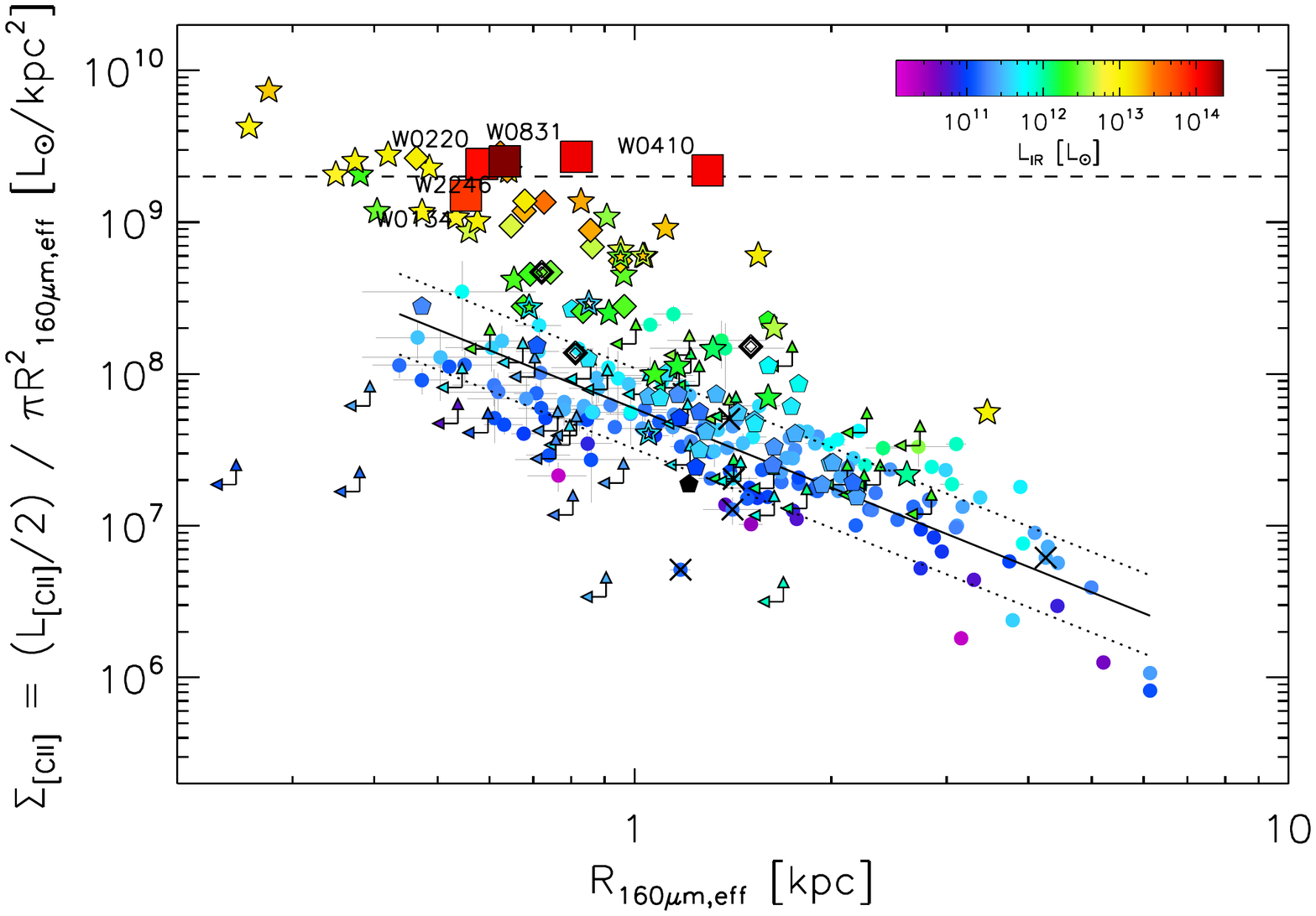}
\vspace{-5.2cm}
\caption{\footnotesize (top-left) \CIIno\, deficit (here defined as \CIIno/\LIR) as a function of the IR luminosity surface density, \SigmaIR, color-coded as a function of the intrinsic 160\,\mic\, radius; (top-right) \CIIno\, deficit as a function of \LIR, color-coded as a function of \SigmaIR; (bottom-left) \CIIno\, EW as a function of \SigmaIR, color-coded as a function of \LCIIno; (bottom-right) \CIIno\, surface density, \SigmaCII, as a function of the intrinsic $\sim$\,160\,\mic\, radius (see Section~\ref{ss:compdata}), color-coded as a function of the IR luminosity. Hot DOGs are shown as squares, along with their names. The symbols of the comparison quasar samples are as in Figure~\ref{f:deltav} (filled stars and diamonds). In addition, galaxy companions found in the neighborhood of some quasars in the sample of \citet{Decarli2018} are shown as open stars \citep[from][]{Neelman2019}. The local, \textit{z}\,$<$\,0.1 (U)LIRG sample from GOALS \citep{Armus2009} is shown as small circles and limits \citep{DS2017}. Black crosses indicate local galaxies in which an AGN contributes $\geq$\,50\% to the bolometric luminosity. In general, black symbols denote galaxies that do not have information regarding the property shown in the color bar. In all panels, the solid line is a fit to the local (U)LIRGs (the dotted lines show the 1$\sigma$ standard deviation around the fit; see \citealt{DS2017} for details), and the dotted-dashed line is the extrapolation towards lower and higher values of the x-axes. The dashed line in the top-left and bottom-right panels represents a constant \SigmaCII\,$\simeq$\,2\,$\times$\,10$^9$\,\lsd, that seems typical of the Hot DOG sample, independently of their \LIR. The two arrows in the top-left panel refer to the discussion in Section~\ref{ss:powersource}. Note that in this figure we are using \LIR\, to show the \CIIno\, deficits and luminosity surface densities, rather than the more traditional approach of using \LFIR\, that is commonly found in the literature.}
\label{f:ciideficits}
\end{figure*}

\subsection{Global ISM conditions}\label{ss:ism}

\subsubsection{The \CIIno\, deficit}\label{sss:ciidef}

By virtue of originating from multiple ISM phases (ionized and neutral gas), the \CIIno\, line is the current workhorse used for the study of the gas content and kinematics of galaxies on or above the MS at high redshift. In combination with the IR luminosity, the so-called \CIIno\, deficit is one of the most widely used far-IR diagnostics with which to infer the state of a galaxy's ISM \citep[e.g.,][]{Malhotra2001, Brauher2008, Stacey2010}. Defined as the ratio of \CIIno-to-IR or to-far-IR luminosity, low values of these ratios (large deficits) are a characteristic signature of nearby LIRGs and ULIRGs \citep{GC2011, DS2013, Farrah2013}. While there is a clear trend for more IR luminous sources to display progressively lower \LCIIno/\LIR\, ratios ($\lesssim$\,10$^{-3}$) with respect to normal star-forming galaxies ($\simeq$\,5\,$\times$\,10$^{-3}$), the most clear and tight correlations of the \CIIno\, deficit relate to the dust temperature and the IR luminosity surface density, \SigmaIR\, \citep{DS2013, Lutz2016, DS2017, Smith2017}, as well as the star formation efficiency \citep[\LIR/\MHH;][]{GC2011}.

\begin{figure*}
\hspace{1.75cm}\includegraphics[scale=0.34]{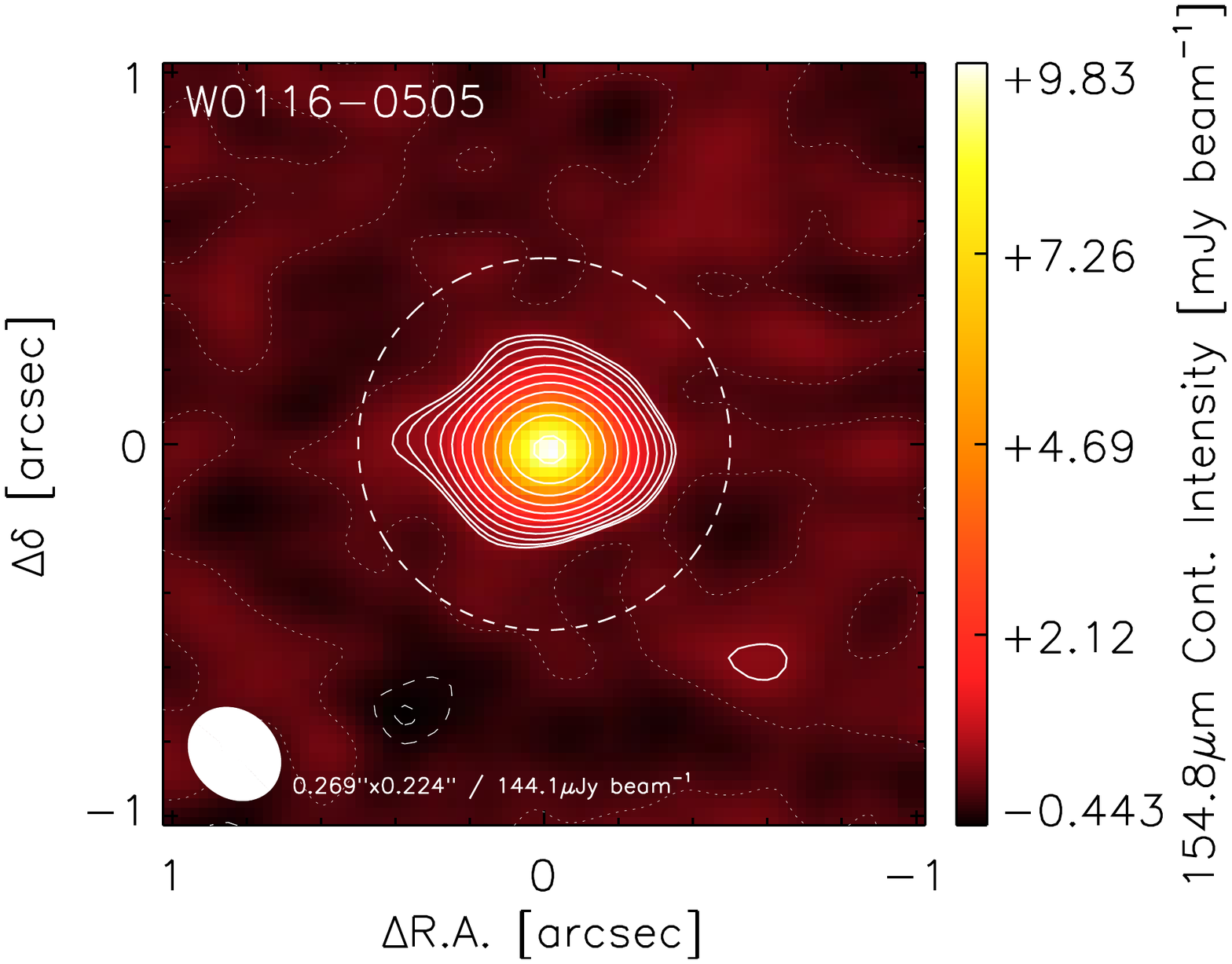}
\includegraphics[scale=0.34]{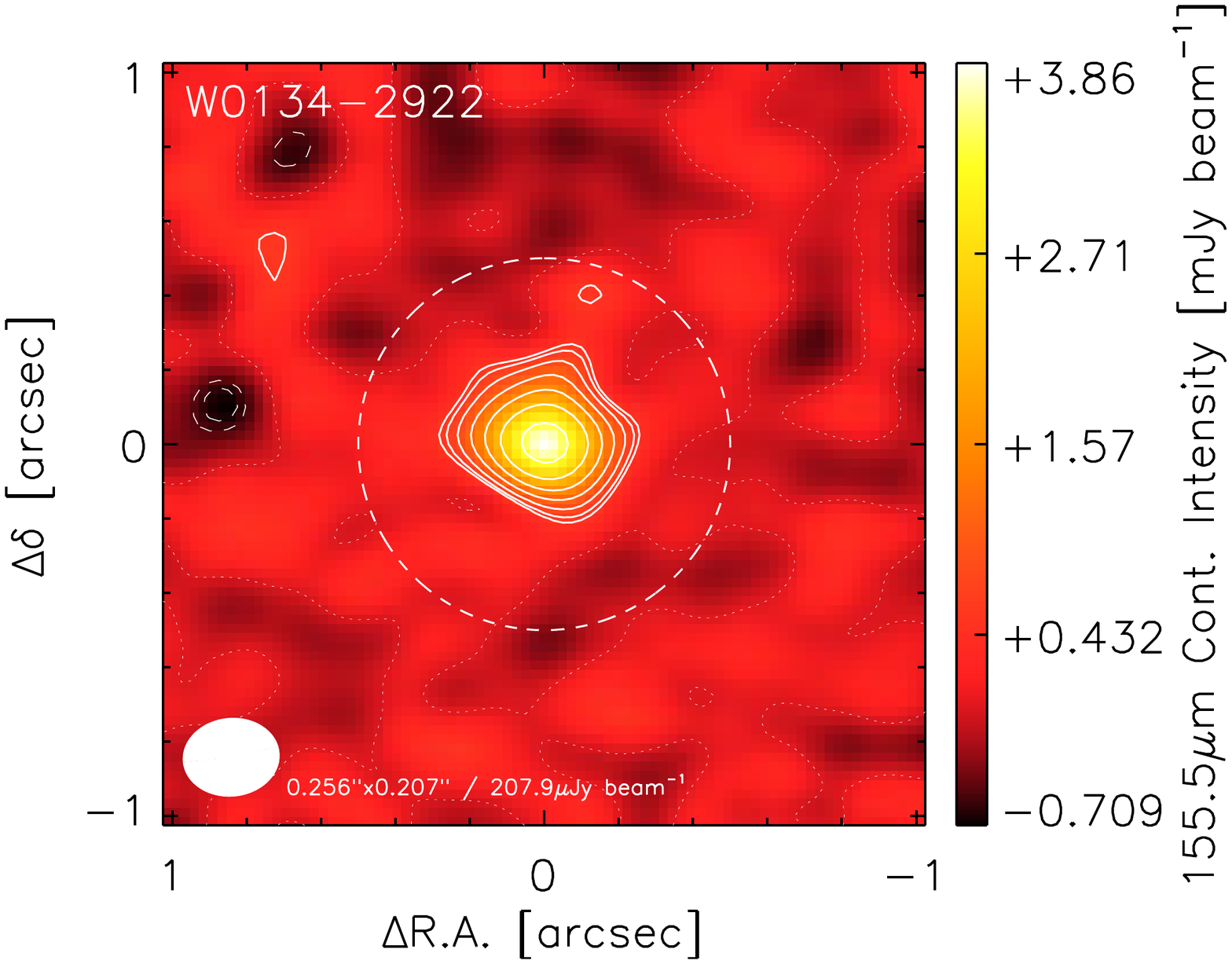}
\vspace{-4.cm}

\hspace{1.75cm}\includegraphics[scale=0.34]{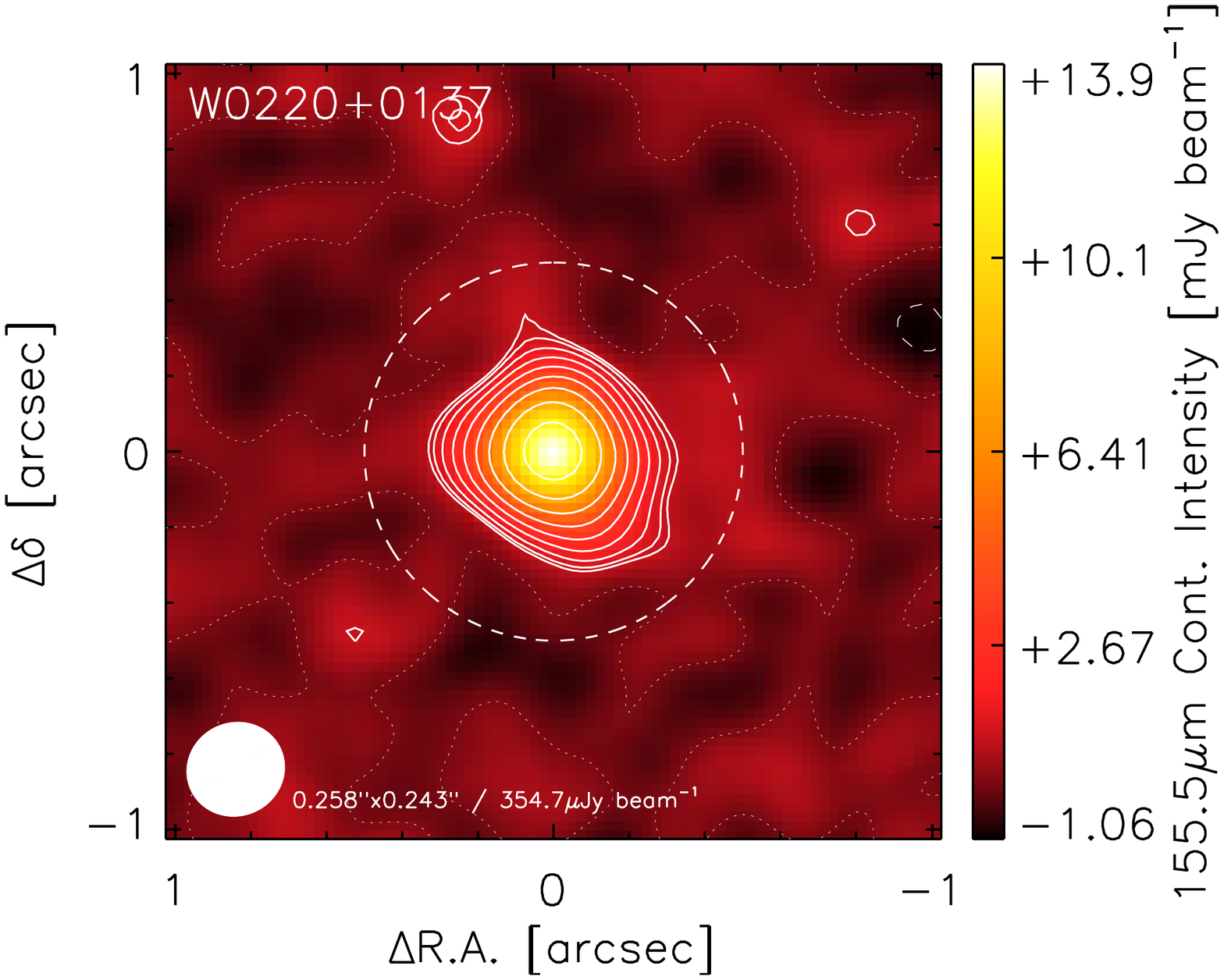}
\includegraphics[scale=0.34]{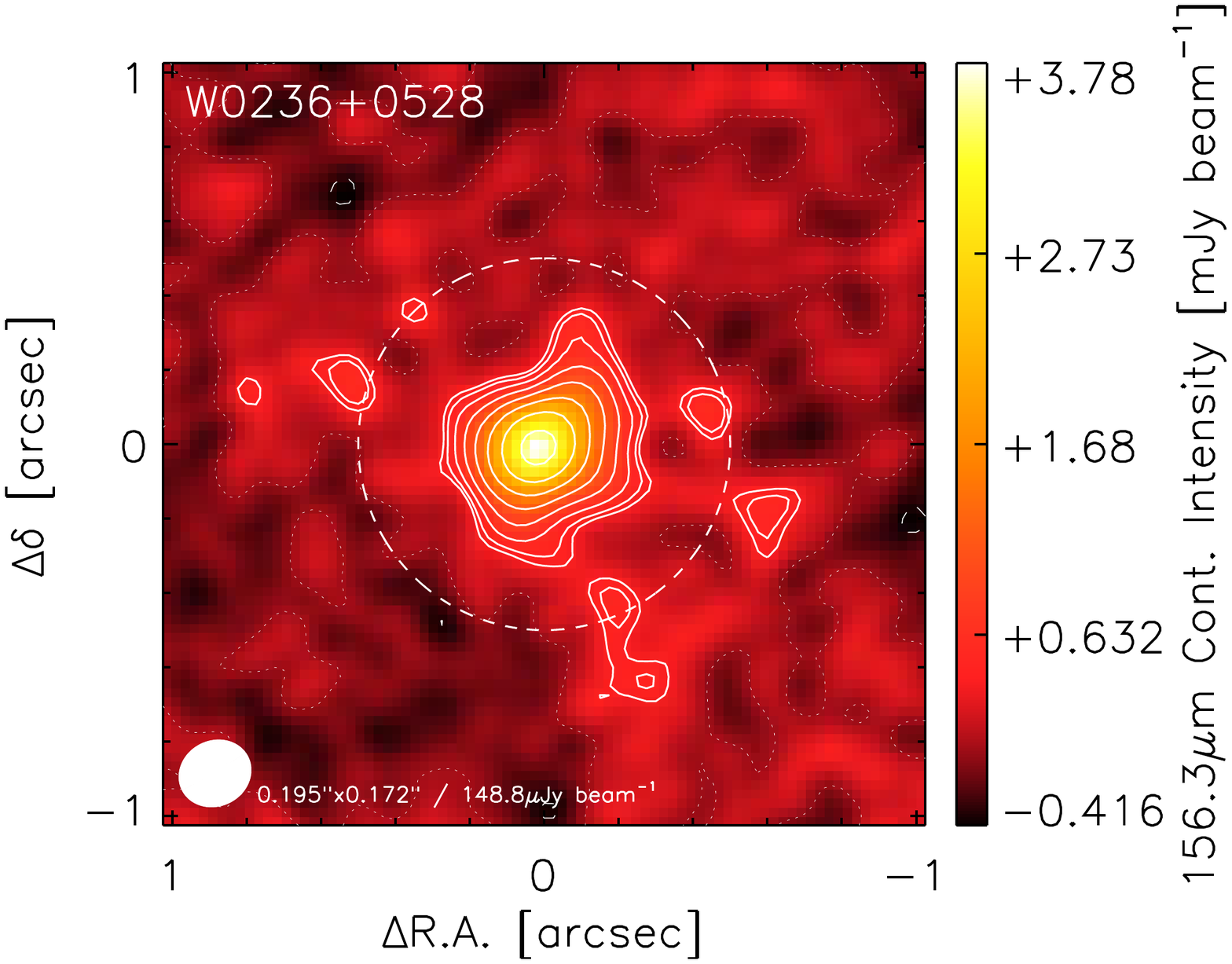}
\vspace{-4.cm}

\hspace{1.75cm}\includegraphics[scale=0.34]{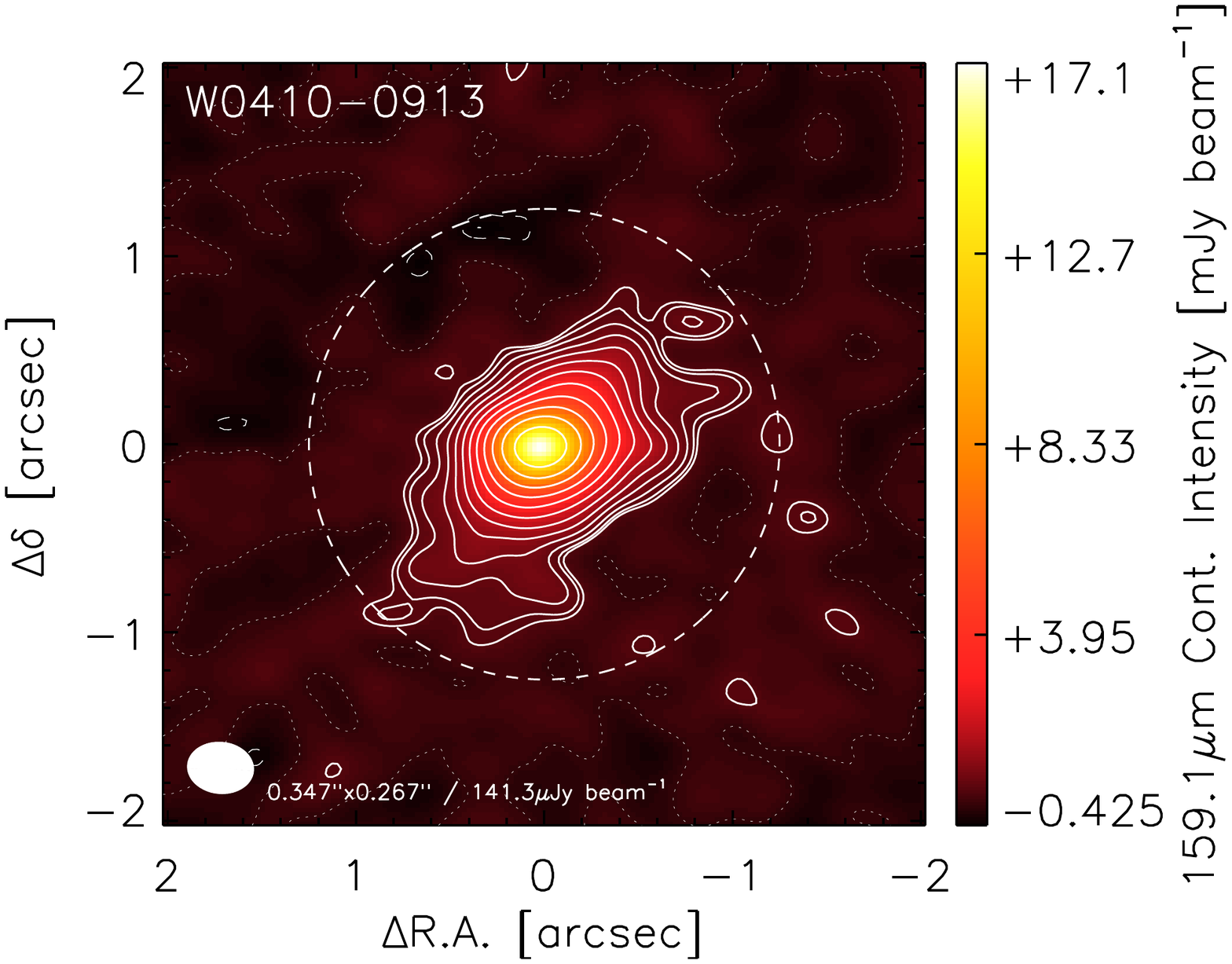}
\includegraphics[scale=0.34]{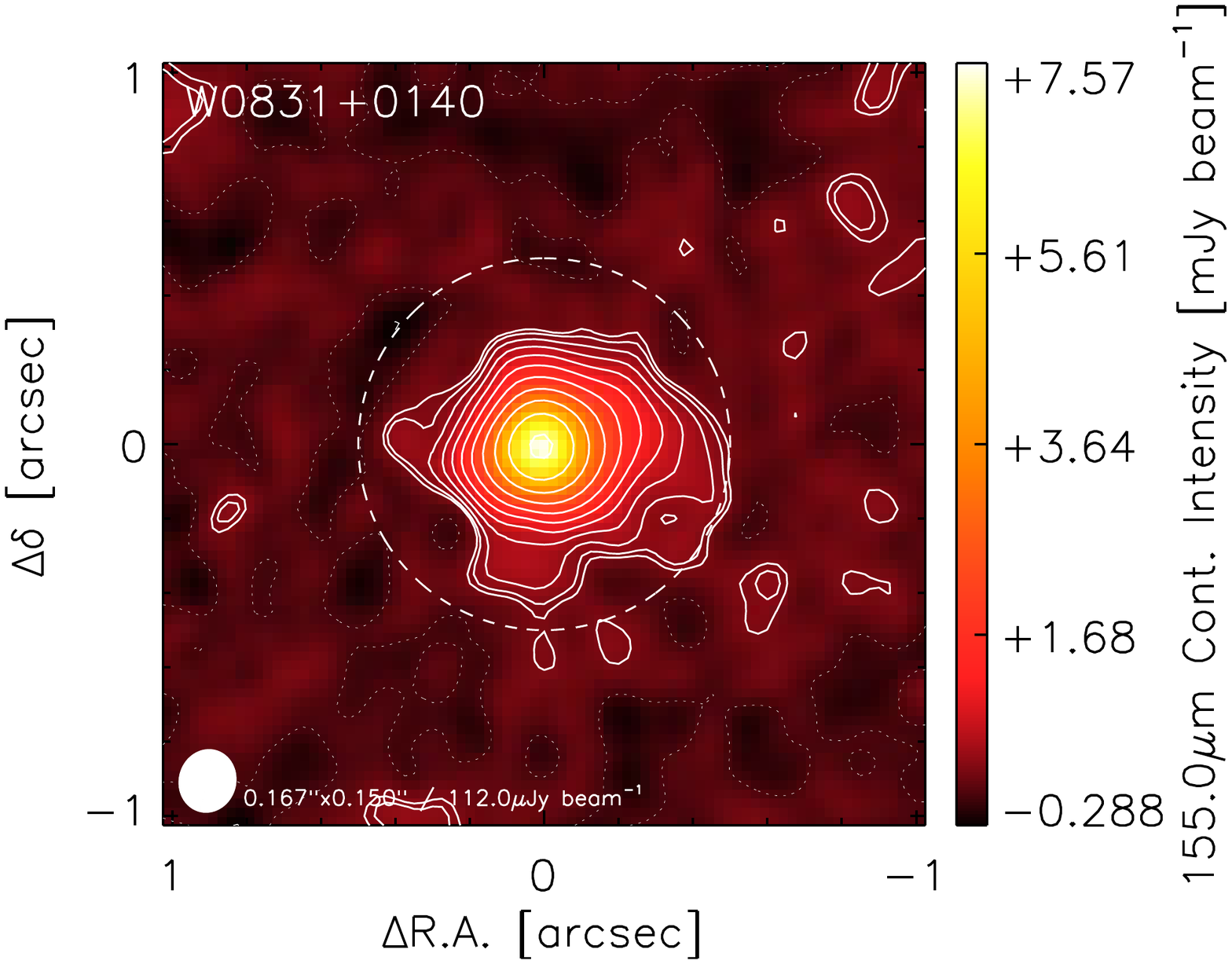}
\vspace{-4.cm}

\hspace{5.5cm}\includegraphics[scale=0.34]{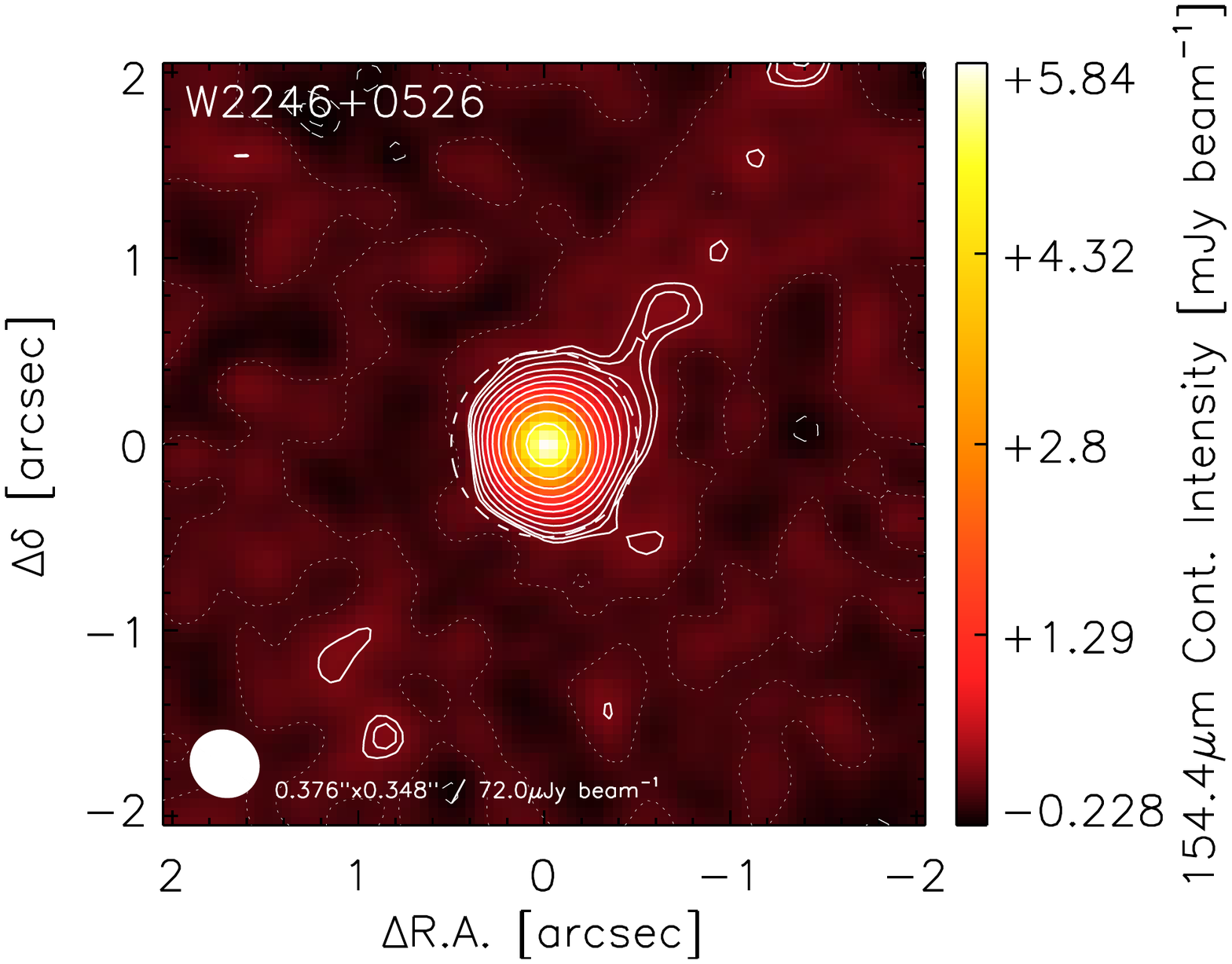}
\vspace{-3.5cm}
\caption{\footnotesize Dust continuum flux density maps of the Hot DOG sample. The beam is shown as a white ellipse on the bottom-left, together with its size and the depth (r.m.s.) of the collapsed, line-subtracted cube. The dashed white circumference denotes the aperture used to extract the spectra shown in Figure~\ref{f:spectra}. The dotted contour shows the zero-flux intensity, and the solid contours represent the [2.5, 3.0, 2$^{(4+n)/2}$]-$\sigma$ levels, where $\sigma$ is the r.m.s. of the map and n\,=\,[0, 1, 2\dots]. Symmetric, negative levels are shown as dashed contours. Field of views (FoVs) of 2\arcsec\, and 4\arcsec\, on a side are displayed depending on the galaxy. Roughly, 1\arcsec\,$\simeq$\,7\,kpc.}
\label{f:contflux}
\end{figure*}

\begin{figure*}
\hspace{1.75cm}\includegraphics[scale=0.34]{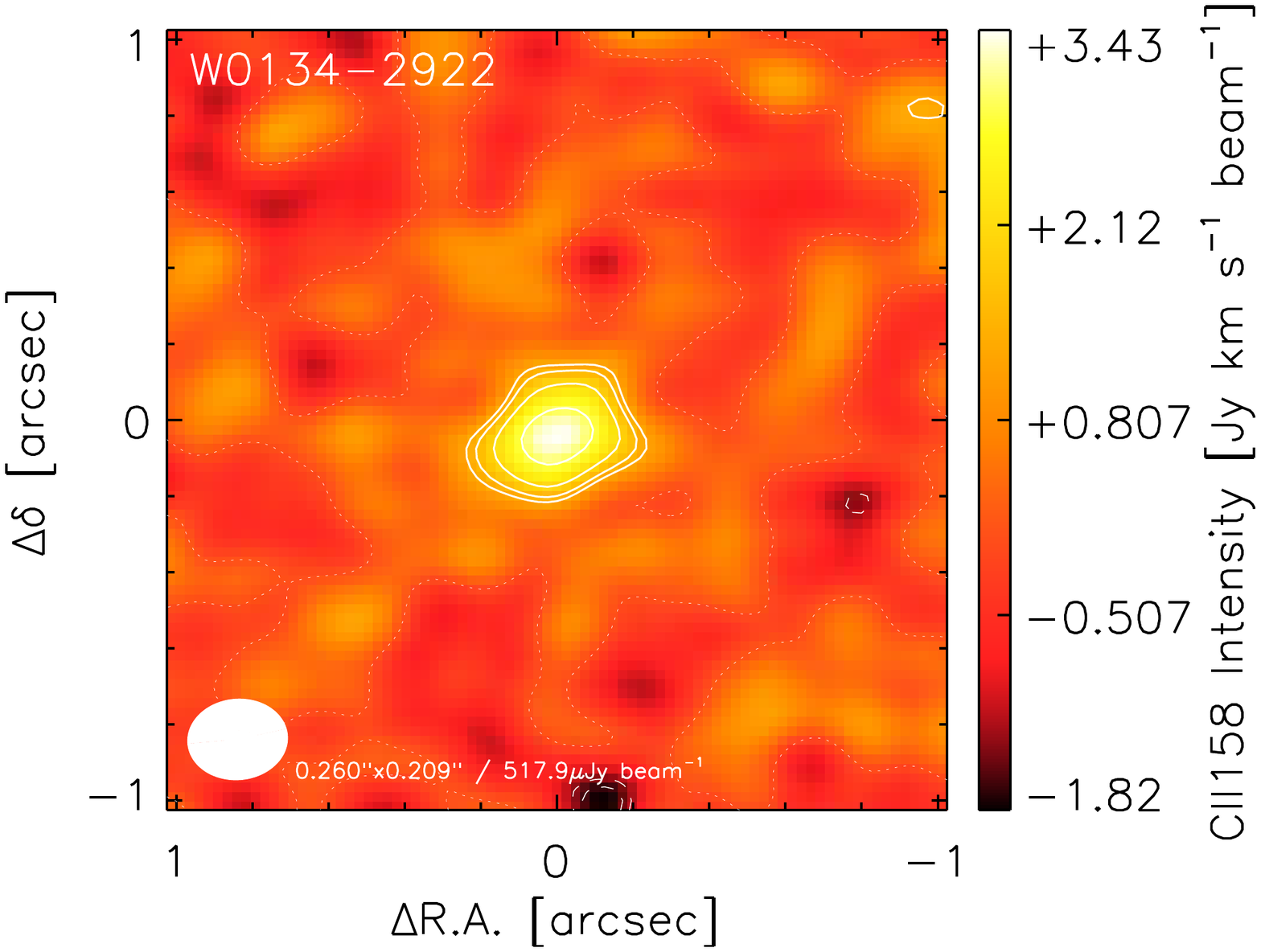}
\includegraphics[scale=0.34]{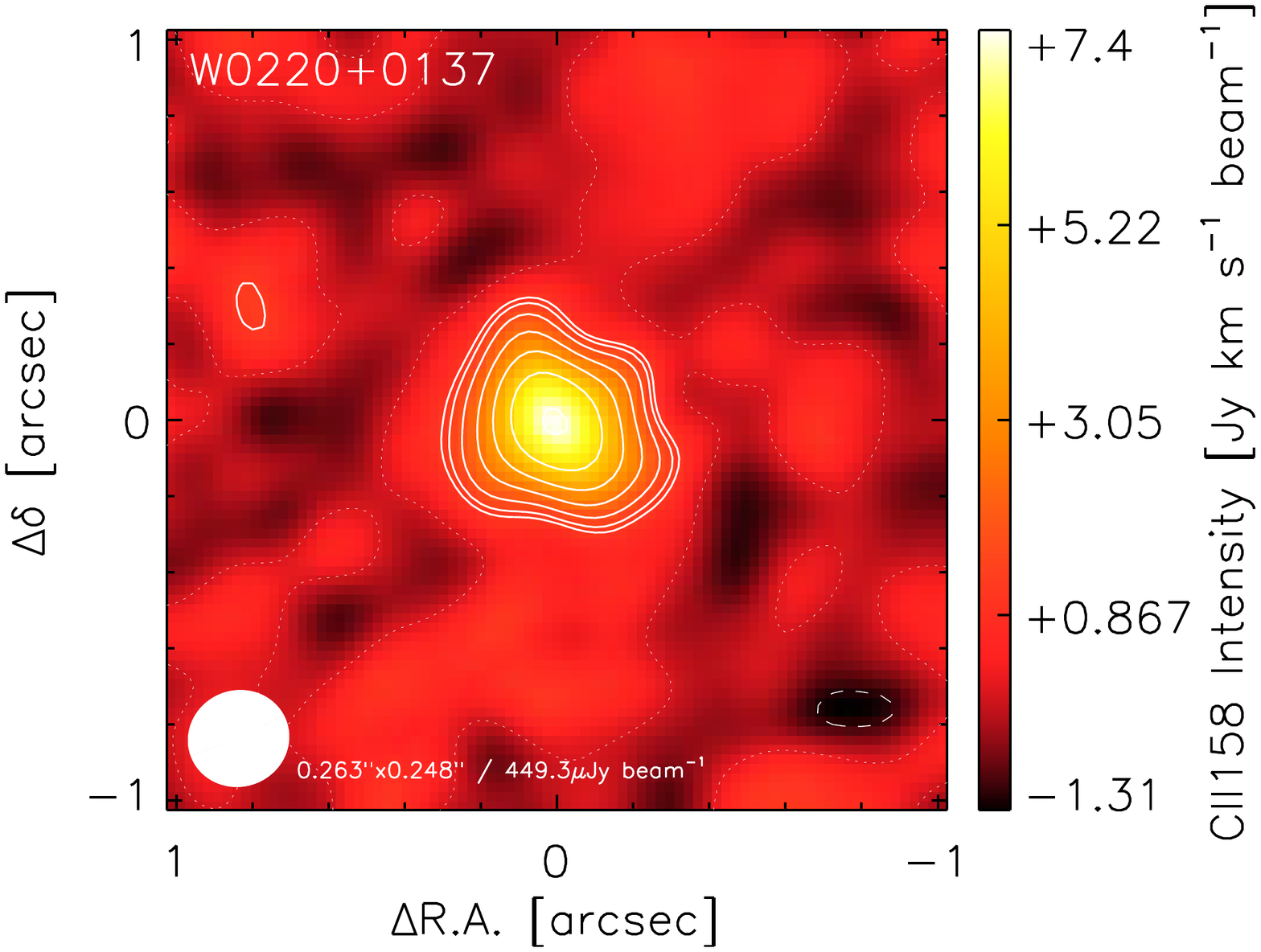}
\vspace{-4.cm}

\hspace{1.75cm}\includegraphics[scale=0.34]{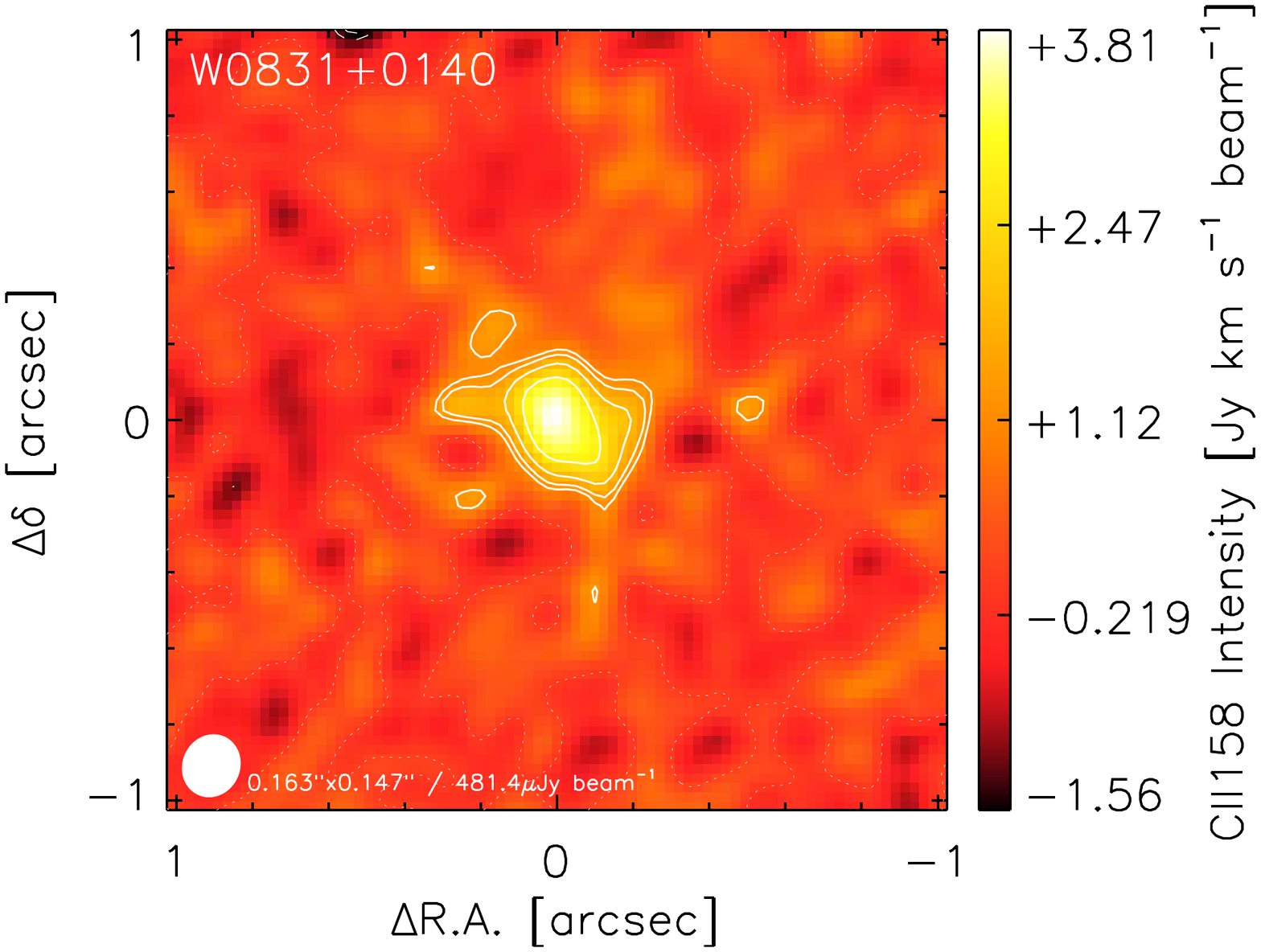}
\includegraphics[scale=0.34]{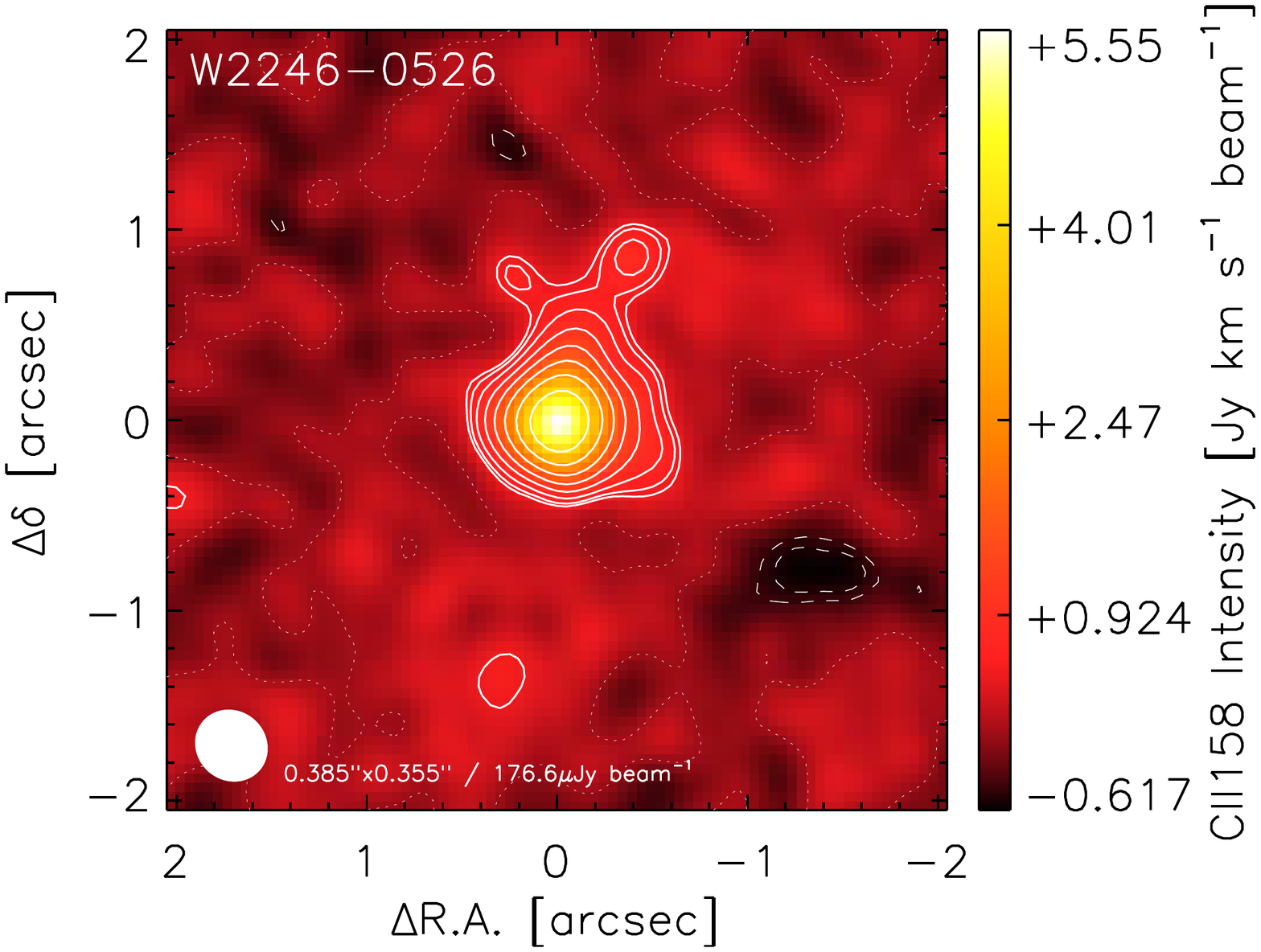}
\vspace{-3.5cm}
\caption{\footnotesize \CIIno\, intensity (moment 0) maps of the Hot DOGs in which the line has been detected. The beam is shown as a white ellipse on the bottom-left, together with its size and the depth (r.m.s.) of the reference SPW used to create the map. Contour levels are displayed as in Figure~\ref{f:contflux}. The FoVs are also the same as in Figure~\ref{f:contflux} to facilitate the comparison.}
\label{f:lineflux}
\end{figure*}

The top left and right panels in Figure~\ref{f:ciideficits} show the \CIIno/\LIR\, ratio as a function of \SigmaIR\, and \LIR, respectively, for our sample of Hot DOGs as well as for the comparison populations of high-\textit{z} QSOs and the GOALS sample. In the top-left panel, high-redshift QSOs seem to follow the tight trend displayed by nearby (U)LIRGs, in which higher \SigmaIR\, sources display lower \LCIIno/\LIR\, ratios regardless of their physical size (see color coding of the symbols) ---a trend whose extrapolation to the highest \SigmaIR\, intercepts the location where the Hot DOGs lie (see discussion in Section~\ref{ss:powersource}). Interestingly, while most OpQs and IRQs (star and diamonds symbols, respectively) seem to follow the local correlation within $\sim$\,2$\sigma$ of the dispersion, they tend to be systematically above it.


\subsubsection{\CIIno\, equivalent widths, \SigmaCII, and core sizes}\label{sss:ciiews}

The bottom-right panel of Figure~\ref{f:ciideficits} shows that the dust continuum emission of Hot DOGs is as compact as in \textit{z}\,$\simeq$\,4.7 IRQs but neither population reach sizes as small as those displayed by some \textit{z}\,$\gtrsim$\,6 OpQs. The most extreme of these cases show physical sizes (FWHM\,=\,2\,$\times$\,$R_{\rm eff}$) as small as $\sim$\,0.5\,kpc, more compact than the majority of local (U)LIRGs. It is interesting to note that, despite the wide range in dust continuum (and \CIIno) physical sizes, our luminous Hot DOGs seem to exhibit consistently large concentrations of \CIIno\, at any $R_{\rm eff}$, with \LCIIno\, surface densities as large as \SigmaCII\,=\,2\,$\times$\,10$^9$\,\lsd\, (or $\simeq$\,1\,$\times$\,10$^9$\,\lsd\, if \CIIno\, sizes are used; see below). Moreover, the entire sample of Hot DOGs seem to be within $\sim$\,50\% of that value. Some high-\textit{z} OpQs and IRQs reach the same level of \LCIIno\, density, but the lower end of the distribution extends down to much smaller values, \SigmaCII\,$\simeq$\,10$^8$\,\lsd.

In Hot DOGs with secure \CIIno\, detections, the size of the line emitting core (see Section~\ref{ss:size}) is clearly resolved in one case, W0831--0140, and marginally resolved in the rest (FWHM$_{\rm int}$\,$\simeq$\,\,FHWM$_{\rm beam}$; see Table~\ref{t:linefluxes}). The circularized, intrinsic \CIIno\, FWHMs are distributed around a narrow range of physical sizes, 1.8--2.4\,kpc, with an average close to $\simeq$\,2\,kpc. In terms of the continuum emission, the sizes range from unresolved (W0116--0505), to barely resolved (W0134--2922, W0220+0137 and W2246--0526) to securely resolved (W0236+0528, W0410--0913 and W0831+0140). The circularized dust sizes span a range between 1.1--2.6\,kpc, with a mean of 1.6\,kpc. In all cases, when both the line and continuum are detected, the size of the \CIIno\, emission is larger than that of the continuum. \citet{Venemans2020} report \CIIno\, and dust continuum sizes for the sample of 27 OpQs at \textit{z}\,$\gtrsim$\,6 introduced in \cite{Decarli2018}. They find ranges between $\simeq$\,0.7–6\,kpc and 0.5–7\,kpc, respectively. We note that the ratio of line-to-continuum sizes in Hot DOGs is very narrow, $\simeq$\,1.35--1.85, with a mean of 1.61\,$\pm$\,0.10, and remarkably similar to the averages found for the sample of \textit{z}\,$\gtrsim$\,6 OpQs, 1.68\,$\pm$\,0.11, and for the \textit{z}\,$\simeq$\,4.7 IRQs, 1.56\,$\pm$\,0.08. This suggests that the line and continuum sizes in high-redshift QSOs, regardless of their selection, could be linearly proportional, and may be pointing to a common physical origin for the size difference. However, to derive any statistically robust conclusion higher angular resolution observations are needed.

\subsection{Spatial distribution and extended emission}\label{ss:morphology}

As described in Section~\ref{s:analysis}, the intrinsic sizes derived from the 2D Gaussian fits are only representative of the core emission of the galaxy and, as can be seen in Figures~\ref{f:contflux} and \ref{f:lineflux}, some sources that are barely resolved in line and/or continuum emission based on this measurement, still show faint structures with clumpy morphologies extended towards particular directions. The most luminous of these clumps are galaxy neighbors located at distances 1--5\arcsec\, from the central Hot DOG host \citep[e.g., W2246--0526;][]{DS2016}. A blind search for companion galaxies in the ALMA \CIIno\, cubes will be presented in a separate work by \cite{GL2021}. The most extended Hot DOG in the sample is W0410--0913, which displays low surface brightness continuum emission extending up to 1--1.5\arcsec\, north-east and south-west from the central quasar, equivalent to $\simeq$\,10\,kpc (projected). Unfortunately, the \CIIno\, line was mostly missed in this object due to the redshift offset with respect to its $z_{\rm UV}$, as described in Section~\ref{ss:lineoffsets}. While fainter, W0236+0528 and W0831+0140 show a number of continuum emission knots surrounding the central galaxy. These clumps have flux density peaks with SNR\,$\simeq$\,3 and are surrounded by large areas of overall positive emission, tentatively suggesting that they may be real.

\begin{figure}
\includegraphics[scale=0.4]{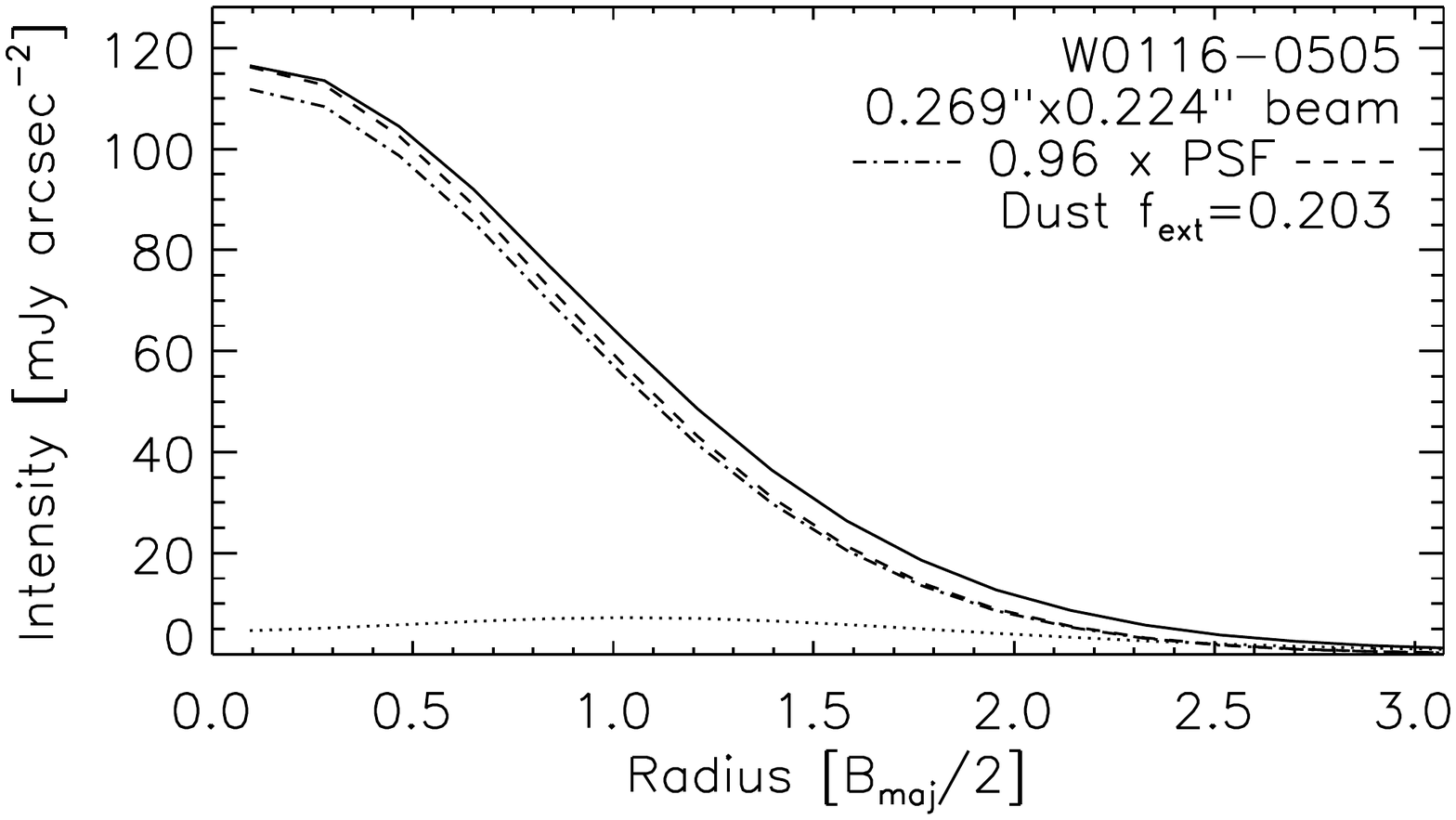}
\vspace{-6.3cm}

\includegraphics[scale=0.4]{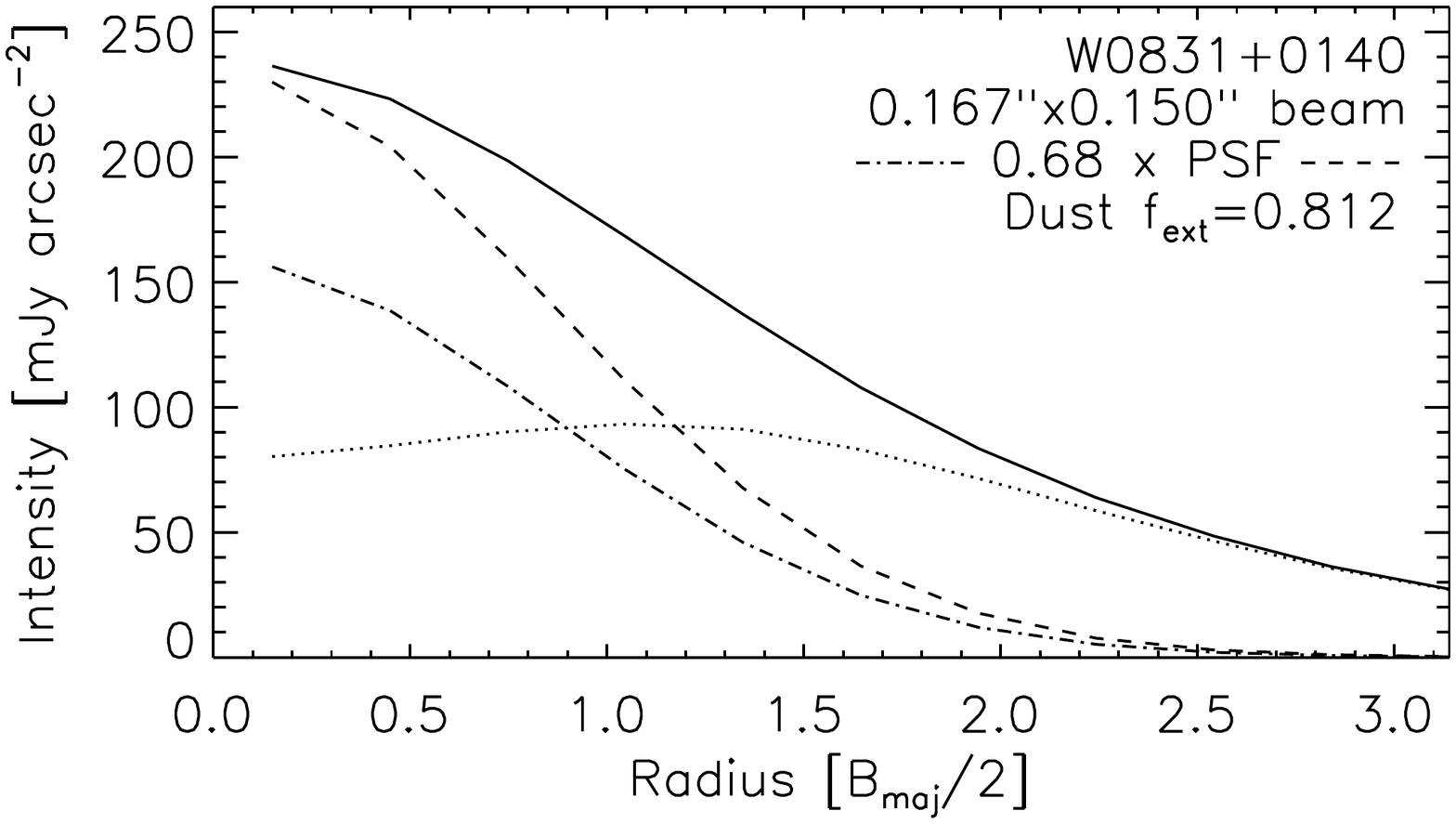}
\vspace{-6.3cm}

\includegraphics[scale=0.4]{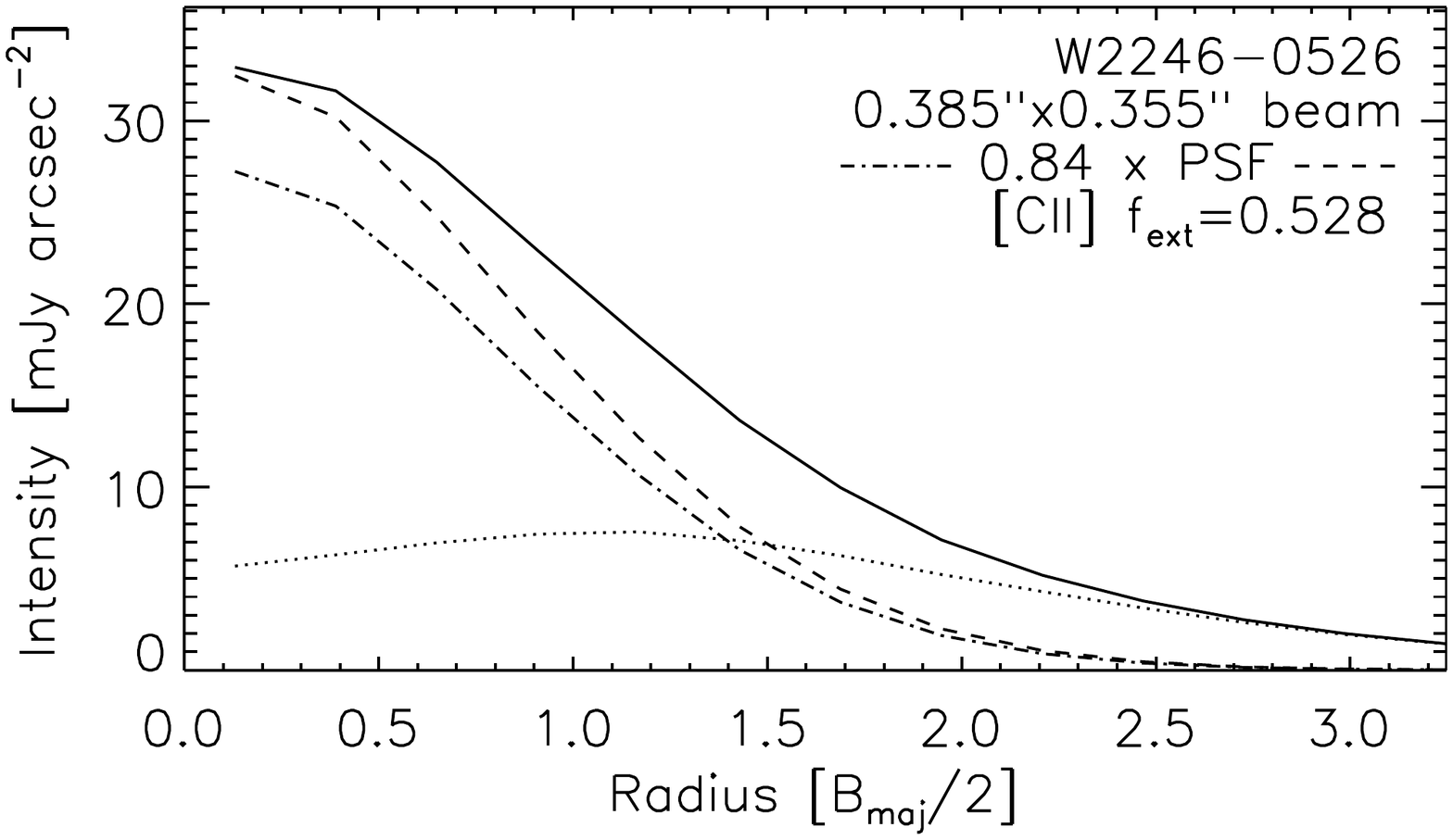}
\vspace{-6.3cm}

\includegraphics[scale=0.4]{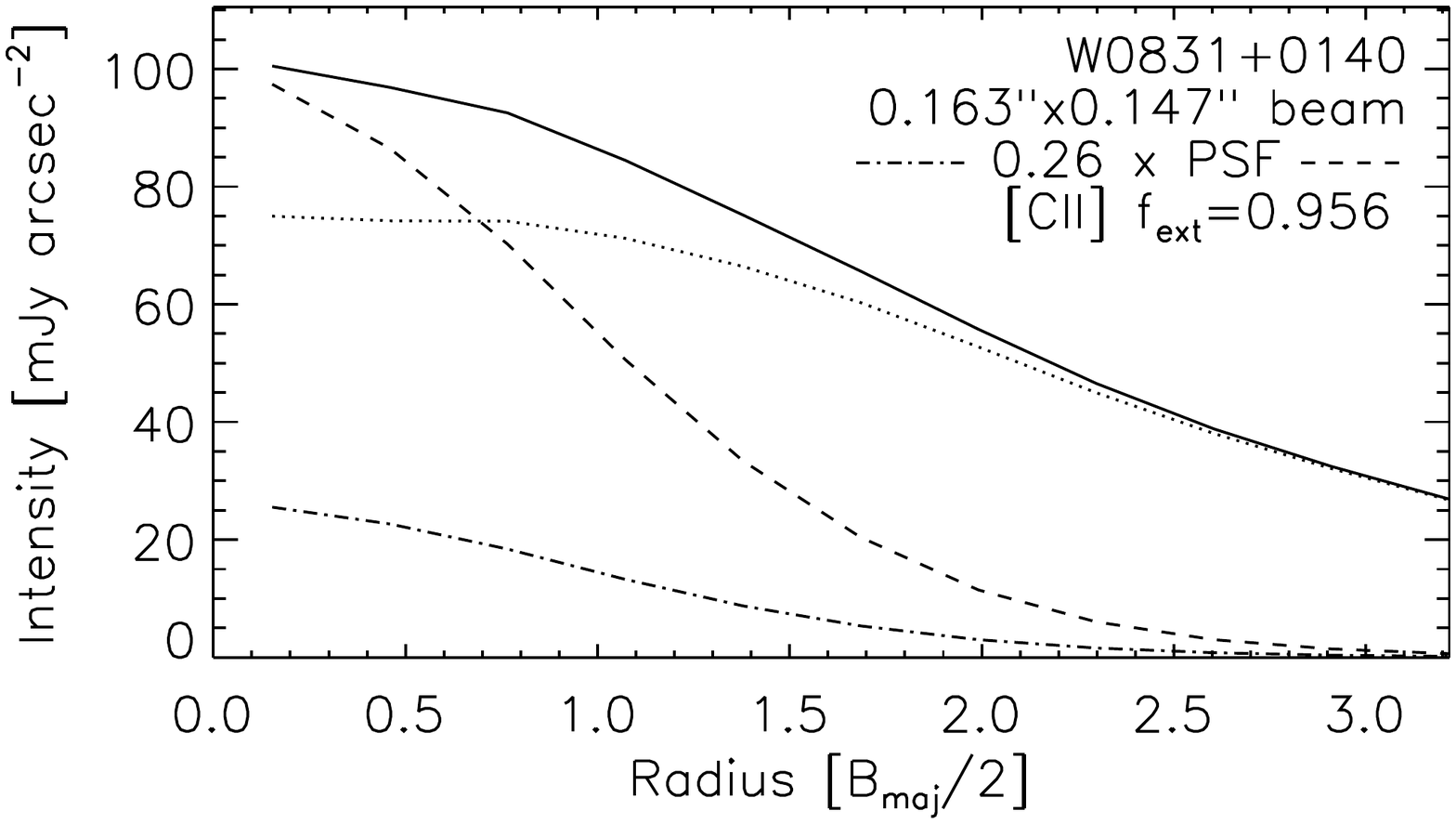}
\vspace{-6.3cm}
\caption{\footnotesize (top two panels) Azimuthal profiles of the dust continuum flux density for the Hot DOGs with the smallest (top, W0116--0505) and largest (bottom, W0831+0140) \FEE$_{\rm ,cont}$. (bottom two panels) Azimuthal profiles of the \CIIno\, flux for the Hot DOGs with the smallest (top, W2246--0526) and largest (bottom, W0831+0140) \FEE$_{\rm ,[CII]}$. The black line is the profile of the observed emission. The dashed line is the profile of the clean PSF. The dotted-dashed line is the scaled PSF that minimizes the scatter of the residuals within B$_{\rm maj}$/2\,=\,1 (see Section~\ref{s:analysis} for details and values in Tables~\ref{t:linefluxes} and \ref{t:contfluxes}). The dotted line is the profile of the residuals (observed -- scaled-PSF). The PSF scaling value, \FEE\, and the radius used to perform the aperture photometry are noted at the top right of each panel (see dotted circumferences in Figure~\ref{f:contflux}).}
\label{f:spatprofs}
\end{figure}

In addition to the sizes, Tables~\ref{t:linefluxes} and \ref{t:contfluxes} present the fraction of extended emission (\FEE; line and continuum) underlying the central, unresolved point-source. In all Hot DOGs, the \FEE$_{\rm ,[CII]}$ is $\gtrsim$\,50\%, and up to $\sim$\,95\% in the case of W0831+0140, indicating that most of the ionized gas is extended over several kpc and is likely associated with star formation in the host galaxy (see discussion in Section~\ref{s:discussion}). On the other hand, the dust continuum shows \FEE$_{\rm ,cont}$ ranging from very compact ($\sim$\,20\%, W0116--0505) to very extended sources ($\sim$\,80\%, W0831+0140). These values are always smaller than \FEE$_{\rm ,[CII]}$, implying that the continuum emission is arising from a volume closer to the central AGN than that responsible for \CIIno. Because we are measuring the rest-frame continuum at $\sim$\,160\,\mic, this implies that either: a) the AGN could be contributing down to the coldest components of the distribution of dust temperatures in the galaxy ($\sim$\,20\,K), something rarely seen in the local Universe (e.g., not even in ultra-luminous quasars like Mrk\,231), and/or b) the temperature distribution of the dust heated by the AGN is typical, with the emission falling rapidly at $\lambda$\,$\gtrsim$\,40\,\mic\, \citep[e.g.,][]{Elvis1994, Nenkova2008b, Mullaney2011}, but its total luminosity completely overwhelms that of the underlying galaxy, and thus still contributes significantly to the dust continuum at longer wavelengths. We note that the physical resolution of the current ALMA data is $\sim$\,1.5--2\,kpc, and therefore much higher resolution observations are needed to further isolate the gas and dust structures located under the sphere of influence of the SMBH; roughly within a radius of a few hundred pc for the estimated masses of these BHs \citep{Wu2018, Tsai2018}.

\begin{figure*}
\hspace{1.9cm}\includegraphics[scale=0.34]{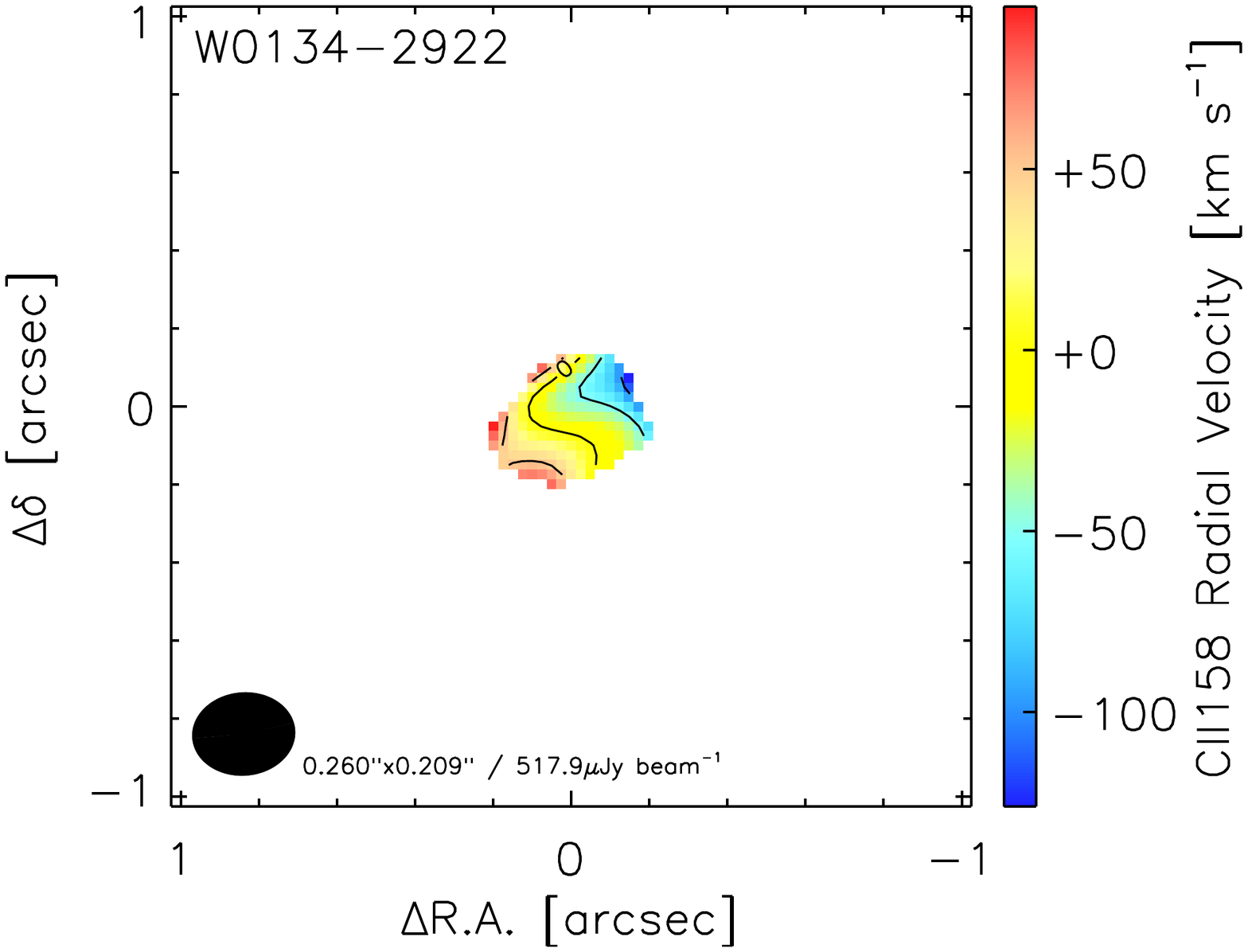}
\includegraphics[scale=0.34]{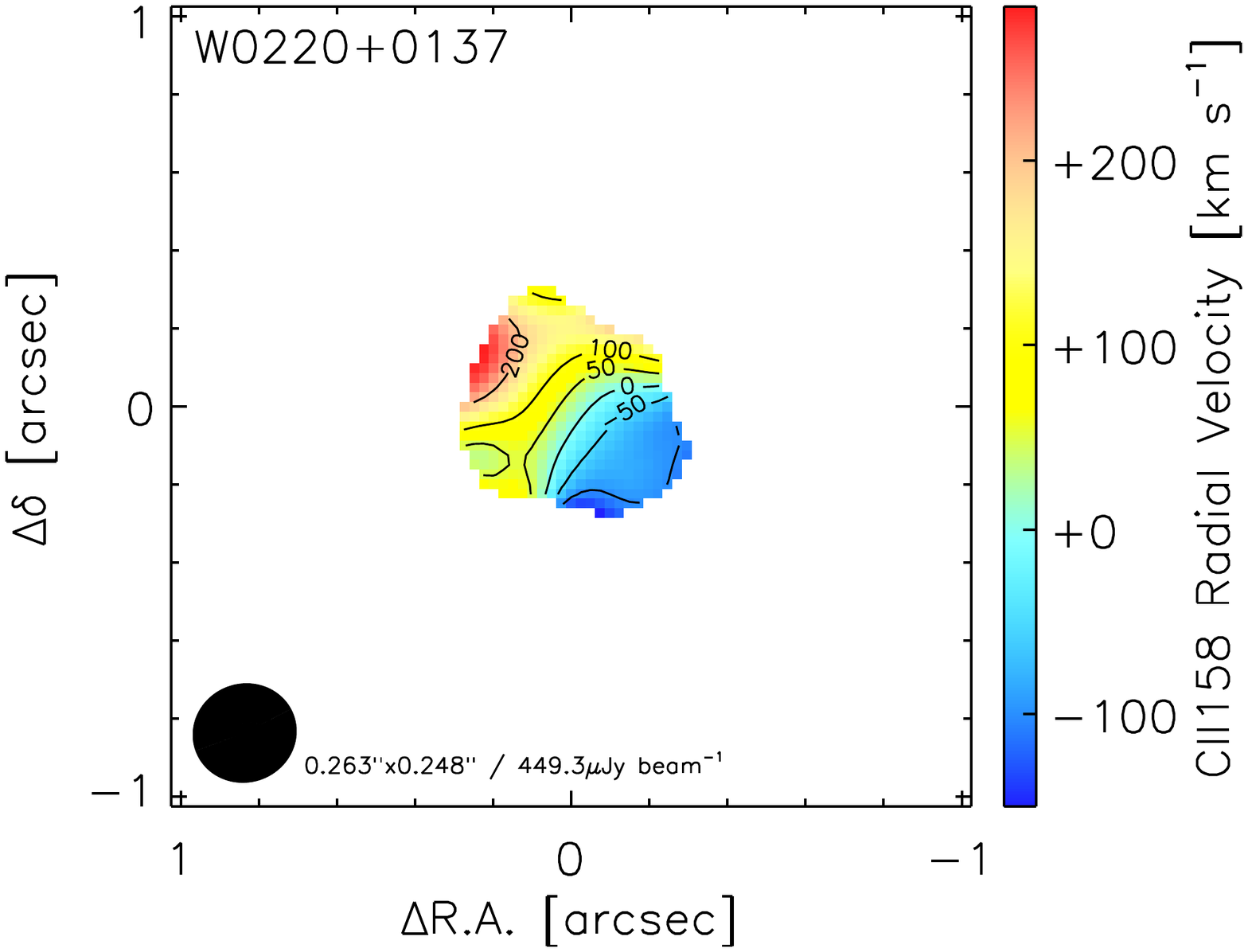}
\vspace{-4.cm}

\hspace{1.9cm}\includegraphics[scale=0.34]{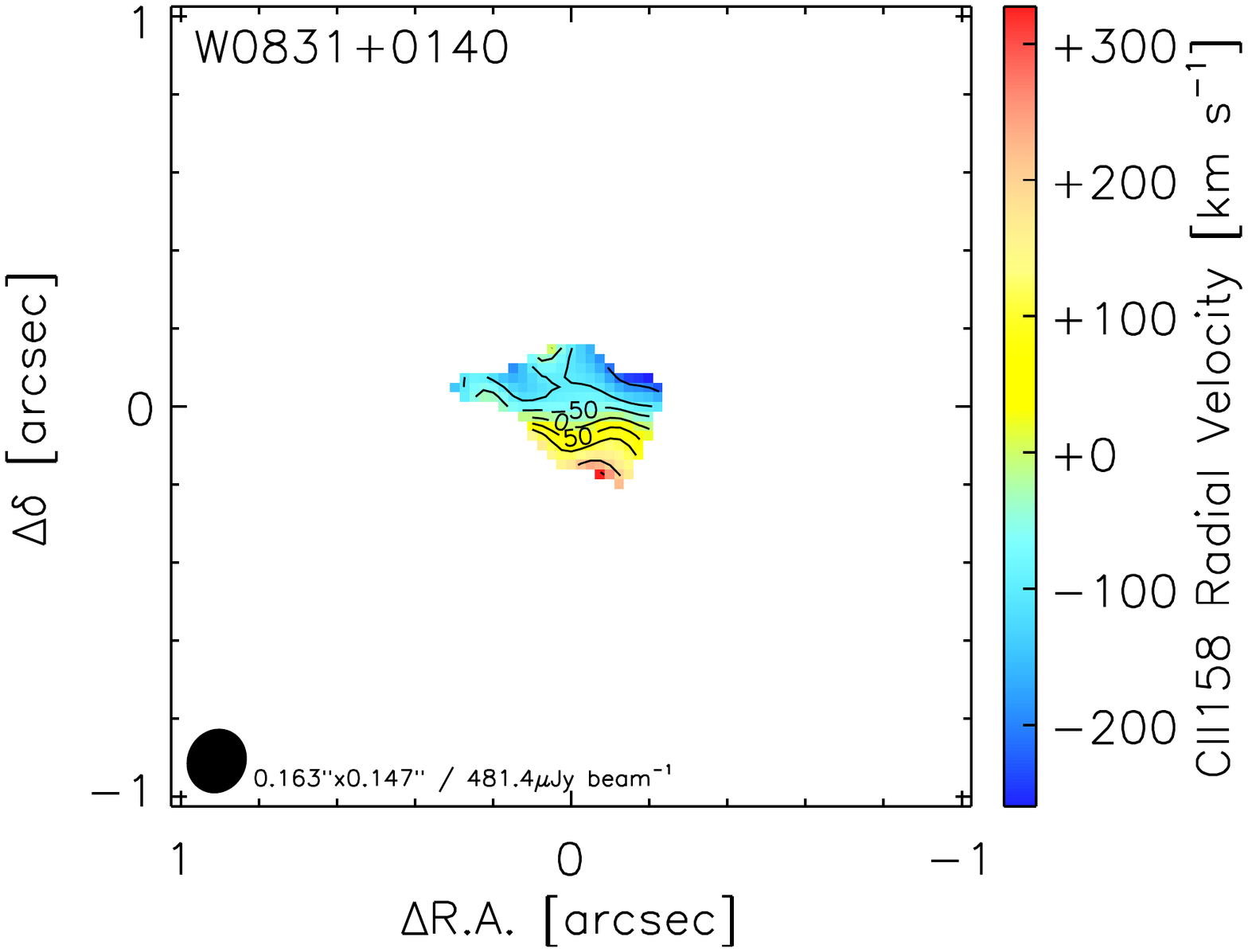}
\includegraphics[scale=0.34]{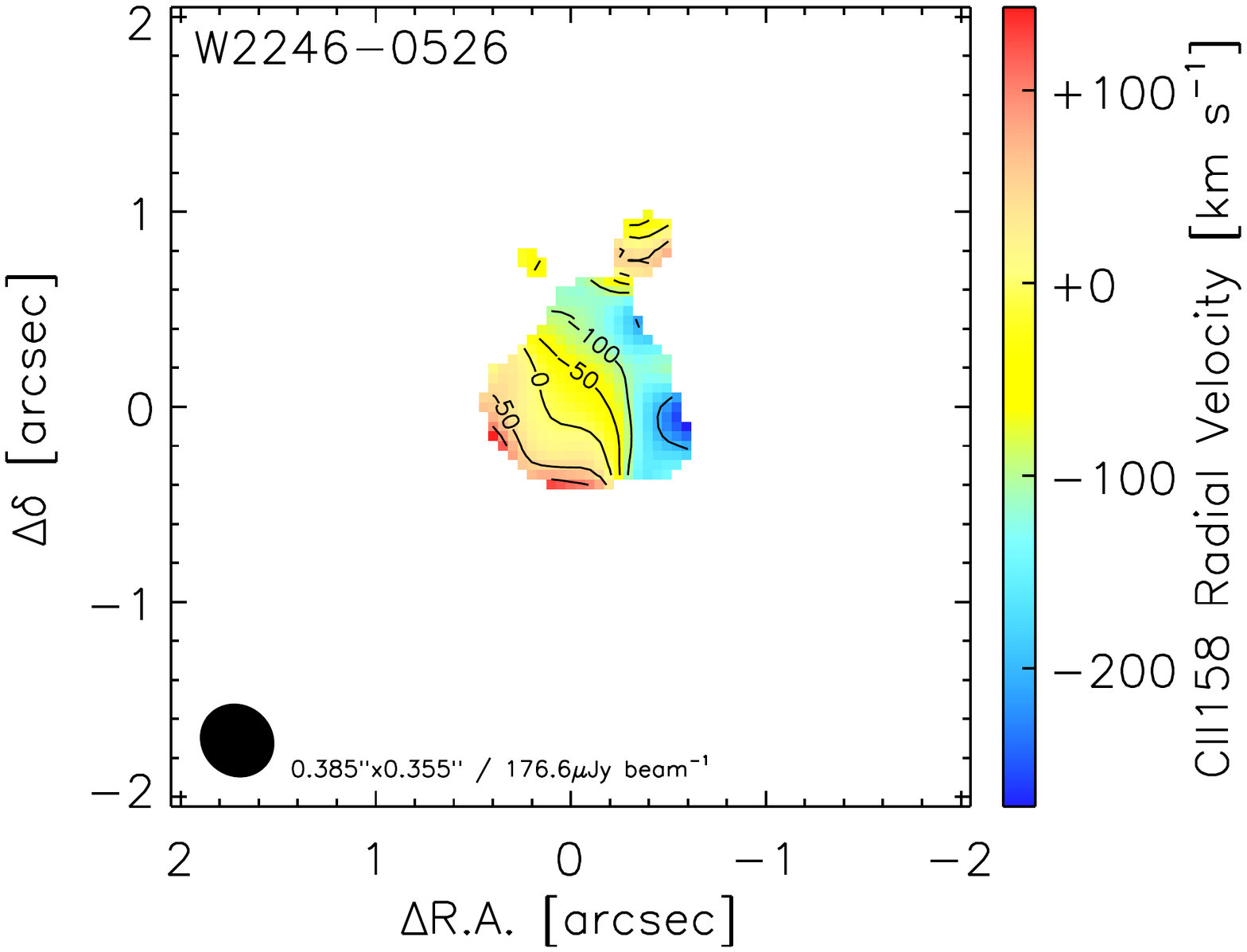}
\vspace{-3.5cm}
\caption{\footnotesize \CIIno\, projected radial velocity (moment 1) maps down to a 3$\sigma$ level (see contours in Figure~\ref{f:lineflux}). The beam is shown as a black ellipse on the bottom-left, together with its size and the depth (r.m.s.) of the reference SPW used to create the map. Contour levels are drawn at [0, $\pm$50, $\pm$100, $\pm$200, $\pm$300\dots]\,\kmns. The FoVs are the same as in Figure~\ref{f:contflux} to facilitate the comparison.}
\label{f:velfield}
\end{figure*}

\begin{figure*}
\hspace{1.9cm}\includegraphics[scale=0.34]{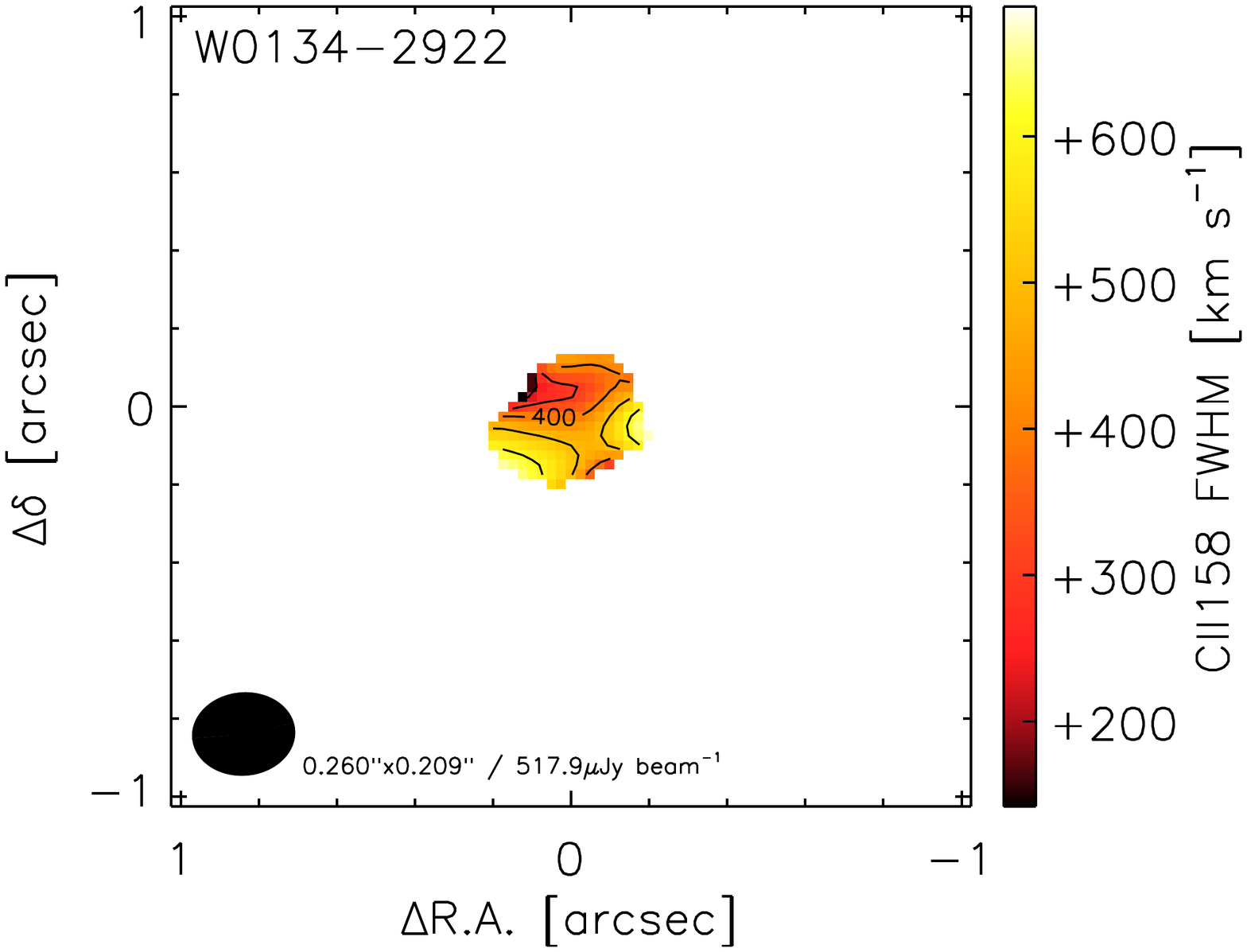}
\includegraphics[scale=0.34]{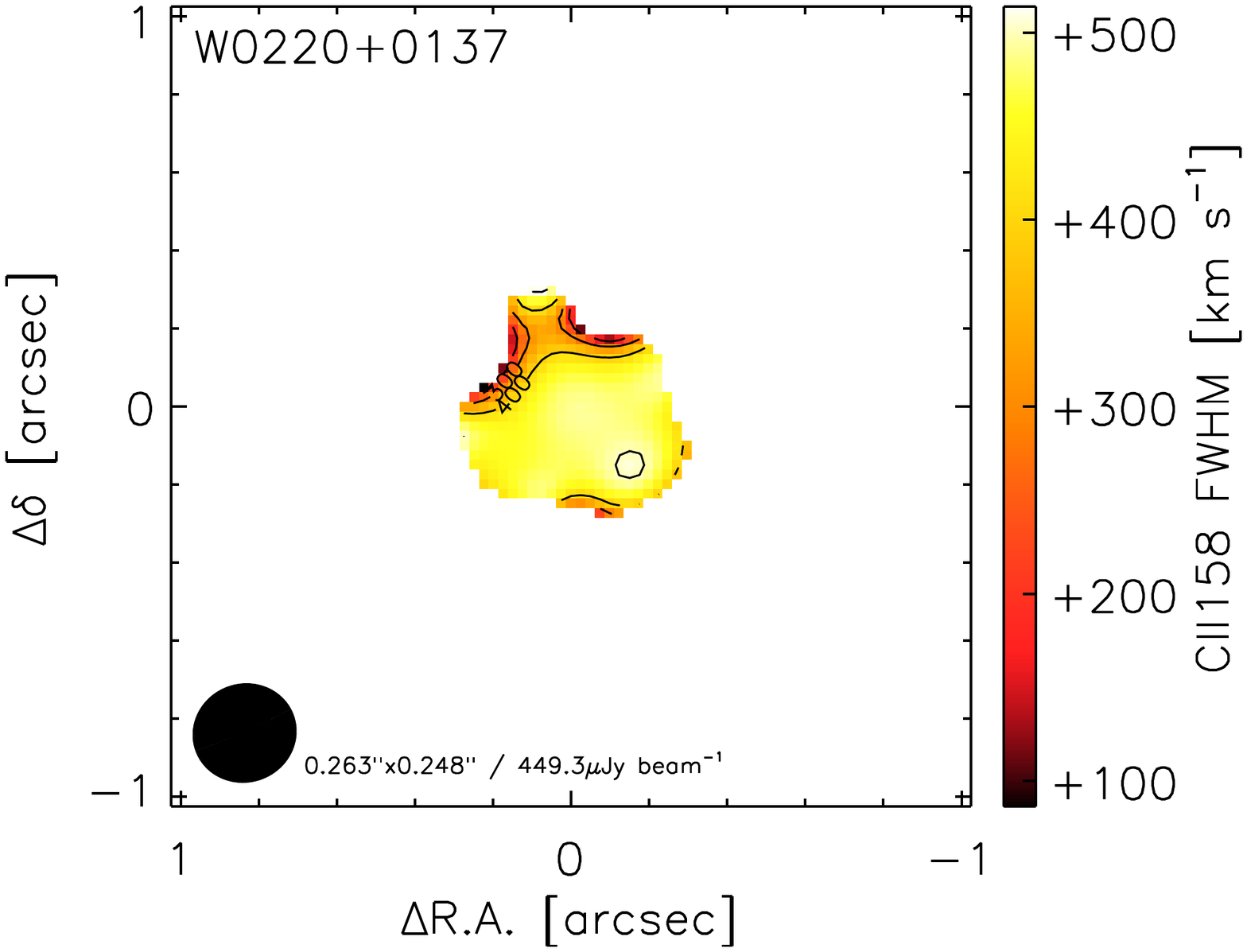}
\vspace{-4.cm}

\hspace{1.9cm}\includegraphics[scale=0.34]{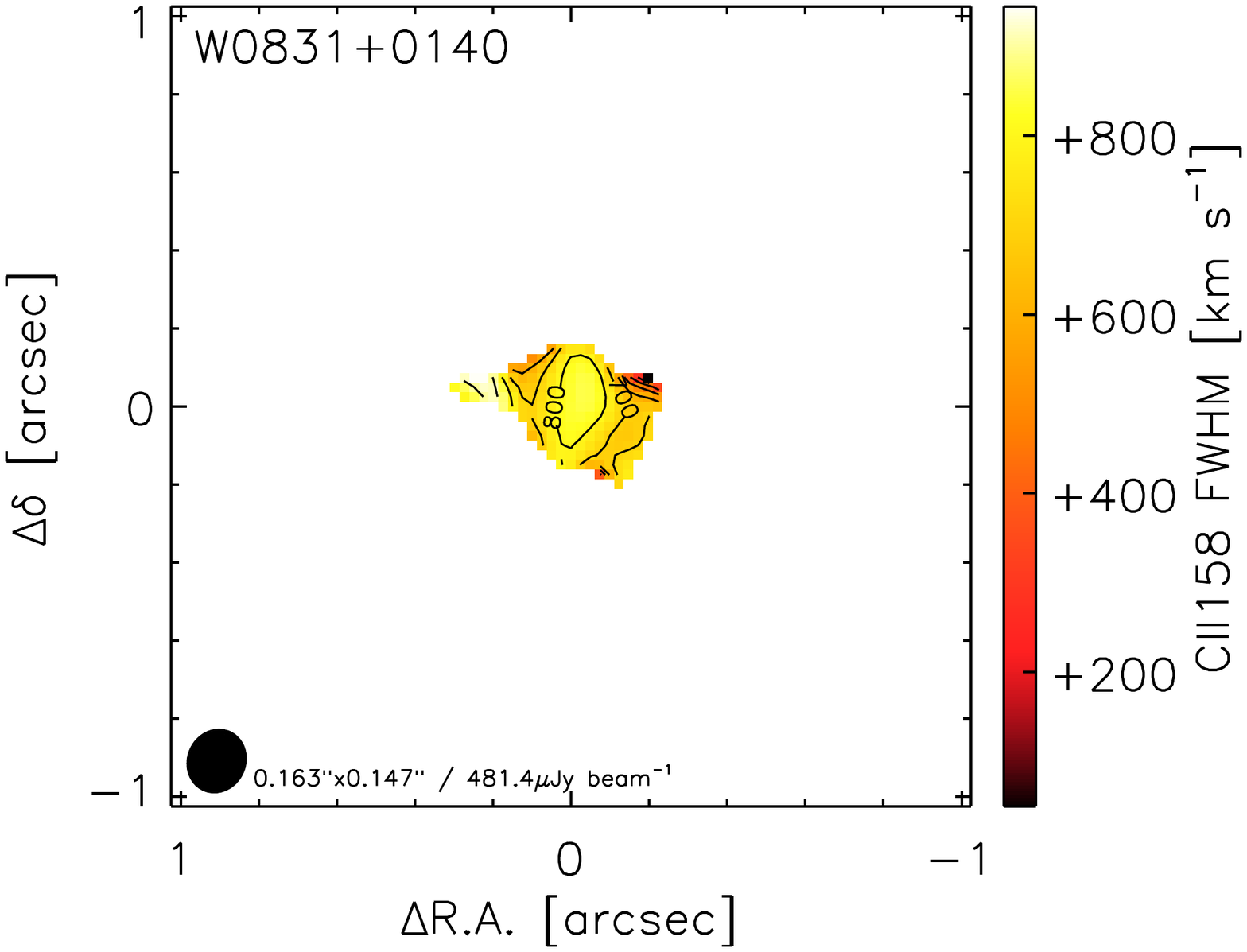}
\includegraphics[scale=0.34]{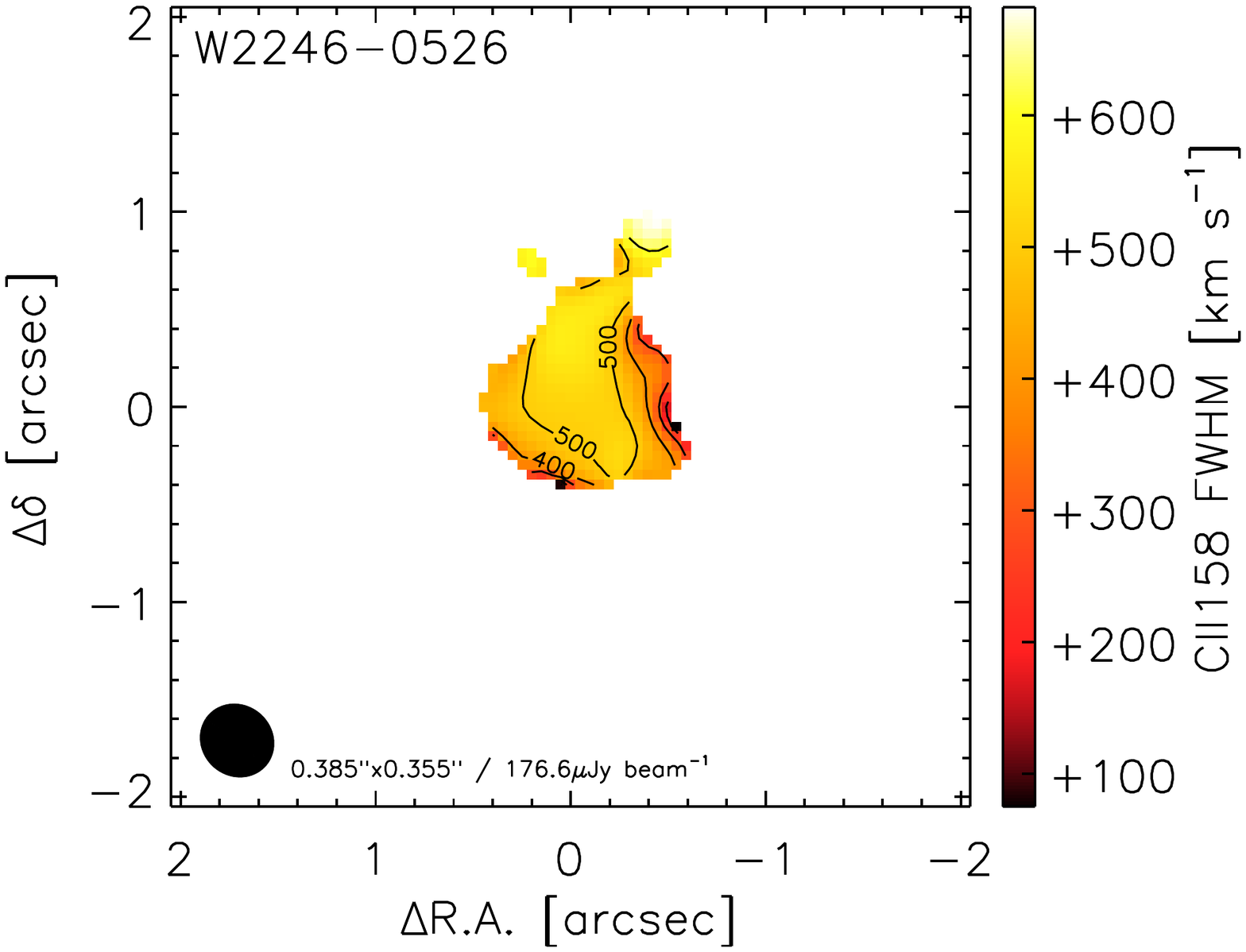}
\vspace{-3.5cm}
\caption{\footnotesize \CIIno\, FWHM (moment 2) maps down to a 3$\sigma$ level (see contours in Figure~\ref{f:lineflux}). The beam is shown as a black ellipse on the bottom-left, together with its size and the depth (r.m.s.) of the reference SPW used to create the map. Contour levels are drawn at [0, 25, 50, 100, 200, 300\dots]\,\kmns. The FoVs are the same as in Figure~\ref{f:contflux} to facilitate the comparison.}
\label{f:veldisp}
\end{figure*}

Figure~\ref{f:spatprofs} shows the radial density profiles of the continuum and line emissions, for the most compact and extended sources in the Hot DOG sample (see Section~\ref{s:analysis} for details). The continuum emission of W0116--0505 is very compact. Although the core of the galaxy is effectively unresolved (Table~\ref{t:contfluxes}), about 20\% of the total flux still arises from a very extended area of low surface brightness emission. W0831+0140 shows both the most extended continuum emission and the most extended line emissions, with \FEE\,$\gtrsim$\,80\% in both cases. However, we note that W0410--0913 has the largest continuum core, which suggests that W0831+0140 shows the largest \FEE$_{\rm ,cont}$ because the beam size of these observations is half that of W0410--0913. Indeed, if we use a \textit{uv}-tapering of 0.5\arcsec\, to place all the cubes on a common beam size, the \FEE$_{\rm ,cont}$ of W0410--0913 becomes the largest of all the Hot DOGs ($\simeq$\,50\%) and that of W0831+0140 is reduced to $\simeq$\,30\%.

\section{Kinematics}\label{s:kinematics}

\subsection{Velocity fields and dispersion maps}

Figure~\ref{f:velfield} presents the moment 1 maps of the Hot DOG sample for the sources in which the \CIIno\, line has been clearly identified. As can be seen, the kinematics are very diverse: W0134--2922 shows a very small velocity gradient; W0220+0137 displays a smoothly varying structure more consistent with a rotating disk; W0831+0140 shows a complex velocity distribution; and W2246-0526 shows relatively slow rotation, with $\Delta v$\,$\simeq$\,150\,\kmns\, \citep[c.f.,][]{DS2016}. While W0410--0913 is not shown, the detected, blue-shifted component of the line reaches velocities up to $\sim$\,600\,\kmns\, (assuming the \CIIno\, systemic velocity is the same as that derived from $z_{\rm CO}$).

The case of W0831+0140 is of particular interest. In this source, the complex velocity field is likely caused by the presence of multiple narrow-velocity components that can be seen in the spectral cube, suggesting that this source could actually be a system of clumps or galaxies in the process of merging. However, higher angular resolution and SNR observations are needed to accurately disentangle these components spatially and in velocity space. In fact, the sources are in general either not well resolved (W0134--2922) or do not have velocity fields smooth enough to perform rigorous kinematic modeling (W0831+0140). The exceptions are W0220+0137 and to a more limited extent, W2246--0526. We show the results from the kinematic modeling, using the tool $^{\rm 3D}$Barolo, in Appendix~\ref{a:kinmod}).

\begin{figure}
\includegraphics[width=\hsize]{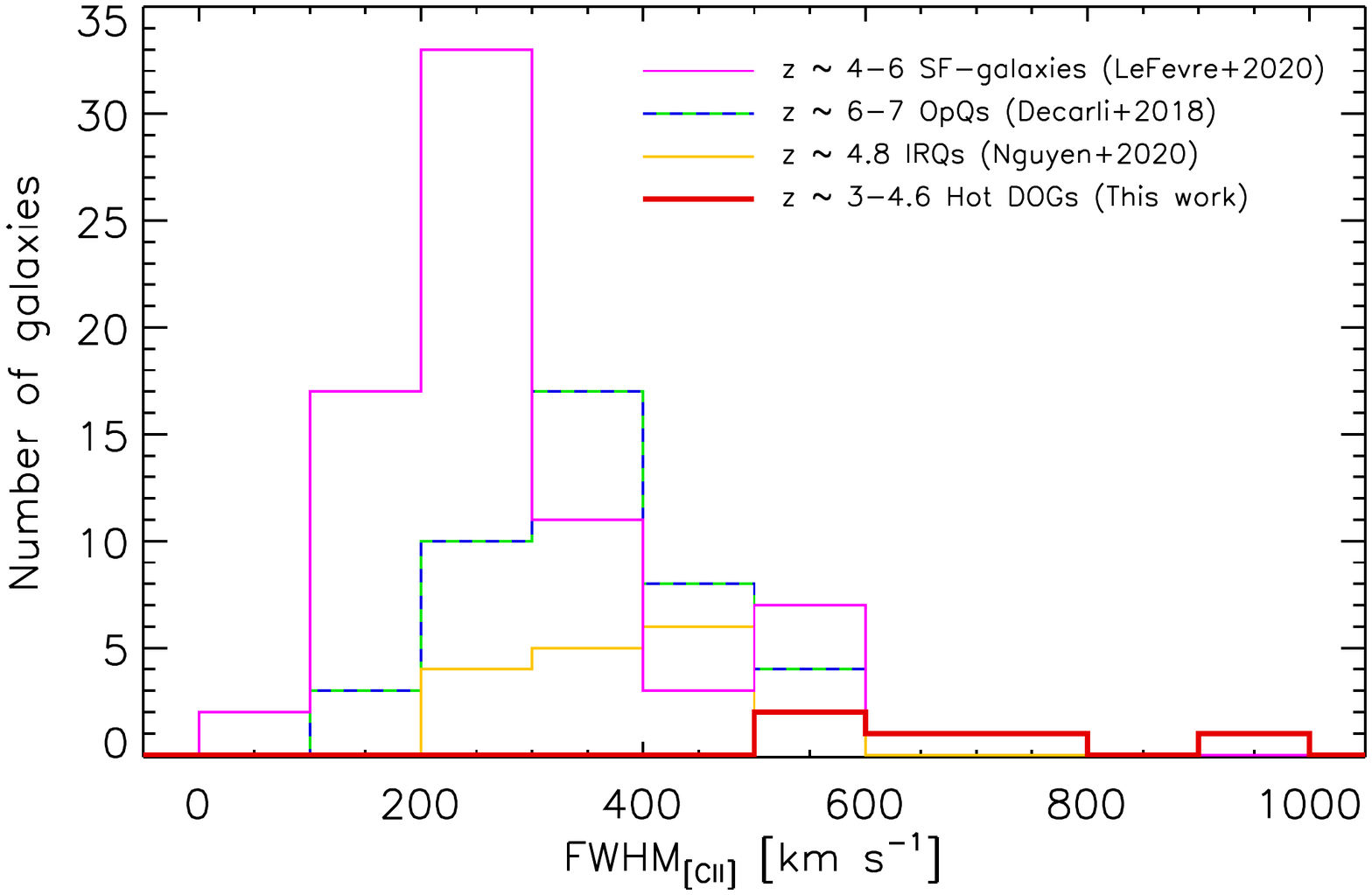}
\vspace{-6.cm}
\caption{Histogram of the \CIIno\, FWHM for the sample of Hot DOGs (red line). The \textit{z}\,$\simeq$\,4.7 IRQs from \citet{Trakhtenbrot2017} and \citet{Nguyen2020} are shown in yellow. The compilation of \textit{z}\,$\gtrsim$\,6 OpQs from \citet{Decarli2018} is displayed as a green/blue dashed line. The 4\,$\lesssim$\,\textit{z}\,$\lesssim$\,6 star-forming galaxies from \citet{LeFevre2020} are shown in pink.}
\label{f:fwhm}
\end{figure}

The moment 2 maps of the Hot DOG sample (see Figure~\ref{f:veldisp}) show relatively uniform values of the velocity dispersion across the area where the \CIIno\, line is detected. Moreover, the measured peak and integrated FWHM values are always in excess of 500\,\kmns, thus suggesting a highly turbulent gas component across the host galaxies. Histograms of the integrated FWHM$_{\rm [CII]}$ for our sample of luminous Hot DOGs and for the comparison quasar samples are presented in Figure~\ref{f:fwhm}. For reference, the distribution of MS galaxies at \textit{z}\,$\sim$\,4--6 is shown as well. Quasars have on average higher \CIIno\, velocity dispersions than star-forming galaxies, and in turn Hot DOGs have systematically larger dispersions than OpQs and IRQs. A two-sample Kolmogorov-Smirnov test performed over the Hot DOG and the combined quasar samples yields a value of the statistic \textit{D}\,=\,0.95 with a significance of \textit{p}\,$\simeq$\,1\,$\times$\,10$^{-4}$, implying that the two distributions are very likely to be drawn from different parent populations. Individual tests comparing the Hot DOGs with each OpQ and IRQ sample also provide \textit{D}\,$\geq$\,0.9 and \textit{p}\,$\leq$\,3\,$\times$\,10$^{-4}$. We explore the implications of this result in the discussion in Section~\ref{s:discussion}.

\subsection{Dynamical masses from modeling}

The kinematic analysis described in Appendix~\ref{a:kinmod} allows for a proper derivation of the dynamical masses of W0220+0137 and W2246--0526 host galaxies (see Equations~\ref{e:mdynrot}, \ref{e:mdyndisp}), which we can compare to the stellar masses estimated in Section~\ref{s:ms} and to BH masses published in the literature (see Table~\ref{t:masses}).

\begin{table}
\caption{Star formation, BH and dynamical properties}
\centering
\scriptsize
\label{t:masses}
\begin{tabular}{ccccc}
\hline\hline
Galaxy & SFR & \Mstar & \MBHdot & \Mdyn \\ 
 & [\Msun\,yr$^{-1}$] & [10$^{11}$\,\Msun] & [10$^{9}$\,\Msun] & [10$^{10}$\,\Msun] \\
(1) & (2) & (3) & (4) & (5) \\
\hline
W0116--0505 & 202$^{+122}_{-122}$     & 4.83$^{+3.30}_{-2.12}$  & $\leq$\,3.2 & \dots \\
W0134--2922 & 431$^{+894}_{-364}$     & 2.00$^{+1.36}_{-0.86}$  & \dots & \dots \\
W0220+0137  & 803$^{+992}_{-713}$     & 1.21$^{+0.82}_{-0.51}$  & $\leq$\,40 & 2.8\,$^{+1.9}_{-1.1}$ \\
W0236+0528  & 493$^{+295}_{-295}$     & 0.84$^{+0.57}_{-0.36}$  & \dots & \dots \\
W0410--0913 & 3314$^{+1988}_{-1988}$  & 1.61$^{+1.09}_{-0.70}$  & $\leq$\,316 & \dots \\
W0831+0140  & 2863$^{+3581}_{-2540}$  & 1.52$^{+1.04}_{-0.67}$  & $\leq$\,5.0 & \dots \\
W2246--0526 & 688$^{+1310}_{-590}$    & 3.04$^{+2.07}_{-1.34}$  & 4.0\,$^{+6.0}_{-2.4}$ & 5.5\,$^{+2.3}_{-1.6}$ \\
\hline\hline
\end{tabular}
\tablefoot{\scriptsize (1) Galaxy name; (2) Star formation rate of the galaxy host, derived from far-IR scaling relations (Section~\ref{s:ms} for details); (3) Stellar mass of the galaxy host, derived from rest-UV through near-IR photometry (see text and Appendix~\ref{a:sedfit} for details); (4) BH mass, derived from rest-UV and optical spectroscopy (\citealt{Finnerty2020}; except W2246--0526, \citealt{Tsai2018}); (5) Dynamical mass, derived from kinematic modeling of the \CIIno\, line (this work; see Appendix~\ref{a:kinmod}). The errors quoted in the SFR and \Mstar\, are not standard deviations, but refer to the entire range of values spanned by the individual estimates used to calculate the averages.}
\end{table}

The best-fit kinematic model for W0220+0137 yields a \Mdyn\,=\,2.8\,$^{+1.9}_{-1.1}$\,$\times$\,10$^{10}$\,\Msun\, regardless of whether the galaxy is assumed to be rotation-supported (i.e., where most of the mass is a thin disk with ordered rotation) or dispersion-supported. In turn, W2246--0526 results in \Mdyn\,=\,5.6\,$\pm$\,$^{+2.3}_{-1.6}$\,$\times$\,10$^{9}$\,\Msun\, assuming the former, and \Mdyn\,=\,5.5\,$^{+2.3}_{-1.6}$\,$\times$\,10$^{10}$\,\Msun\, considering the latter. Since the modeling of W2246--0526 strongly favors a dispersion-supported system, we refer from here on to the value derived from this assumption as the most likely \Mdyn. This value is within the uncertainties of the \Mdyn\, obtained by \cite{DS2018} based on simple considerations of the \CIIno\, line width and the size of the emitting region, \Mdyn\,=\,8\,$\pm$\,4\,$\times$\,10$^{10}$\,\Msun.

In Section~\ref{s:ms} we derive the stellar masses of the Hot DOG hosts from a variety of assumed SFR histories. The values obtained for W0220+0137 and W2246--0526 are \Mstar\,=\,1.2\,$\pm$\,0.6\,$\times$\,10$^{11}$\,\Msun\, and \Mstar\,=\,3.0\,$\pm$\,1.6\,$\times$\,10$^{11}$\,\Msun, respectively. Notwithstanding that the values of \Mstar\, and \Mdyn\, still agree at a $\simeq$\,3$\sigma$ level (given the large uncertainties in both measurements), the stellar mass of the Hot DOG hosts appears to exceed their dynamical mass by a factor of $\sim$\,5, suggesting \Mstar\, might be significantly over-estimated. Indeed, it should be noted that the stellar masses are based on photometric data \citep[for details, see the supplementary material in][]{DS2018} that have much lower angular resolution than the ALMA observations used to derive the dynamical masses. Specifically, while the observed angular sizes measured from the \CIIno\, line for both W0220+0137 and W2246--0526 are around $\simeq$\,0.3--0.4\arcsec, the \textit{Spitzer} and \textit{WISE} broad-band photometry have an angular resolutions of $\simeq$\,2--6\arcsec; around an order of magnitude larger in projected area than the ALMA data. Thus, the near-IR photometry could include emission from companion galaxies and extended structures not captured by the ALMA-derived \Mdyn. This caveat is not only important for Hot DOGs but also when comparing stellar and dynamical masses of any quasar or galaxy population at high redshift. Sub-arcsecond, rest-frame near-IR observations with \textit{JWST} will be needed to estimate the stellar masses of Hot DOG's hosts that allow for a direct comparison with the dynamical masses obtained from ALMA at similar physical scales.

\begin{figure}
\includegraphics[width=\hsize]{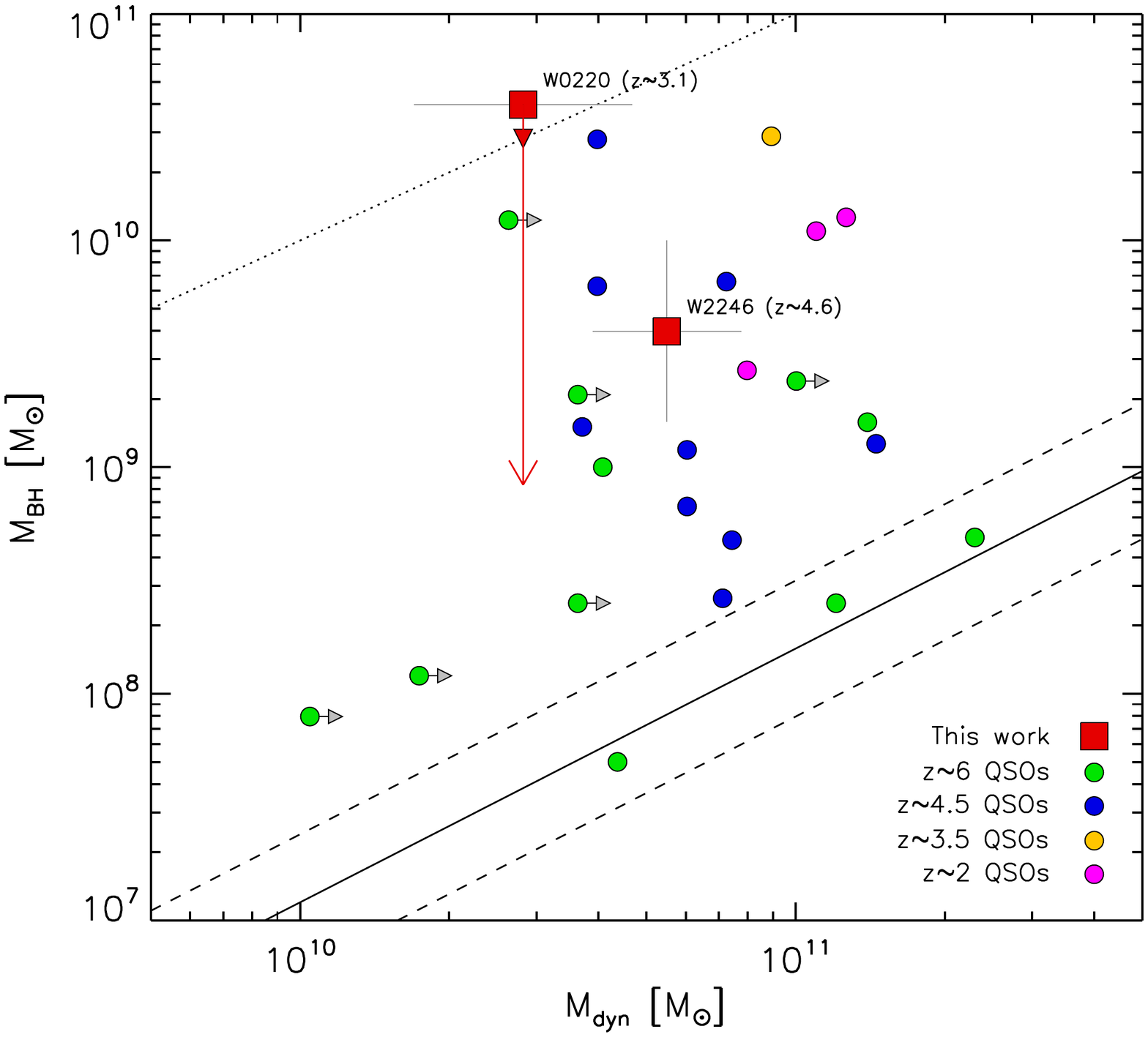}
\vspace{-4.cm}
\caption{SMBH mass, $M_{\rm BH}$, as a function of dynamical mass, \Mdyn, for the Hot DOGs W0220+0137 and W2246--0526. Literature OpQ and IRQ samples, from \textit{z}\,$\sim$\,2 to \textit{z}\,$\sim$\,6, are displayed for comparison \citep[][and references therein]{Bischetti2020}. The black solid line is the $M_{\rm BH}$ versus dynamical bulge mass relation from the nearby galaxy sample of \cite{Haring2004}, and the dashed represents the 1$\sigma$ intrinsic scatter. The dotted line is the $M_{\rm BH}$/\Mdyn\,=\,1 ratio. The SMBH mass of W0220+0137 is an upper limit \citep{Finnerty2020}, and the large red arrow pointing downwards represents the change in $M_{\rm BH}$ if a \lamEdd\,=\,2.8 is assumed, equal to that derived for W2246--0526 (see text for details).}
\label{f:mbhmdyn}
\end{figure}

\cite{Tsai2018} presented a rest-UV spectrum observed with Keck/OSIRIS of W2246--0526 and used the Mg II 2799\AA\, emission line to obtain an estimate of the BH mass. They use the calibration from \cite{WangJ2009} to arrive to a \MBHdot\,=\,4\,$^{+6}_{-2.4}$\,$\times$\,10$^{9}$\,\Msun\, and an Eddington ratio (\lamEdd\,=\,\Lbol/\LEdd\,$\propto$\,\Lbol/\MBHdot) of \lamEdd\,$\simeq$\,2.8. Similarly, \cite{Finnerty2020} used rest-frame optical spectroscopy obtained with Keck/NIRES to derive an upper limit to the mass of the SMBH in W0220+0137. They assumed the \MBHdot--$\sigma_\star$ relation derived by \cite{Kormendy2013}, where $\sigma_\star$ is determined from the narrowest detected emission feature in the spectrum, to obtain \MBHdot\,$\leq$\,4\,$\times$\,10$^{10}$\,\Msun\, and from it a \lamEdd\,$\geq$\,0.06.
 
Figure~\ref{f:mbhmdyn} shows the SMBH mass as a function of the system's dynamical mass for W0220+0137 and W2246--0526 compared to other OpQs and IRQs in the literature, in the \textit{z}\,$\sim$\,2--6 range \citep[see][]{Bischetti2020}. The black hole mass of W2246--0526 is well within the dynamical mass of the system, with \MBHdot\, representing a fractional $\simeq$\,0.07 of \Mdyn. Unfortunately, the black hole mass of W0220+0137 only allows us to place a rough upper limit to its contribution to the dynamical mass, which could be up to a 100\%. However, if we were to assume the Eddington ratio of W0220+0137 to be similar to that of W2246--0526 (i.e., slightly super-Eddington), the contribution of \MBHdot\, to \Mdyn\, would decrease to a fraction of $\simeq$\,0.03 (see long red arrow), more similar to the value obtained for the latter as well. Considering this, these two luminous Hot DOGs seem to host BHs that are at least an order of magnitude more massive than nearby elliptical and spheroid galaxies with comparable dynamical masses, in agreement with \cite{Assef2015}. However, their location in the \MBHdot-\Mdyn\, parameter space is not significantly different from other quasar populations at similar or lower/higher redshifts.

\section{Hot DOGs on the main sequence}\label{s:ms}

\begin{figure*}
\centering
\includegraphics[scale=0.7]{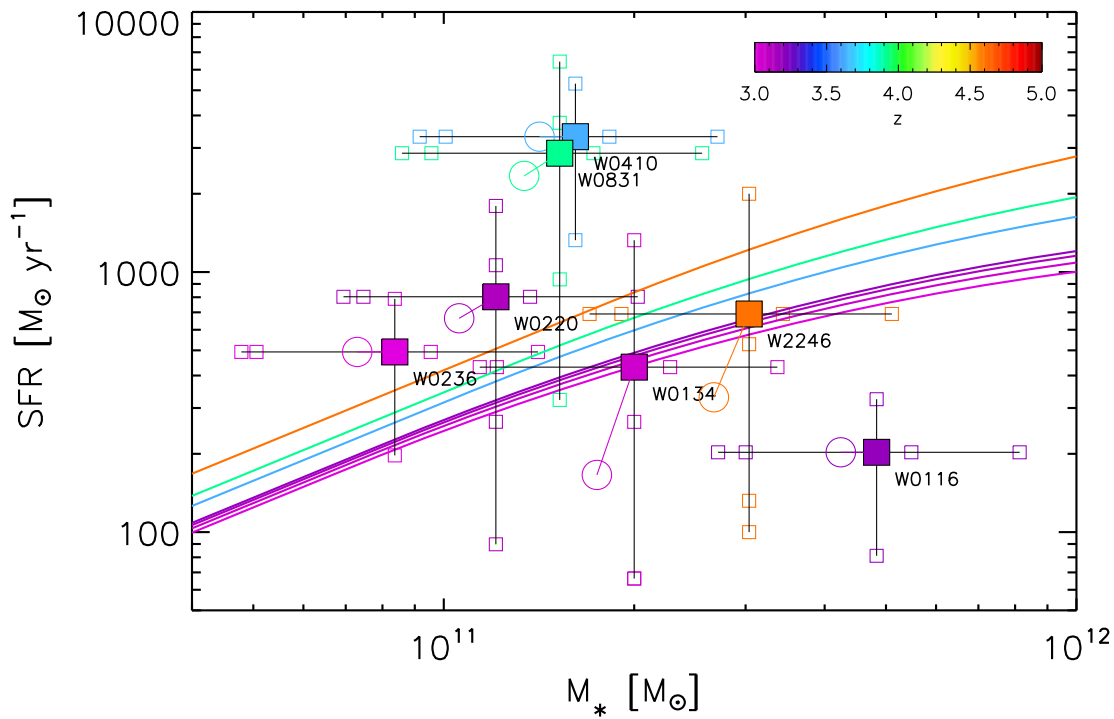}
\vspace{-9.5cm}
\caption{SFR as a function of \Mstar\, for the host galaxies of the Hot DOG sample, color-coded as a function of redshift. The galaxy star-formation MS at the specific redshift of each source is represented by solid lines (from \citealt{Schreiber2015}) also colored to match their redshifts. Average values of SFR and \Mstar\, for each source are shown as solid squares; median values are shown as open circles. The small, open squares represent the individual \Mstar\, and SFR estimates based on the SED modeling and the assumed SFR histories and scaling relations (see text and Appendix~\ref{a:sedfit}). Note that the error bars associated to the mean values do not represent standard deviations but rather extend the full range of individual estimates.}
\label{f:ms}
\end{figure*}

In order to estimate the stellar masses for the Hot DOG sample, we perform a spectral energy distribution (SED) fitting to the UV-through-mid-IR photometry available for each source following \citet{Assef2010, Assef2015}. A complete description of the methodology can be found in the Supplementary Materials section of \cite{DS2018}, which describes the example of W2246-0526. For additional clarity, we also re-describe the methodology in Appendix~\ref{a:sedfit}, where we also present the photometry used for the fitting of each source, as well as the best combination of model templates that best reproduce their SED (see Figure~\ref{f:seds}). To obtain the SFR in the host, we derive \SFRIR\, estimates based on both the far-IR rest-frame dust continuum emission and the \CIIno\, luminosity of each galaxy when available. Both observables are corrected by multiplying by the fractions of extended emission (\FEE, see Tables~\ref{t:linefluxes} and \ref{t:contfluxes}) to account only for the resolved component of the flux and thus avoid any potential AGN contamination. For the former, continuum-based method, we use two SED templates: one is the starburst galaxy M82 and the other is a typical Sd spiral galaxy, which we scale to the rest-frame 158\,\mic\, dust continuum to obtain the SFR. For the latter, we use \LCIIno\, and two representative values of the \CIIno-deficit seen in local, non AGN-dominated (U)LIRGs (\LCIIno/\LIR\,=\,[5,0.2]\,$\times$\,10$^{-3}$, see Figure~\ref{f:ciideficits}) to estimate the star-forming \LIR, which in turn is converted to a SFR using the calibration from \cite{Murphy2011}: SFR/\LIR\,=\,1.48\,$\times$\,10$^{-10}$\,[\Msunny]/[\Lsun], for a \cite{Kroupa2001} IMF. Table~\ref{t:masses} presents the average \Mstar\, and SFRs for the Hot DOG sample.

Figure~\ref{f:ms} shows the position of the Hot DOG hosts on the SFR-\Mstar\, plane, together with the star-formation MS (from \citealt{Schreiber2015}) at the specific redshift of each galaxy (solid lines, color-coded as a function of redshift). The individual estimates of \Mstar\, and SFR, based on the assumptions described above, are also shown (as small open squares). Note that the error bars around the average values (displayed as big solid squares) represent the full extent of the range of estimates and not the standard deviation. In addition, we also show the median values as open circles. As can be seen, EL Hot DOGs are spread over a large region of the parameter space. Three of the four \textit{z}\,$\sim$\,3 Hot DOGs (purple/magenta squares) and W2246--0526 are consistent with being on the MS. W0831+0140 and especially W0410--0913, at \textit{z}\,$\sim$\,3.9 and \textit{z}\,$\sim$\,3.6, respectively, seem to be well above it. Interestingly, these two Hot DOGs have the largest dust continuum emission core sizes and show among the largest \FEE, which would agree with potential wide-spread star formation in the hosts. W0116--0505, on the other hand, is well below the MS.

Given the small number of sources and the large uncertainties inherited from the wide range of \Mstar\, and SFR values obtained from the SED modeling and far-IR scaling relations, respectively, it is difficult to provide any detailed interpretation regarding the overall population of EL Hot DOGs or any particular object. Perhaps the only inference that can be done is that, when considered as a whole population, they do not all seem to be lying systematically below or above the star-formation MS. This is rather surprising considering the active accretion their SMBHs are experiencing, which should be accompanied by active star formation in the host, thus elevating the galaxies significantly above the MS if both the host and the SMBH were roughly co-evolving, as expected \citep{Madau2014}. However, when considering sources individually, it is possible that they can deviate from the MS significantly, above or below, and therefore the average position of the population on the SFR-\Mstar\, plane would not have much meaning. \citet{Riguccini2015, Riguccini2019} found a similar result for a sample of DOGs at \textit{z}\,$\sim$\,1--3, which display a large scatter around the MS with no specific trend based on the galaxies being dominated by star formation, or having a significant contribution from a moderately obscured or Compton-thick AGN.

A larger number of Hot DOGs and a better sampling of their photometric SEDs are necessary to obtain more precise stellar masses and SFRs, and thus be able to investigate whether individual cases such as those of W0410--0913, W0831+0140 or W0116--0505 are significant.

\section{Discussion}\label{s:discussion}

\subsection{What emission is the SMBH in extremely luminous Hot DOGs powering?}\label{ss:powersource}

In Section~\ref{ss:ism} we described where our sample of Hot DOGs lie in a number of diagrams relating the luminosity of the \CIIno\, line to that of the underlying dust continuum and its physical size (Figure~\ref{f:ciideficits}). In this section we discuss how the presence of an extremely luminous (EL) AGN (\Lbol\,$\geq$\,10$^{14}$\,\Lsun) may be able to alter the regular ISM signatures of the host galaxy, and provide a possible interpretation for the position of the most luminous Hot DOGs in the parameter space delimited by the ALMA data.

The dust continuum emission signature of AGNs, and in particular of Hot DOGs, is most noticeable at rest-frame mid-IR wavelengths, where the contribution of the highest-temperature components of the dusty structure reprocessing most of the SMBH's power dominates the SED \citep{Wu2012, Tsai2015}. However, as the AGN becomes extremely luminous and starts to bolometrically account for nearly the entire power output of a galaxy, it also emerges as the dominant source even at $\lambda$\,$\gtrsim$\,100\,\mic. This is all the more likely if the coldest component of the temperature distribution of the obscuring medium is still warmer than the overall dust temperature of the underlying galaxy \citep[see, e.g.,][]{Wu2012, Wu2014} and/or if a significant amount of the AGN dust is optically thick in the far-IR. Were this to happen, the measured size of the dust continuum would progressively appear more compact, as it would be representative of the more dominant central point-source. These two effects (emission and size) would also likely come in hand with the fact that, assuming the powerful far-IR AGN is also a significant X-ray emitter (\LX\,$\gtrsim$\,10$^{12}$\,\Lsun), its hard ionizing spectrum could reduce the C$^+$ abundance in the warm ionized medium (WIM) in the vicinity of the SMBH ($\lesssim$\,few hundred pc), converting some fraction of C$^+$ to C$^{2+}$ and higher ionization states \citep{Langer2015}. This would have the effect of suppressing the surrounding \CIIno\, luminosity up to almost two orders of magnitude in the most extreme cases, generating a cavity of higher ionization gas. All these effects combined would strongly drive galaxies that are bolometrically dominated by an AGN to have high \SigmaIR\, and very low \CIIno\, EWs, i.e., towards the lower-right region of the parameter space in the bottom-left panel of Figure~\ref{f:ciideficits}, significantly below the region where EL Hot DOGs actually lie, and below the extrapolation from local (U)LIRGs.

There are a number of observables suggesting that some of these effects may actually not be as important as expected. The bottom-right panel in Figure~\ref{f:ciideficits} shows the luminosity surface brightness of \CIIno, \SigmaCII, as a function of the intrinsic radius of the dust continuum emission. EL Hot DOGs do not seem to have particularly small sizes when compared to the most extreme, nearby systems. This implies that even if the unresolved AGN in Hot DOGs is two orders of magnitude more luminous than that in local (U)LIRGs, its relative contribution to the far-IR continuum, which would make the size smaller, is somehow compensated by a proportional increase of emission by the resolved host (or underlying extended gas structure in general), in order to keep the system physically resolved. This is in agreement with the extended component of the Hot DOGs accounting a significant fraction of the dust continuum emission (from \FEE\,$\sim$\,20\% up to $\sim$\,70--80\% in the most extreme cases; see Table~\ref{t:contfluxes} and Figure~\ref{f:spatprofs}). The \CIIno\, emission also has a substantial extended component fraction, larger than that of the continuum in all cases.

Instead of the size, the main difference with respect to the local, IR galaxy population is actually \SigmaCII, which in Hot DOGs is consistently around an order of magnitude higher than local (U)LIRGs for their given size range, and only matched by some of the most compact \textit{z}\,$\gtrsim$\,6 OpQs and a few \textit{z}\,$\simeq$\,4.7 IRQs (see bottom-right panel of Figure~\ref{f:ciideficits}). This excess in \CIIno\, luminosity per area also seems to argue against the scenario in which an energetic AGN would deplete the surrounding abundance of C$^+$ by pumping higher ionization states and thus suppressing the \CIIno\, luminosity. Or alternatively, it could imply that Hot DOGs are not be powerful X-ray sources \citep{Stern2014,Stern2015}.

The location of the EL Hot DOGs in the two bottom panels of Figure~\ref{f:ciideficits} is then likely set by a combination of an IR and \CIIno\, luminosity excess with respect to the extrapolation of local (U)LIRGs, and not (or to a much lesser degree) to a difference in galaxy size. While the \SigmaIR\, excess can be easily explained by the contribution of the dominant AGN to the bolometric luminosity of the galaxy, the origin of the \SigmaCII\, excess is less evident, and can be interpreted in at least two (non exclusive) ways:
\smallskip

(1) At higher redshifts, the gas content of galaxies progressively accounts for a larger fraction of their baryonic mass \citep{Scoville2017, Tacconi2018, Walter2020}. If the total gas reservoir of galaxies, $M_{\rm gas}$, roughly scales with \LCIIno\, \citep{Zanella2018}, then EL Hot DOGs should have an order of magnitude more gas mass than local (U)LIRGs of similar size (Figure~\ref{f:ciideficits}, bottom-right panel). In addition, in order to explain the narrow range of relatively low \CIIno\, EWs (bottom-left panel), the luminosity of the continuum under the line needs to increase roughly proportionally, which would be consistent with the dust emission being also a tracer of the gas content in the host (or at least of the ionized gas probed by \CIIno). In this case, the AGN has no impact on the properties of a Hot DOG other than boosting the total \LIR\, of the galaxy (at shorter than far-IR wavelengths). In other words, if we were to assume that the typical Hot DOG host is a galaxy similar to a maximal, star-forming local (U)LIRG (\SigmaIR\,$\simeq$\,5\,$\times$\,10$^{11}$\,\lsd\, and $R_{\rm eff}$\,$\simeq$\,0.5--1\,kpc; see Figure~\ref{f:ciideficits}, top-left panel) but containing an order of magnitude more gas mass (and thus a proportionally higher \LCIIno/\LIR\, ratio), when corrected by the two orders of magnitude increase in total luminosity due to the AGN (i.e., along the arrows in the same Figure), the resulting galaxy would be put back on the extrapolation of the local IR population (dotted-dashed line).
However, the actual ISM content and relative contribution of the energy sources (SFR vs. AGN) in the resulting Hot DOG may be very different from a galaxy that is based on the extrapolation of a purely star-forming ULIRG towards higher \SigmaIR\, following the local correlation.
\smallskip

(2) An alternative possibility is the existence of an additional heating source, other than star formation, that enhances the \CIIno\, emission. Evidence of excess C$^+$ with respect to the maximum line cooling efficiency expected from pure photo-dissociation regions has been seen in purely star-forming galaxies at \textit{z}\,$\simeq$\,1--2 \citep{Brisbin2015}. In the local Universe, Seyfert galaxies hosting powerful AGN-driven radio jets and winds inject turbulence into the surrounding ISM, contributing to the excitation of a larger mass of the C$^+$ reservoir \citep{Guillard2015, Appleton2018, SP2019} (in contrast with the depletion scenario based on ionization described above, from \citealt{Langer2015}). In addition, low velocity shocks driven by galaxy collisions can also turbulently heat the gas on large scales in dense environments \citep{Appleton2013, Appleton2017}. Given that Hot DOG hosts do not seem to be forming stars at a much higher rate than normal MS galaxies at the same redshifts (see Figure~\ref{f:ms} and Section~\ref{s:ms}), significant turbulent heating could be provided by the energy input from the central SMBH \citep{DS2016} and/or by the dynamical friction consequence of the accretion of neighbor galaxies (see next section). Both cases would be in concordance with the large velocity dispersion of the \CIIno\, line seen in all EL Hot DOGs, FWHM\,$\gtrsim$\,500\,\kmns\, (Section~\ref{s:kinematics}), and with the fact that they seem to live in over-dense, merger-driven environments \citep[e.g.,][]{Jones2014, Assef2015, Jones2017, DS2018}.
\smallskip

Because of the size constraint, the most likely scenario is a combination of these two possibilities. That is, the typical Hot DOG galaxy host could be a scaled-up version of a local LIRG ($\sim$\,10$^{11-12}$\,\Lsun). That is, with up to an order of magnitude larger total \LIR\, ($\sim$\,10$^{12-13}$\,\Lsun), continuum emission at $\sim$\,160\,\mic\, ($\nu L_{\rm \nu 160\mu m}$\,$\sim$\,10$^{12-13}$\,\Lsun), and \LCIIno\, ($\sim$\,10$^{9-10}$\,\Lsun; see Table~\ref{t:linefluxes}). This would be in agreement with the position of Hot DOGs in the SFR-\Mstar\, plane, lying on or slightly above the star-forming galaxy MS at their respective redshifts (see Figure~\ref{f:ms}). In this scenario, the host galaxy would contribute equally to the line and dust continuum emission under it, thus maintaining the EW (Figure~\ref{f:ciideficits}, bottom-left). In addition, the AGN could also account for a fraction of the continuum and line flux, the first originating from the coldest dust component heated by the SMBH, and the latter potentially by its injection of turbulence into the surrounding ISM, in addition to galaxy-galaxy collisions. And while contributing to both continuum and line, the AGN would dominate neither, nor would Hot DOGs appear as mostly unresolved, but rather a combination of a central source plus an extended host (see \SigmaIR\, profiles presented in Section~\ref{ss:morphology} and Figure~\ref{f:spatprofs}), which is kinematically and morphologically disturbed by the interaction with neighbor galaxies (as some of the moment 0 maps suggest in Figures~\ref{f:contflux} and \ref{f:lineflux}). Yet, because of the larger dynamic range and higher dust equilibrium temperatures, the AGN would boost at least another order of magnitude the total IR luminosity of the galaxy, dominating the emission at shorter wavelengths, and thus accounting for $\geq$\,90\% of the \LIR. Hot DOGs are very strong emitters in the rest-frame mid-IR (by selection) and therefore if this picture is correct, future observations of \textit{z}\,$\lesssim$\,1.5 Hot DOGs with the \textit{JWST} should display mostly unresolved emission at observed-frame $\lambda$\,$\gtrsim$\,20\,\mic.


\subsection{An evolutionary perspective}\label{ss:cosmicview}

Within the current paradigm of galaxy evolution, the most luminous quasars are a relatively short phase during the merger process of two massive, gas-rich, star-forming galaxies \citep[e.g.,][]{Sanders1988a, Hopkins2008a}. As the galaxies coalesce, gas and dust are piled up onto the nascent common nucleus, triggering both massive star formation and boosting SMBH accretion. This dusty, obscured phase is assumed to be traced by IRQs. Once the feedback from the quasar sets in, the radiation pressure exerted on the surrounding ISM by the central AGN clears out the dusty cocoon, making the quasar detectable in the optical (OpQ). However, while this scenario can account for ULIRGs in the nearby Universe, the physical processes and environment through which quasar activity proceeds at high redshift is likely not as straightforward. At high redshift, galaxies have significantly higher gas fractions \citep{Scoville2017, Tacconi2018}, and EL quasars, likely being hosted by galaxies at the knots of the cosmic web, may be subject to intermittent, yet sustained accretion from in-falling companion galaxies \citep{DS2018, Bischetti2018, Bischetti2020}, and not just to a single major-merger event.

In Section~\ref{s:kinematics} we presented the velocity and dispersion maps of the Hot DOG sample. We found that EL Hot DOGs have very diverse dynamical states: from sources that are barely rotating, to resolved hosts with ordered, circular motions, to complex, disturbed systems that are likely the result of ongoing mergers. These findings are very similar to the diversity shown by ultra-luminous IR quasars in the local Universe. In a sample of eight IRQs, \cite{Tan2019} found rotating disks, disturbed velocities, and some sources with morphologies suggesting they are undergoing a merger process (albeit these results were drawn from observations of the CO(1$\rightarrow$0) emission line rather than \CIIno, thus probing a different gas phase of the ISM). However, in contrast to nearby IRQs, which generally show moderate velocity dispersion (FWHM$_{\rm CO(1-0)}$\,$\simeq$\,250\,\kmns), EL Hot DOGs are characterized by FWHM$_{\rm [CII]}$\,$\gtrsim$\,500\,\kmns, which in a number of cases is uniform across most of the area where the underlying host is detected. This is also reflected in the fact that the $|V_{\rm max}|$/$\sigma_V$ ratios of rotation-dominated local IRQs are in the range of $\sim$\,4--6, while in Section~\ref{ss:vsigma} we found that our Hot DOG sample display a lower range, with $|V_{\rm max}|$/$\sigma_V$\,$\lesssim$\,2.

In terms of the connection between the general galaxy evolution framework described above and the particular context of this study, we have reported on a population of EL, hot dust-obscured quasars that do not share a consistent or characteristic velocity field, but display a common kinematic signature in the form of a high velocity dispersion of the neutral and ionized gas as traced by \CIIno. While a number of non-exclusive scenarios can be invoked to explain this diversity in morphology and dynamical states, they also need to account for the short-lived nature of the Hot DOG phenomenon, which lasts on average, and overall, for less than a few tens of Myr as implied by the limited number of such objects detected in the entire sky ($\lesssim$\,0.01\,deg$^{-2}$ for \Lbol\,$\geq$\,10$^{14}$\,\Lsun; \citealt{Assef2015}). Here we put forward a few potentially important aspects to consider in order to characterize and explain the rise of Hot DOGs:
\smallskip

(1) \textit{The nature of the activation mechanism(s)}: It is possible that there is more than one mechanism triggering the EL Hot DOG phenomenon. Large-volume cosmological simulations predict that smooth accretion of pristine gas from the cosmic web could provide the SMBH of the central galaxy in a proto-cluster with enough gas to make the AGN become extremely luminous, as well as to trigger significant star formation in the host \citep{Dekel2009, Fumagalli2011}. On the other hand, mergers are also able to provide an intermittent supply of material in the form of gas that has been pre-processed in neighbor galaxy companions \citep{Decarli2018, DS2018, Bischetti2020}. Distinguishing between these two mechanisms is very complicated, as the identification and measurement of low-metallicity gas reservoirs around galaxies at high redshift is very challenging \citep[e.g.,][]{Rafelski2012}. Companion galaxies have been found in around 70\% of the Hot DOGs analyzed here \citep{GL2021}. In comparison, the fraction of OpQs at \textit{z}$\gtrsim$\,4.5 with detected nearby companions is significantly lower \citep[$\lesssim$\,30\%;][]{Decarli2018, Trakhtenbrot2017, Nguyen2020}, although recent results by \cite{Bischetti2020} suggest that around $\sim$\,80\% of those quasars at the brightest end of the QSO luminosity function are surrounded by one or more bright companion galaxies. Deeper and more uniform observations of both optical quasars and Hot DOGs are needed to investigate whether this is revealing a real evolution or it is simply a consequence of the selection bias (or probably both).
\smallskip

(2) \textit{Time and spatial scales}: Mergers may be the dominant mechanism triggering EL Hot DOGs, but there is a time sequence connecting the different steps in the accretion process that is correlated with the physical scales over which observables enable its tracking. That is, the identification of the merger nature of a system is usually done through the detection of companion galaxies \citep[e.g.,][]{DS2018, Miller2018, Hodge2019}; the dynamical state of the host galaxy is probed by the kinematics of the gas \citep[e.g.,][]{Carniani2013, Kimball2015, Tadaki2018}; and the presence of an accreting SMBH is revealed by the fact that an obscured AGN dominates the bolometric output of the source \citep{Tsai2015}. Each of these signatures has a characteristic spatial scale attached to it: neighbor galaxies in the vicinity of high-\textit{z} quasars are identified at projected distances greater than at least a few kpc \citep[][]{Petitjean1996, McGreer2014, Trakhtenbrot2017, Decarli2018, Decarli2019, Nguyen2020}; the dynamical state of the host, which is driven by the interaction between the pre-existing bound gas and the in-fall of companions, is characterized by physical scales of $\sim$\,1--2\,kpc that, when tied to the gas velocity dispersion observed in the host galaxies (FWHM\,$\gtrsim$\,500\,\kmns), provides a time-scale of $\lesssim$\, a few Myr; and the funneling of gas into the SMBH happens efficiently only inside of its sphere of influence (within a few hundred pc, considering BH mass estimates of \MBHdot\,$\sim$\,10$^9$\,\Msun; \citealt{Wu2018, Tsai2018, Jun2020a, Finnerty2020}), where it can be transported further in by rotationally supported structures \citep{AA2020}. The lack of synchronicity and delay among all these phases \citep[including with respect to the star-formation process;][]{Hopkins2012} can potentially generate the diversity of dynamical states displayed by Hot DOGs despite all sharing a common origin.
\smallskip

(3) \textit{Recurrence}: Independently of the mechanism dominating the accretion (1), and the time/spatial-scales describing the accretion (2), this process could occur once, \textit{or it may be recurrent}. The former scenario is usually simulated in the local Universe as a single-merger event of two massive, gas-rich galaxies leading to a quasar phase \citep{Hopkins2008a}. This is supported by the fact that the majority of nearby, IR-selected ULIRGs and IRQs display signatures suggesting they are the product of major mergers in the latest stage of the interaction, immediately previous to ---or right after--- coalescence of the two galactic nuclei \citep{Veilleux2002, Veilleux2009, Haan2011, Kim2013, Stierwalt2013}. However, at high-\textit{z}, galaxies may have large enough gas reservoirs stored in clumpy, unstable disks \citep[e.g.,][]{Decarli2016, Falgarone2017, Hodge2019} such that minor mergers could potentially be sufficient to trigger a number of short-lived extremely luminous, highly dissipative phases associated with rapid SMBH and galaxy mass growth. In the particular case of EL Hot DOGs, which seem to be living in over-dense environments \citep{Jones2014, Assef2015, Fan2017, GL2021}, a frequent in-fall of gas in the form of whole galaxies, large clumps, or tidally disrupted material would also help to explain the observed widespread turbulence of their ISM, in addition to the energy and momentum injected by the central AGN. Indeed, kpc-scale turbulence is a clear signature of the existing gravitationally bounded gas in a host galaxy attempting to relax while still being subjected to external accretion and/or internal AGN feedback \citep{Appleton2017}. If that is the case, EL Hot DOGs could be characterized as a recurrent phase (rather than a single event, like nearby ULIRGs/quasars) in which a galaxy at the center of a young proto-cluster \citep[where the red galaxy sequence has not been formed yet; ][]{Penney2019} is experiencing chaotic episodes of accretion, each individually lasting for a time-scale shorter than a dynamical time. This would keep the gas turbulent in the host, promoting cloud-cloud collisions and the removal of angular momentum, which would make the gas more likely to fall under the gravitational influence of the SMBH. Some of the gas could be subsequently funneled to the SMBH accretion disk, a fraction of which could be expelled via feedback, further contributing to the turbulence in the host in a self-sustaining, intermittent process as long as external accretion continues.
\smallskip

This is a potential scenario that would overall agree with the evidence presented in this work. It is also important to note that within this hypothesis of recurrent accretion, and considering that EL Hot DOGs seem to have Eddington ratios close to unity or above \citep[e.g.,][]{Tsai2018, Jun2020a, Finnerty2020}, each individual EL episode (\Lbol\,$\gtrsim$\,10$^{14}$\,\Lsun) would need to be significantly shorter than the characteristic timescale for Eddington-limited BH growth (also known as Salpeter time: \tSalp\,=\,\MBHdot/$\dot{M}_{\rm Edd}$\,$\simeq$\,40\,($\eta$/0.1)\,Myr; where $\eta$ is the conversion efficiency of the accreted mass into electromagnetic radiation, and \lamEdd\, is assumed to be unity). That is, with BH masses on the order of $\sim$\,10$^{9}$\,\Msun\, or slightly above (see Table~\ref{t:masses}), a large number of EL events with typical efficiencies of $\eta$\,=\,0.1 jointly lasting for a period significantly longer than \tSalp\, would lead to final BH masses in excess of those seen in SMBHs hosted by the brightest cluster galaxies (BCGs) today (\MBHdot\,$\simeq$\,10$^{10}$\,\Msun; \citealt{McConnell2013}). Therefore, if the EL Hot DOG phase is to be recurrent, each episode would need to be shorter than a few Myr ($t_{\rm e}$\,$\lesssim$\,\tSalp\,$N_{\rm e}^{-1}$\lamEdd$^{-1}$, where $N_{\rm e}$ is the number of episodes), which would be in agreement with the observed AGN flickering time, of about 10$^{5}$\,yr \citep{Schawinski2015}, and with recent ultra-high resolution simulations of the BH accretion rate down to the central sub-pc scale in high-redshift quasars at the center of proto-clusters \citep{AA2020}.

We note that the picture put forward here may not be valid for lower luminosity Hot DOGs and it is possible that not all accretion events produce an EL, hot obscured phase. To investigate if this scenario holds in general, a larger number of Hot DOGs spanning a wider range in bolometric luminosity is needed; and in particular for less powerful Hot DOGs (10$^{13}$\,$\lesssim$\,\Lbol/\Lsun\,$\lesssim$\,10$^{14}$). A larger sample is also required to perform statistical studies regarding the presence of galaxy companions, the time lags between the different accretion phases, and in general to accurately characterize the most luminous, dust-obscured population of galaxies so far detected in the Universe. Deeper observations are also critical to be able to perform a detailed modeling of their kinematics (see Appendix~\ref{a:kinmod}), and investigate whether simulations of massive galaxy formation \citep[e.g.,][]{Dekel2009, FG2012, Gaspari2013, AA2017a, Lupi2019, AA2020} are able to reproduce the different morphologies, velocity fields, and potential quasar-driven outflows in the most rapidly evolving galaxies at the nodes of the cosmic web. These simulations are critical, as they can be used to infer and predict the duration and recurrence of Hot DOG phase(s), thus allowing to quantitatively translate snap-shots of space into actual, distinctive physical stages of galaxy evolution.

\section{Summary}\label{s:summary}

We have presented ALMA observations of the \CIIno\, emission line for a sample of seven of the most extremely luminous (EL; \Lbol\,$>$\,10$^{14}$\,\Lsun) IR quasars, at redshifts \textit{z}\,$\sim$\,3--4.6, from the population of hot, dust-obscured galaxies (Hot DOGs) discovered by \textit{WISE}. We have compared their physical properties to other high-redshift populations of IR and optically selected QSOs at \textit{z}\,$\simeq$\,4.7 (IRQs) and $\gtrsim$\,6 (OpQs), respectively. The results of the study are summarized as follows:

\begin{itemize}
    
    \item The \CIIno\, line is robustly detected in four out of the seven EL Hot DOGs (W0134–2922, W0220+0137, W0831+0140 and W2246–0526). Based on other ALMA emission line data-sets, we know that for two sources without an identification (W0116–0505 and W0410–0913) the line was likely red-shifted out of the frequency range covered by the reference spectral window and side-band. All Hot DOGs are detected in dust continuum emission at $\simeq$\,158\,\mic.
    
    \item In a significant number of sources the \CIIno\, and/or other far-IR emission lines are systematically red-shifted with respect to the systemic velocities inferred from rest-frame UV spectroscopy, with an average shift $\Delta v$\,$\simeq$\,780\,\kmns.
    
    
    \item The EL Hot DOGs in the sample exhibit the highest IR luminosity surface densities, \SigmaIR, and the lowest \LCIIno/\LIR\, ratios among galaxy populations at any redshift, with values commonly $\geq$\,10$^{13}$\,\lsd\, and $\lesssim$\,10$^{-4}$, respectively. They also tightly follow the extrapolation of the \LCIIno/\LIR\, vs. \SigmaIR\, relation found in nearby (U)LIRGs.
    
    \item All Hot DOGs detected in \CIIno\, show \LCIIno\, surface densities of \SigmaCII\,=\,1--2\,$\times$\,10$^9$\,\lsd\, with a very narrow scatter. Only some of the most compact high-\textit{z} OpQs and IRQs reach such high \CIIno\, surface densities, but these populations also extend to surface areas one order of magnitude smaller. The \CIIno\, equivalent widths of the Hot DOG sample range between $\simeq$\,0.4--0.8\,$\mu$m.
    
    \item The core line emission is spatially resolved clearly in 1/4 of Hot DOGs and marginally resolved in 3/4. In dust continuum, 3/7 are securely resolved, 3/7 are marginally resolved, and 1/7 is unresolved. The average \CIIno\, and continuum emission sizes are $\simeq$\,2.1\,kpc and $\simeq$\,1.6\,kpc, respectively. In all cases when both line and continuum are detected, the core size of the \CIIno\, emission is larger than that of the continuum, and the range of line-to-continuum size ratios is very narrow, 1.35--1.85 with an average of 1.61\,$\pm$\,0.10, remarkably similar to other high-\textit{z} QSO populations, and therefore suggesting the sizes could be linearly proportional. Although the core emissions in most objects are very compact, nearly all Hot DOGs display large extended emission fractions, from \FEE\,$\sim$\,20--80\% in continuum to $\sim$\,50--80\% in \CIIno.
    
    \item The velocity fields of EL Hot DOGs are very diverse: from barely rotating sources, to resolved hosts with ordered, circular motions, to complex, disturbed systems that are likely the result of ongoing mergers. However, all Hot DOGs are characterized by large line-velocity dispersions, FWHM$_{\rm [CII]}$\,$\gtrsim$\,500\,\kmns, which on average are significantly larger than optically selected QSOs at similar or higher redshifts. This is a clear signature of a turbulent ISM that may be a characteristic feature common to the Hot DOG phase, despite their diversity in rotation properties.
    
    \item As a population, the Hot DOG hosts seem to be centered around the star-formation MS. Individually, three of the most nearby Hot DOGs, at \textit{z}\,$\simeq$\,3, and W2246–0526 are consistent with being roughly on the MS, W0410–0913 and W0831+0140 seem to be lying above, and W0116--0505 below. However, uncertainties are large, and additional observations are needed to pin down the position of individual sources on the SFR-\Mstar\, plane and better locate the population as a whole.
    
\end{itemize}

In summary, there is no unique morphological and/or dynamical state that describes the extremely luminous, AGN-powered Hot DOG phase: not all sources have galaxy companions identified in their fields; the morphology of their line and continuum emissions show relatively compact cores on top of very extended structures; and a wide variety of velocity fields are observed. The only common kinematic signature shared among the EL Hot DOG population is the large turbulence observed in their ionized/neutral gas phase as traced by \CIIno, which is likely related to their systematically large \SigmaCII\, with respect to high-\textit{z} OpQs and local (U)LIRGs with similar physical sizes. In the discussion section we explored what may be the source(s) powering their turbulent ISM, and investigated how different scenarios may agree with the particular location of the Hot DOG sample in a number of diagnostic diagrams enabled by the ALMA data. We also tried to place EL Hot DOGs within the merger-driven galaxy evolution framework and speculate on their origin and nature, proposing a possible hypothesis by which the most luminous Hot DOGs could be characterized as recurrent, yet short-lived periods ($\lesssim$\, a few Myr; significantly shorter than the Eddington time) in which the galaxy at the center of a proto-cluster is undergoing rapid accretion in the form of companion galaxies, clumps or tidal structures, which frequently replenish the gas reservoir in the host and maintains its turbulence. This likely chaotic in-fall and lack of organized motions would promote the removal of angular momentum, allowing the SMBH to trap gas under its sphere of influence. There, matter can be subsequently funneled to the accretion disk, simultaneously keeping the quasar obscured. Some of the accreted gas can be expelled away in the form of feedback, further contributing to the turbulence in the host in a self-sustained process. We note, however, that other interpretations are also conceivable. In addition, the proposed scenario may not be valid for lower luminosity Hot DOGs and it is possible that not all the accretion events produce an EL hot, obscured phase.

The confirmation of the proposed picture thus requires higher signal-to-noise ALMA observations of a larger, statistically significant number of Hot DOGs for which a proper kinematic modeling of the galaxies can be performed and compared with detailed, physics-driven simulations. In addition, future \textit{JWST} data will help to elucidate what is the main heating mechanism of the dust at different spatial scales in this outstanding galaxy population.

\section*{Acknowledgments}

We thank the referee for their constructive comments and useful suggestions, which significantly improved the paper. We would like to thank Roberto Decarli and Bram P. Venemans for insightful discussions and early access to ALMA datasets from high redshift quasars. We would like to thank Jordan I. Penney for comments on early drafts of the paper. T.D-S. acknowledges support from the CASSACA and CONICYT fund CAS-CONICYT Call 2018. R.J.A. was supported by FONDECYT grant No. 1191124. H.D.J. was supported by the Basic Science Research Program through the National Research Foundation of Korea (NRF) funded by the Ministry of Education (NRF-2017R1A6A3A04005158). G.C.J. acknowledges ERC Advanced Grant 695671 ``QUENCH'' and support by the Science and Technology Facilities Council (STFC). M.A. and R.J.A. have been supported by the grant CONICYT + PCI + Instituto Max Planck de Astronomia MPG190030. C-W.T. was supported by a grant from the NSFC (No. 11973051). S.E.L. was funded by the Chinese Academy of Sciences President's International Fellowship Initiative. Grant No.2019PM0017/2021PM0076. J.W. acknowledges support from MSTC through grant 2016YFA0400702. This paper makes use of the following ALMA data: ADS/JAO.ALMA\#2013.1.00576.S and ADS/JAO.ALMA\#2015.1.00612.S. ALMA is a partnership of ESO (representing its member states), NSF (USA) and NINS (Japan), together with NRC (Canada), MOST and ASIAA (Taiwan), and KASI (Republic of Korea), in cooperation with the Republic of Chile. The Joint ALMA Observatory is operated by ESO, AUI/NRAO and NAOJ. This publication makes use of data products from the Wide-field Infrared Survey Explorer, which is a joint project of the University of California, Los Angeles, and the Jet Propulsion Laboratory/California Institute of Technology, funded by the National Aeronautics and Space Administration. Portions of this research were carried out at the Jet Propulsion Laboratory, California Institute of Technology, under a contract with the National Aeronautics and Space Administration (80NM0018D0004). This research has made use of the NASA/IPAC Extragalactic Database (NED), which is operated by the Jet Propulsion Laboratory, California Institute of Technology, under contract with the National Aeronautics and Space Administration, and of NASA's Astrophysics Data System (ADS) abstract service.\\

\appendix\label{s:appendixDM}

\section{Kinematic modeling}\label{a:kinmod}

The continuum-subtracted \CIIno\, cubes of each source (considering natural weighting --robust parameter equal to 2-- and no $uv$-tappering), containing only the line SPW, were spatially trimmed to the central 200\,$\times$\,200 spaxels, centered on the galaxy. After trimming, these cubes were fit with tilted ring models using the code $^{\rm 3D}$Barolo \citep[hereafter BBarolo;][]{DiTeodoro2015} in order to determine the dynamical masses of each source. First, the line emission in each cube was identified using the \texttt{SEARCH} algorithm in BBarolo, which is based on the code DUCHAMP \citep{Whiting2012}. After automatically determining the r.m.s. noise level of the cube, this algorithm searches for pixels with intensities above a user-provided SNR ratio (\texttt{SNR$_{\rm upper}$}). A three-dimensional (i.e., RA, DEC, velocity) search is then conducted around these peaks for emission above a second user-provided SNR ratio (i.e., \texttt{SNR$_{\rm lower}$}), thus creating a 3-D signal mask.
\smallskip

Each cube is then collapsed over all channels containing line emission using the CASA toolkit task \texttt{im.moments}. The resulting integrated intensity (i.e., moment 0) map is fit with a 2D Gaussian using the CASA toolkit task \texttt{im.fitcomponents}, yielding a peak intensity, integrated flux density, FWHM of the major and minor axes, position angle, and central position. In addition, a velocity dispersion (i.e., moment 2) map is created using the same task, including only pixels identified through the \texttt{SEARCH} algorithm.
With these maps in hand, the main function of BBarolo (i.e., \texttt{3DFIT}) is used to fit a tilted ring model to the line emission identified through the \texttt{SEARCH} routine. While this task is able to automatically predict many parameters, and is well-tested for low-SNR and low-resolution observations, its performance is dependent on the initial parameter estimates and the overall model geometry. For our fits, the width of each ring is set to the FWHM of the minor axis of the restoring beam divided by 2.5, while the maximum radius of the model is set to the morphological major axis FWHM. The central position of all model rings is fixed to the morphological central position, while the initial guess for the velocity dispersion is the maximum value in the moment 2 map, and a thin disk is assumed ($z_{\circ}=0.01''$). The inclination is allowed to vary between $10-80^{\circ}$, with an initial guess of $45^{\circ}$. BBarolo provides an initial guess for the rotational velocity and position angle.
\smallskip

Using these initial morphological and kinematic estimates, BBarolo creates a model of the first ring by populating a physical volume with discrete clouds, such that the rotational velocity, position angle, etc. of the estimates are recreated. This model is then converted into an observational (i.e., RA, Dec, velocity) cube, convolved with the synthesized beam of the data cube , normalized by setting the azimuthally-averaged flux in the model and data cubes equal, and compared to the model cube. Each parameter is varied, until the absolute residual (i.e., $|$model--observation$|$) is minimized. This process is then repeated with the next ring, and so on.
    \smallskip
  
In the first run, a number of parameters were fit (i.e., rotational velocity, velocity dispersion, systemic velocity, inclination, and position angle). The three latter variables are then set to the average values between all rings, and the fit algorithm is run again, with only the rotational velocity and velocity dispersion as variables.
\smallskip

\begin{figure*}
\includegraphics[width=\hsize]{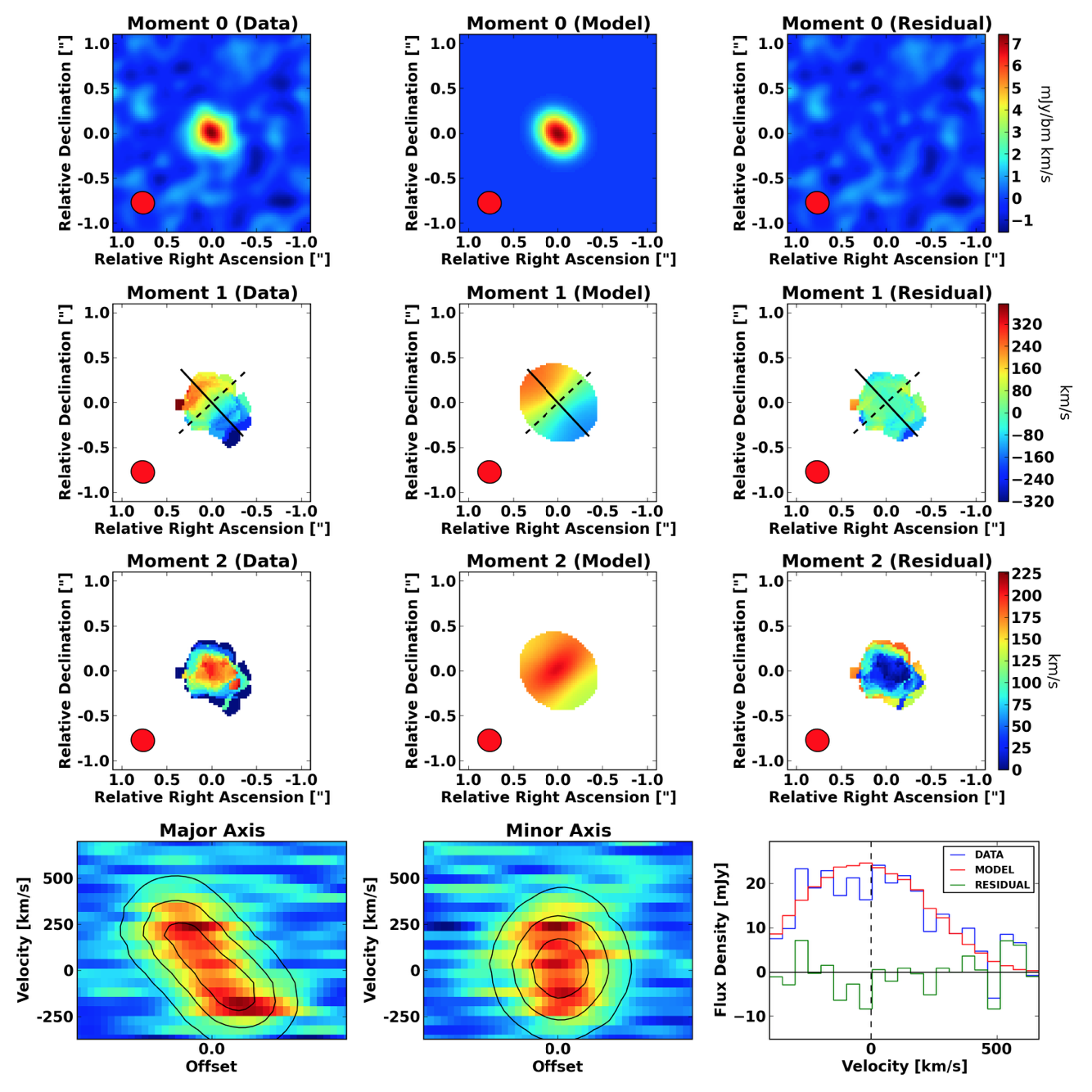}
\caption{Dynamical modeling of W0220+0137 using BBarolo. The first three rows show the moment 0 (integrated intensity), moment 1 (velocity field), and moment 2 (velocity dispersion field) for the data and model cubes (created using the CASA toolkit task im.moments), as well as the difference between them. The solid lines in the second row represent the kinematic major axis, while the dashed lines represent the minor axis. In order to avoid the low-signal portion of the model, the model moment maps are only shown over the spaxels where the moment zero intensity is $\ge$\,1\% of the maximum value. The bottom row shows (from left to right) the major axis postion-velocity diagram (PVD), minor axis PVD, and extracted spectra. For each PVD, the data is shown by the background color, while the overlaid contours represent the model. The data, model, and residual spectra in the bottom-right panle are depicted by the red, blue, and green lines, respectively.}
\label{f:W0220barolo}
\end{figure*}

\begin{figure*}
\includegraphics[width=\hsize]{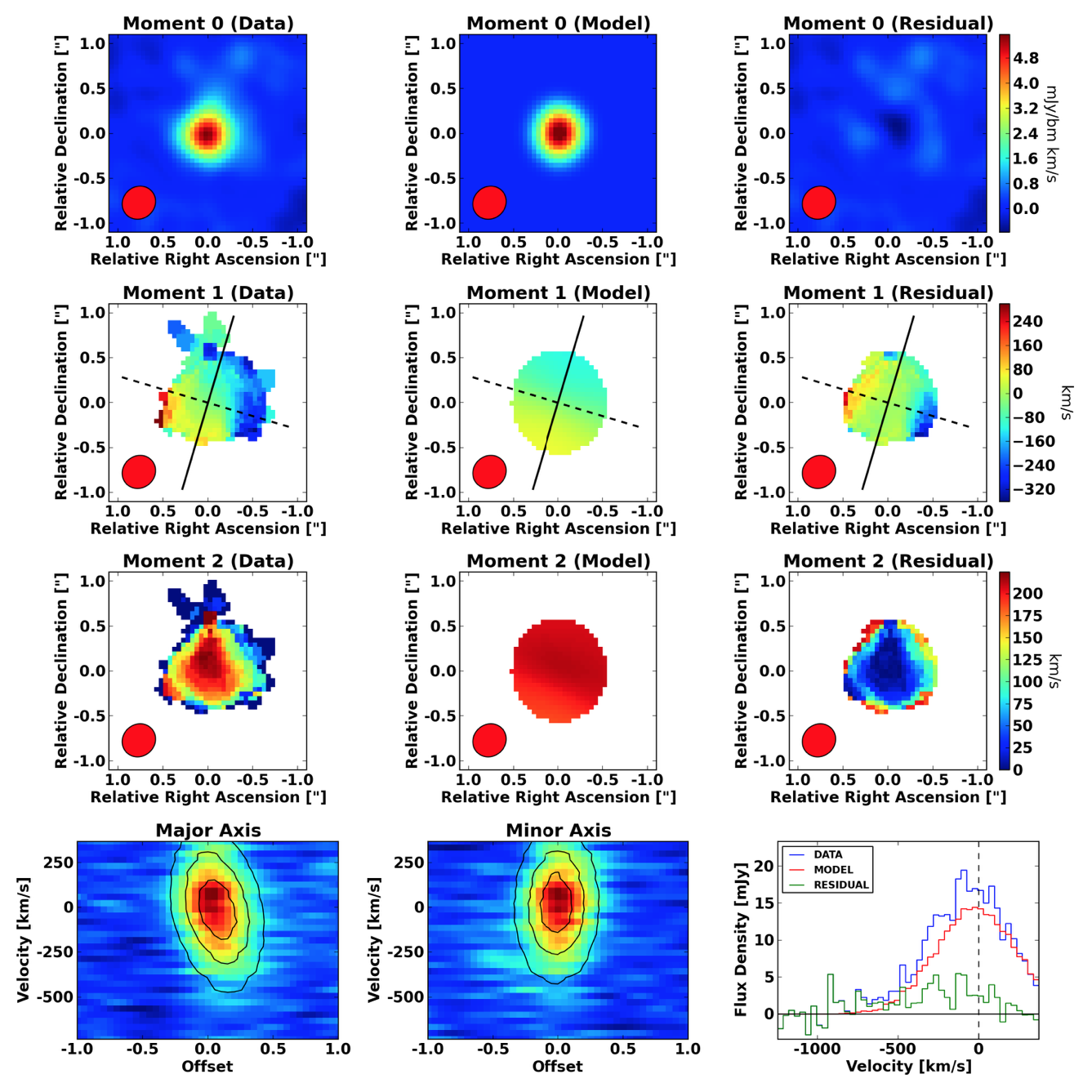}
\caption{Same as Figure~\ref{f:W0220barolo} but for W2246--0526.}
\label{f:W2246barolo}
\end{figure*}

The final outputs of this process are morphological parameters (e.g., central position, integrated flux) and kinematic parameters (e.g, inclination, position angle, velocity dispersion and rotation curves, systemic velocity/redshift). We use these outputs to generate two estimates of the dynamical mass of the system. First, we consider a rotation-dominated system:

\begin{equation}
M_{dyn}^{rot}=\frac{v^2r}{G}
\label{e:mdynrot}
\end{equation}

\noindent
where $r$ and $v$ are the radius and velocity of the outermost ring, respectively, and $G$ is the gravitational constant. Second, we consider a dispersion-dominated system:

\begin{equation}
M_{dyn}^{disp}=\alpha\frac{\sigma_v^2r}{G}
\label{e:mdyndisp}
\end{equation}

\noindent
where $\sigma_v$ is the velocity dispersion of the outermost ring and $\alpha$ is a constant which typically takes a range of values from 2.5 to 8.5 depending on geometry. Here we follow \cite{Stott2016} and adopt a value of $\alpha$\,=\,3.4. From Figures~\ref{f:W0220barolo} and \ref{f:W2246barolo} it is clear that the two systems analyzed here seem to be dispersion-dominated and therefore we use equation~\ref{e:mdyndisp}2 to estimate their dynamical masses.

As shown in Figure~\ref{f:W0220barolo}, the model is able to reproduce the kinematic properties of W0220+0137 to an excellent level, with very small moment 0 residuals. The only significant residuals appear in the moment 1 and 2 maps at the outskirts of the galaxy, where the SNR is low. The spectral residuals (bottom-right panel) are small as well. The galaxy has a well-defined rotation (moment 1 model panel), with circular velocities that range between [--100, 300]\,\kmns. Despite the strong coherent rotation, the velocity dispersion is still large, with a peak of around FWHM\,$\simeq$\,500\,\kmns. The morphological and kinematic position angles agree (41\,$\pm$\,16\deg\, and 43\,$\pm$\,15\deg, respectively), as well as the inclination on the plane of the sky (51\,$\pm$\,14\deg\, and 48\,$\pm$\,13\deg, also respectively).
\smallskip

Figure~\ref{f:W2246barolo} shows the modeling results for W2246--0526. The intrinsic kinematics of the galaxy are not well recovered by the model, as evidenced by the large residuals in all moment maps, including the moment 0, where the model cannot reach the peak intensity of the spectrum (bottom-right panel). The primary evidence for rotation is only visible in the position-velocity diagram along the visual major axis of the galaxy, and the velocity gradient, clear in the moment 1 map, is not captured in the modeled rotation. The model, instead, seems to favor the existence of a large intrinsic velocity dispersion over the entire galaxy of FWHM\,$\simeq$\,500\,\kmns, in agreement with the results from \cite{DS2016}. The inclination of the galaxy is constrained between 31\,$\pm$\,17\deg\, and 47\,$\pm$\,17\deg, and the position angle between 150\,$\pm$\,50\deg\, and 163\,$\pm$\,8\deg, corresponding to the morphological and kinematic estimates, respectively.\\
\\

\section{SED modeling}\label{a:sedfit}

This section describes the methodology used to infer the star formation rate (SFR) and stellar masses (\Mstar) of the quasar-host galaxies used in Section~\ref{s:ms} to place the EL Hot DOGs in context with the MS of star-forming galaxies at the same redshifts. The analysis starts with the construction of the spectral energy distributions (SEDs) of galaxies from the optical to the mid-IR, based on HST, ground-based optical, \textit{Spitzer} and \textit{WISE} broad-band imaging. The SED coverage depends on the availability of photometric data for each source. Details about the telescopes and filters are provided in Table~\ref{t:seds}, together with measurements and associated uncertainties. In all cases we collected the photometry with the smallest aperture available to avoid including emission from the neighbor galaxies. While the angular resolution of HST allows us to easily separate all companions, the \textit{Spitzer} and \textit{WISE} angular resolutions are coarser, and thus the fluxes could be contaminated. We use the AllWISE catalog to investigate the mid-IR WISE photometry, which is extracted via a point-spread-function (PSF) fitting procedure that also provides the goodness of fit of the model. In all WISE bands the galaxies in the sample are unresolved and no deblending was needed. Therefore, any contamination of extended structures beyond 6\arcsec\, to the mid-IR fluxes is negligible. Regarding the near-IR bands, the Spitzer photometry was obtained with a 6\arcsec-diameter circular aperture using Sextractor \citep{Griffith2012}. While Sextractor performs a deblending process in crowded fields, extended emission could potentially contaminate the photometry. We note, however, that no other source of emission is evident in the IRAC imaging within the apertures. Minding these caveats, errors associated to uncertainties in the photometry are propagated via the Monte Carlo analysis performed to estimate the uncertainty in the $H$-band photometry of the host (see below).

\begin{sidewaystable}
\caption{Photometric data used for the SED fitting}
\centering
\scriptsize
\label{t:seds}
\begin{tabular}{ccccccccccccccc}
\hline\hline
Galaxy      &       \multicolumn{2}{c}{$J$}       &           \multicolumn{2}{c}{$H$}   &      \multicolumn{2}{c}{$K_s$}        &        \multicolumn{2}{c}{IRAC1}      &       \multicolumn{2}{c}{IRAC2}       &       \multicolumn{2}{c}{W3}       &      \multicolumn{2}{c}{W4}        \\ 
(1)         & \multicolumn{2}{c}{(2)} &           \multicolumn{2}{c}{(3)}   &   \multicolumn{2}{c}{(4)}           &  \multicolumn{2}{c}{(5)}            &  \multicolumn{2}{c}{(6)}            &  \multicolumn{2}{c}{(7)}            &  \multicolumn{2}{c}{(8)}            \\
\cline{2-15}
            & $\lambda_{\rm eff}$ & f$_{\nu}$ [Jy] & $\lambda_{\rm eff}$ & f$_{\nu}$ [Jy] & $\lambda_{\rm eff}$ & f$_{\nu}$ [Jy] & $\lambda_{\rm eff}$ & f$_{\nu}$ [Jy] & $\lambda_{\rm eff}$ & f$_{\nu}$ [Jy] & $\lambda_{\rm eff}$ & f$_{\nu}$ [Jy] & $\lambda_{\rm eff}$ & f$_{\nu}$ [Jy] \\
            & [$\mu$m] & $\times$10$^{-5}$         & [$\mu$m] & $\times$10$^{-5}$         & [$\mu$m] & $\times$10$^{-5}$         & [$\mu$m] & $\times$10$^{-5}$         & [$\mu$m] & $\times$10$^{-5}$         & [$\mu$m] & $\times$10$^{-3}$         & [$\mu$m] & $\times$10$^{-2}$         \\
\hline
W0116--0505 & \dots    & \dots                    & \dots    & \dots                    & \dots    & \dots                    & 0.847    & 5.288\,$\pm$\,0.264      & 1.072    & 9.206\,$\pm$\,0.460      & 2.521    & 2.792\,$\pm$\,0.152      & 5.230    & 1.193\,$\pm$\,0.092      \\ 
W0134--2922 & 0.303    & 1.028\,$\pm$\,0.051      & \dots    & \dots                    & \dots    & \dots                    & \dots    & \dots                    & 1.120    & 9.089\,$\pm$\,1.356      & 2.604    & 5.315\,$\pm$\,0.266      & 5.401    & 1.789\,$\pm$\,0.097      \\
W0220+0137  & 0.297    & 0.788\,$\pm$\,0.109      & \dots    & \dots                    & 0.520    & 2.444\,$\pm$\,0.263      & 0.859    & 2.262\,$\pm$\,0.190      & 1.098    & 4.877\,$\pm$\,1.029      & 2.555    & 2.087\,$\pm$\,0.113      & 5.299    & 1.103\,$\pm$\,0.074      \\
W0236+0528  & \dots    & \dots                    & 0.414    & 1.152\,$\pm$\,0.221      & \dots    & \dots                    & 0.898    & 2.614\,$\pm$\,0.188      & 1.136    & 5.988\,$\pm$\,0.299      & 2.671    & 2.673\,$\pm$\,0.165      & 5.540    & 1.114\,$\pm$\,0.086      \\
W0410--0913 & \dots    & \dots                    & \dots    & \dots                    & 0.465    & 4.488\,$\pm$\,1.373      & 0.767    & 2.724\,$\pm$\,0.243      & 0.970    & 5.101\,$\pm$\,0.255      & 2.282    & 2.870\,$\pm$\,0.167      & 4.733    & 1.137\,$\pm$\,0.091      \\
W0831+0140  & \dots    & \dots                    & 0.334    & 0.921\,$\pm$\,0.199      & 0.438    & 2.675\,$\pm$\,0.538      & 0.670   & 1.309\,$\pm$\,0.643      & 0.925    & 5.625\,$\pm$\,1.368      & 2.150    & 3.232\,$\pm$\,0.162      & 4.460    & 1.089\,$\pm$\,0.161      \\
W2246--0526 & \dots    & \dots                    & \dots    & \dots                    & 0.384    & 0.896\,$\pm$\,0.283      & 0.634    & 3.778\,$\pm$\,0.191      & 0.802    & 3.291\,$\pm$\,0.165      & 1.886    & 2.447\,$\pm$\,0.171      & 3.913    & 1.326\,$\pm$\,0.171      \\
\hline\hline
\end{tabular}
\tablefoot{\scriptsize (1) Galaxy name; (2) $J$-band flux density: SOAR/OSIRIS (W0134--2922) and MMT/SWIRC (W0220+0137); (3) $H$-band flux density: SOAR/OSIRIS (W0236+0528) and WIYN/WHIRC (W0831+0140); (4) $K_s$-band density: P200/WIRC (W0220+0137), WIYN/WHIRC (W0410-0913), WIYN/WHIRC (W0831+0140) and P200/WIRC (W2246-0526); (5) \textit{Spitzer}/IRAC1 flux density; (6) \textit{Spitzer}/IRAC2 flux density; (7) \textit{WISE}/W3 flux density; (8) \textit{WISE}/W4 flux density. Effective wavelengths are in rest-frame. Flux densities are in observed-frame.}
\end{sidewaystable}

\begin{figure*}
\hspace{0.5cm}\includegraphics[scale=0.395]{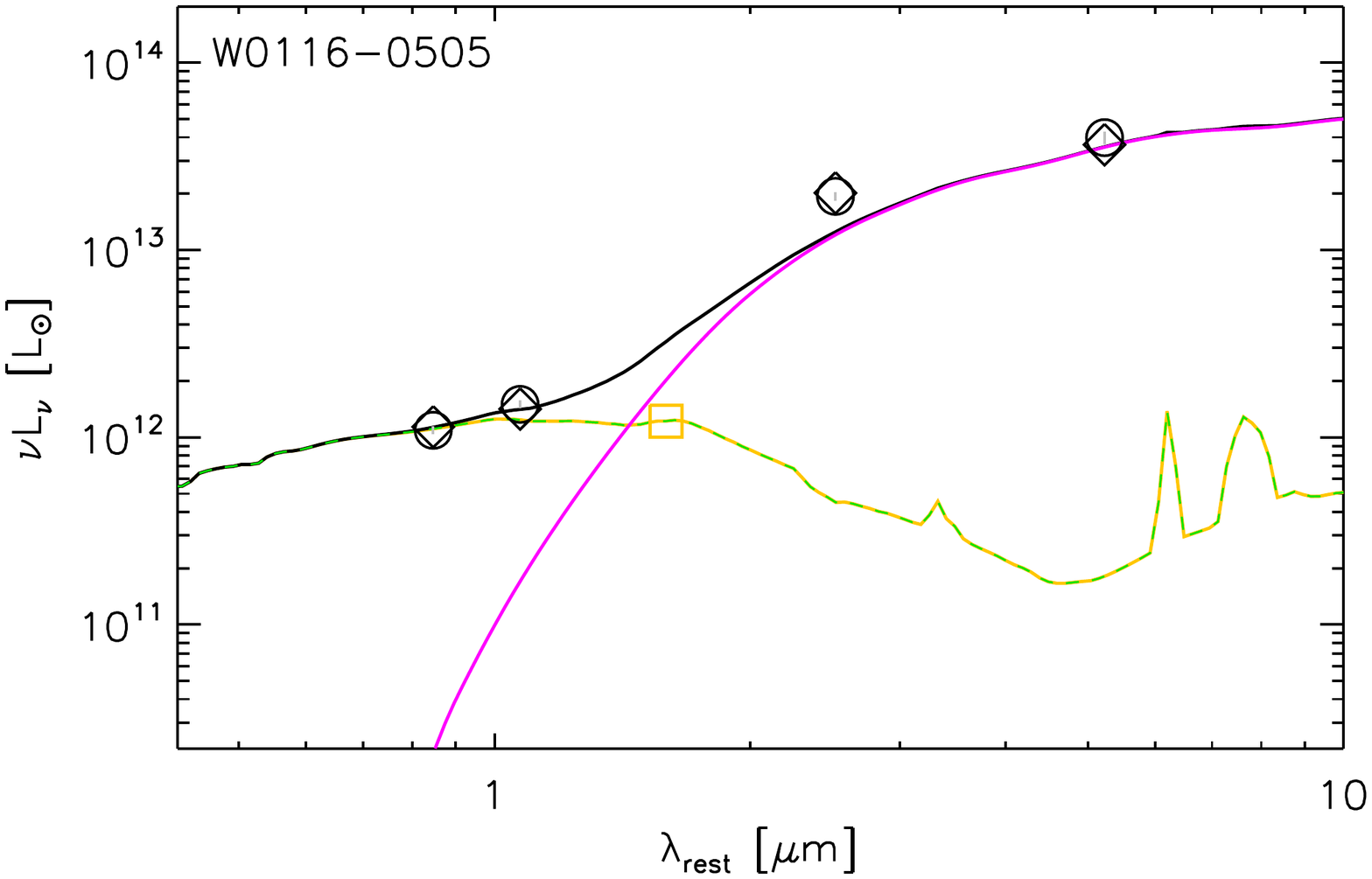}
\includegraphics[scale=0.395]{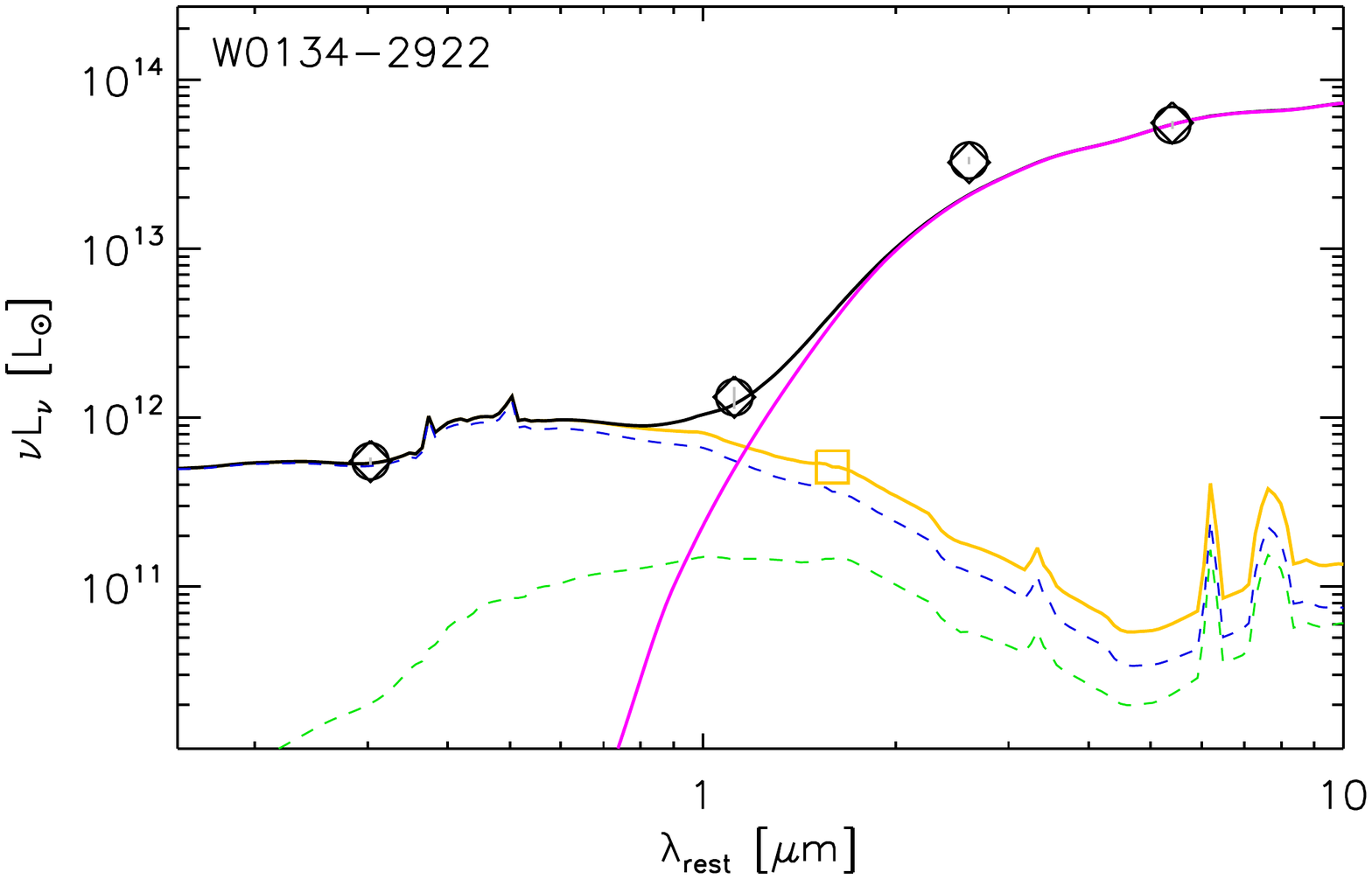}
\vspace{-5.5cm}

\hspace{0.5cm}\includegraphics[scale=0.395]{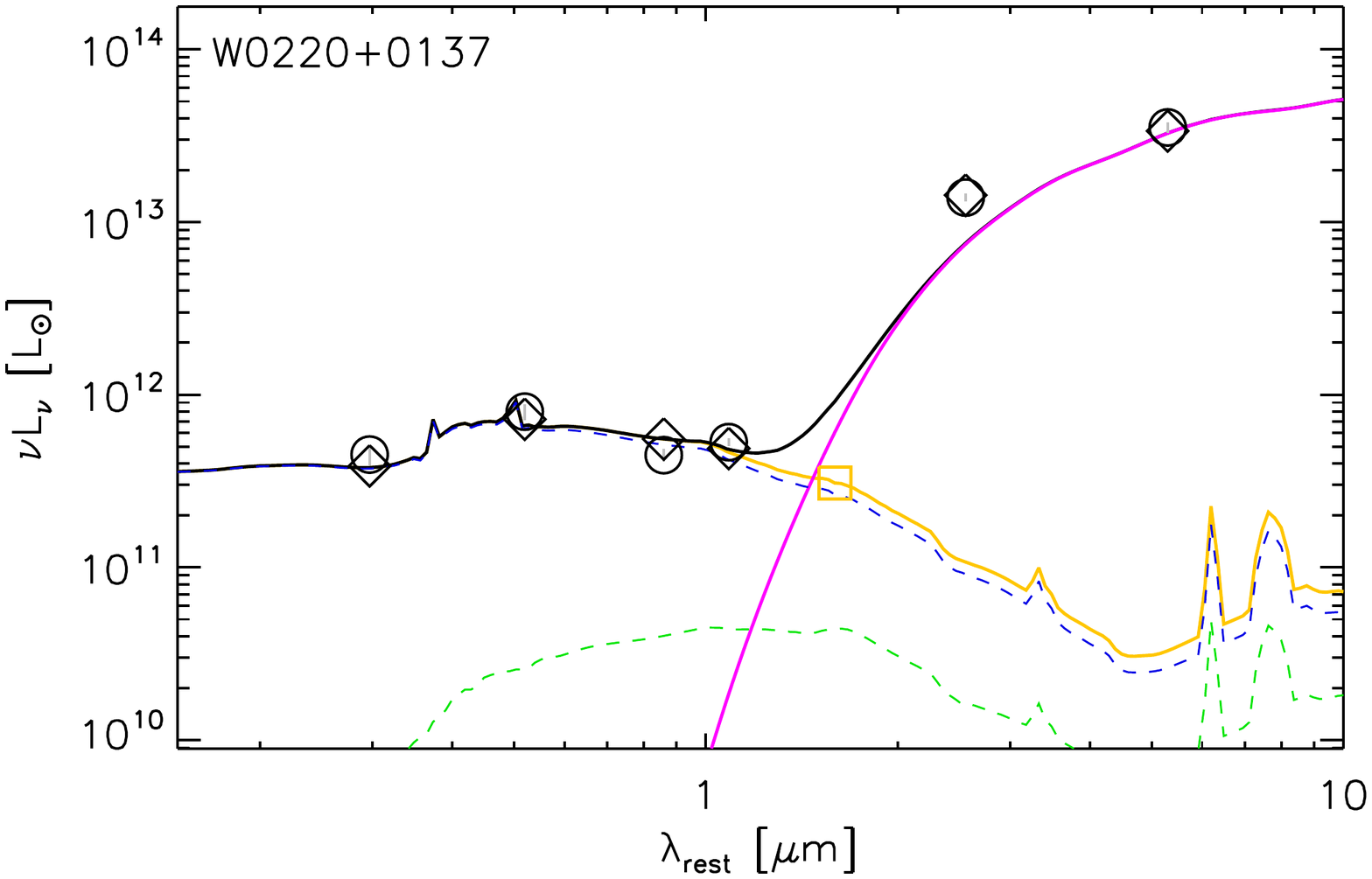}
\includegraphics[scale=0.395]{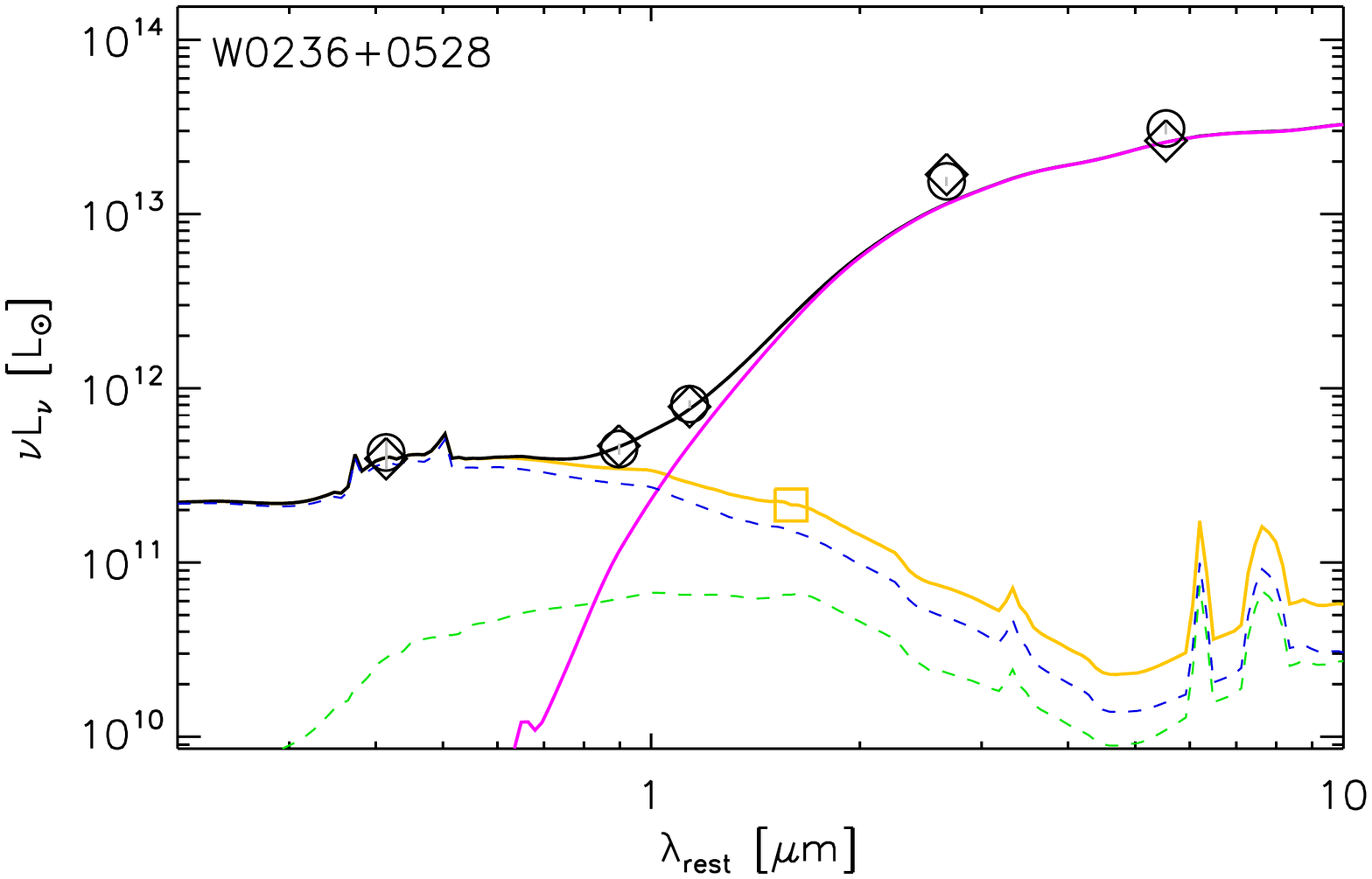}
\vspace{-5.5cm}

\hspace{0.5cm}\includegraphics[scale=0.395]{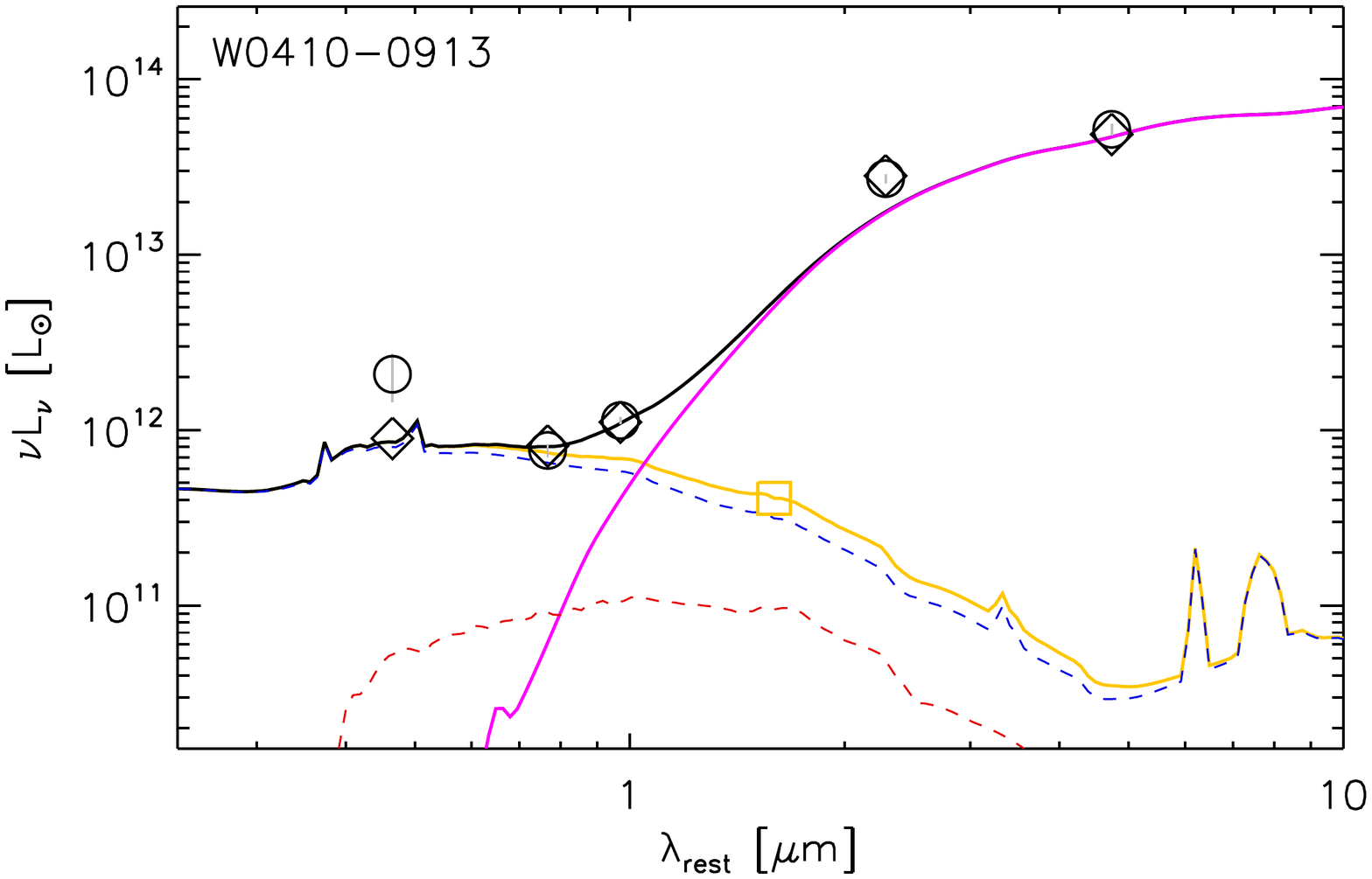}
\includegraphics[scale=0.395]{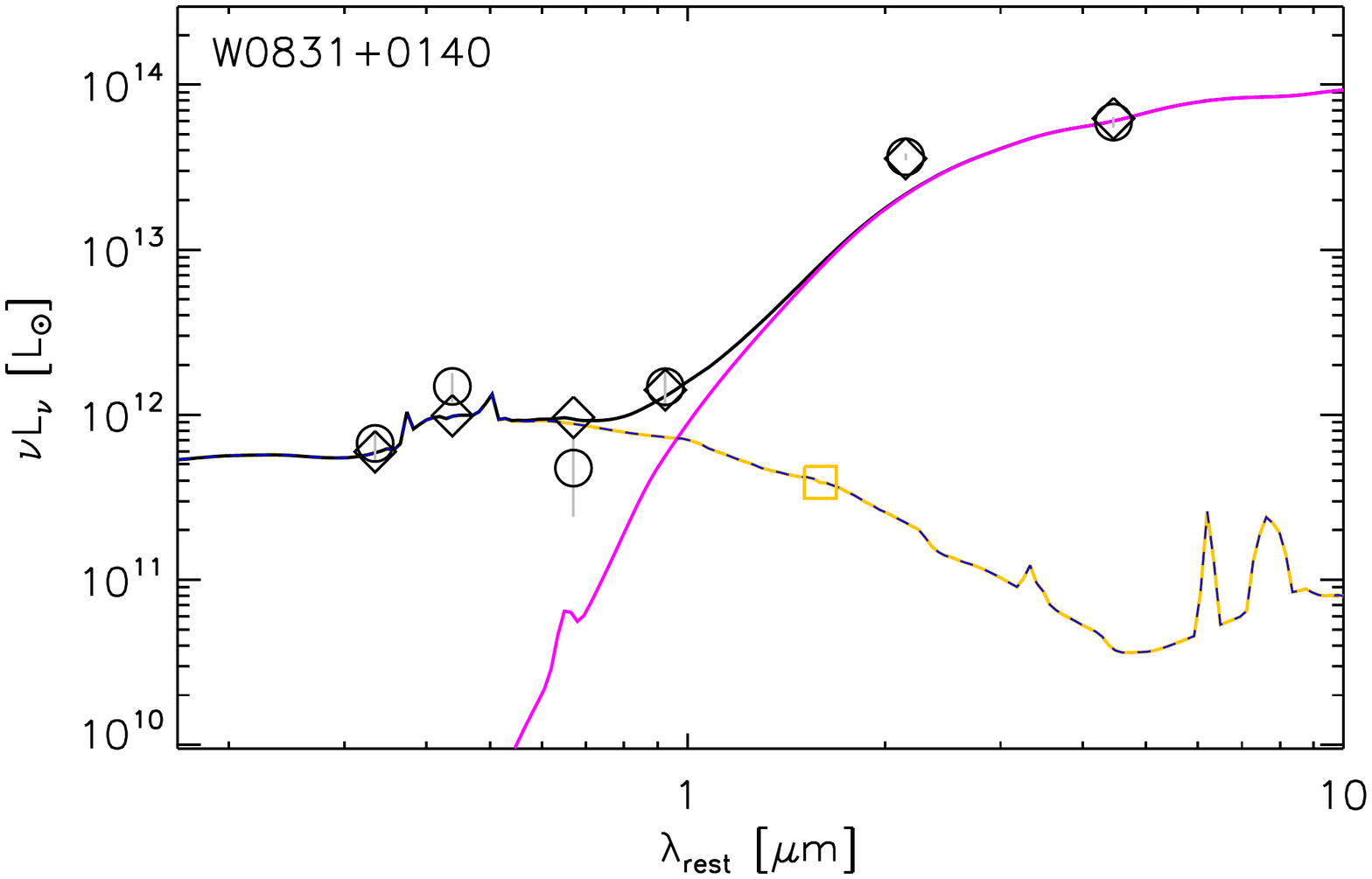}
\vspace{-5.5cm}

\hspace{4.5cm}\includegraphics[scale=0.395]{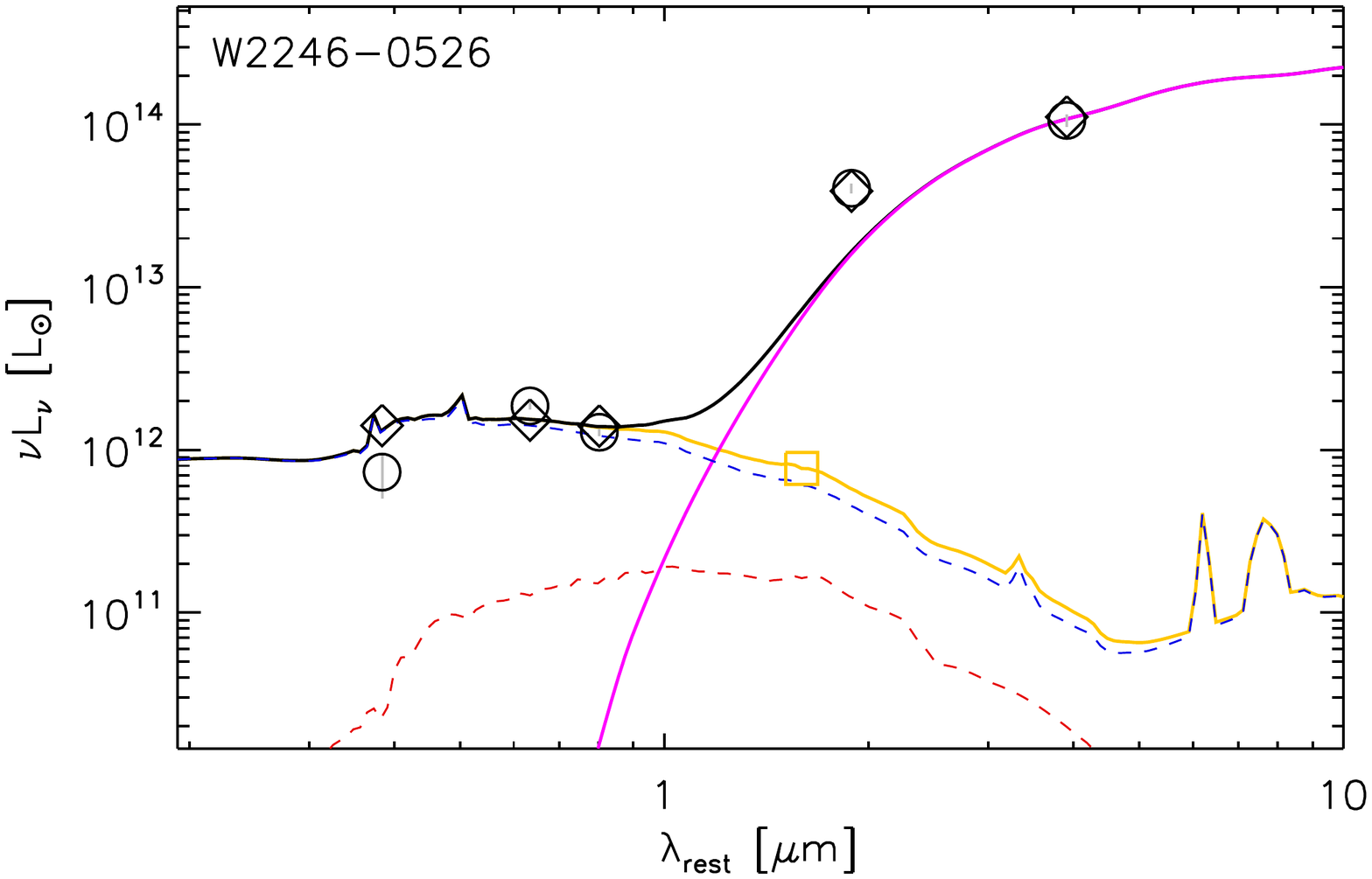}
\vspace{-5.5cm}
\caption{\footnotesize Results from the SED fitting. The photometric data of the sources are shown as black open circles. The AGN component is shown as a pink, solid line, while the E, Sbc and Im galaxy templates are displayed as red, green and blue dashed lines, respectively. The SED of the host galaxy (E\,+\,Sbc\,+\,Im) is shown as a yellow solid line, while the total SED of the source (including the AGN component) is displayed as a black solid line, with the synthetic photometry obtained from it (filter-convolved, color-corrected) shown as black open diamonds. The $H$-band flux of the host, used to obtain its stellar mass via scaling with a [M/L]$_{H}$ ratio (see text), is shown as a yellowe open square.}
\label{f:seds}
\end{figure*}

We follow the approach of \cite{Assef2015} and use the low spectral resolution templates from \cite{Assef2010}, spanning the wavelength range from 0.03 to 30\,$\mu$m, to fit the photometric SEDs of the Hot DOGs and obtain an estimate of the host galaxy luminosity after removing the AGN contribution. There are four empirical templates in the suite: one used to fit the AGN component and three representative of galaxy hosts dominated by old (E template), intermediate (Sbc template), and young (Im template) stellar populations. The three galaxy templates allow for enough spectral versatility such that it is not necessary to account for additional dust extinction at the time of fitting, other than the intrinsic reddening the templates already have. On the other hand, the AGN template is allowed to have additional obscuration, which is modeled using a dust attenuation law that is a combination of a Small Magellanic Cloud (SMC) reddening law at shorter wavelengths and a Milky Way reddening law at longer wavelengths. Intergalactic medium absorption is also considered. Figure~\ref{f:seds} shows the best-fit SED model for each Hot DOG in the sample. The galaxy hosts dominate the rest-frame optical wavelengths in most cases, while the AGN becomes relevant above $\gtrsim$\,1\,$\mu$m. We note that, although broad and narrow optical emission lines are detected in the spectra of many Hot DOGs, the photometry is unlikely to be severely biased by these features, as the equivalent widths of rest-frame optical emission lines in Hot DOGs are small \citep{Jun2020a, Finnerty2020} compared to the width of the broad-band filters. Using \textit{HST} imaging, \cite{Assef2020} has recently identified a few Hot DOGs displaying a rather blue UV/optical slope, most noticeable at $\lambda_{\rm rest}$\,$\lesssim$\,3000\AA. The origin of this excess blue light is likely emission from the accretion disk of the central SMBH, which has been scattered into our line of sight by gas or dust. Two of the sources in the study of \cite{Assef2020} are in the Hot DOG sample analyzed in this work (W0116--0505 and W0220+0137). The contribution of this component to the rest-frame near-IR emission is, however, small ($\lesssim$\,10\% at $H$-band; see their Fig. 1) and therefore has a limited impact on the estimation of the stellar masses of Hot DOGs given the significantly larger uncertainties associated to the SED modeling assumptions.

After the best SED fit is found, the three galaxy templates are added together to obtain the rest-frame $H$-band flux density of the host, free from AGN contamination. In order to estimate the uncertainty associated with this value, we run 1000 Monte Carlo simulations for each Hot DOG, sampling from the errors of the photometric data-points assuming they follow a normal distribution, and calculate the $H$-band flux density for each instance. The uncertainty is obtained as the standard deviation of the simulated results. Because the galaxy (and AGN) templates are purely empirical, the SED fitting procedure does not provide information about the SFR history (SFRH) or any other intrinsic physical property of the stellar populations of the source. Therefore, in order to obtain the stellar masses of the host galaxies, an additional step is required; that is, a [M/L]$_{H}$ ratio. To derive it, we use the code \textit{ezgal} \citep{Mancone2012} and the Galaxev stellar population SED models \citep{Bruzual2003}, which include a contribution from thermally pulsing-asymptotic giant branch (TP-AGB) stars, with a metallicity Z\,=\,0.008 ($\equiv$\,0.4\,\Zsun, where \Zsun\, is the metallicity of the Sun) and a Chabrier intial mass function (IMF) \citep{Chabrier2003}. We consider exponentially declining SFRHs that depend only on two parameters: the time of formation, $t_{\rm f}$, defined with respect to the current age of the galaxy, $t$, where $\Delta t$\,=\,$t-t_{\rm f}$, and the characteristic decaying time-scale of the SFR, $\tau$. The SFRH is then parametrized as SFR($t^\prime)$\,$\propto$\,e$^{-(t^\prime-t_{\rm f})/\tau}$ with $t^\prime$\,$\in$\,[$t_{\rm f}$, $t$]. The integral of the SFRH provides the \Mstar\, of the model together with the resulting SED, from which the [M/L]$_{H}$ is derived. In turn, the scaling with the data yields the normalization factor with which to obtain an absolute value of SFR($t$).

Because it is extremely difficult to infer any information regarding the SFRH of galaxies based on sparse photometric data, for this work we decide to consider four simple limiting cases that are selected to represent reasonable lower and upper boundaries for $\Delta t$ and $\tau$, given the redshifts of the Hot DOGs, which correspond to between $t$\,$\simeq$\,1.3 and 2.2\,Gyr after the Big Bang. Specifically, we assume that star formation began $\Delta t$\,=\,\{500, 100\}\,Myr before the age of each galaxy, $t$, and two time-scales, of $\tau$\,=\,\{0.1, 1\}\,Gyr. Note that the combination of $\Delta t$\,=\,500\,Myr and $\tau$\,=\,0.1\,Gyr closely mimics a past starburst event, while the combination of $\Delta t$\,=\,100\,Myr and $\tau$\,=\,1\,Gyr is equivalent to having an almost constant SFR. The \Mstar\, values derived based on these four combination of parameters are used in Section~\ref{s:ms} to place the Hot DOGs along the x-axis in the star-formation main sequence (Figure~\ref{f:ms}). In turn, the SFRs obtained from these SFRHs and scalings are not used in the calculation of the average SFR of each source in order to avoid the intrinsic correlation existing between SFR and \Mstar\, for each assumed SFRH.

\bibliographystyle{./aasjournal}
\bibliography{./bib,compiled}{}

\end{document}